\documentclass[longauth]{aa}  

\usepackage{graphicx}
\usepackage{rotating}
\usepackage{txfonts}
\usepackage{booktabs}
\usepackage{tabularx}
\usepackage{array}
\usepackage{multirow}
\usepackage{orcidlink}
\usepackage{lscape}

\newcolumntype{P}[1]{>{\centering\arraybackslash}p{#1}}

\newcommand{\hii}{\ion{H}{ii}}
\usepackage{hyperref}
\begin{document}

   \title{ALMAGAL II. The ALMA evolutionary study of high-mass protocluster formation in the Galaxy}

   \subtitle{ALMA data processing and pipeline}

   \author{\'A.~S\'anchez-Monge\inst{\ref{icecsic}, \ref{ieec}}\orcidlink{0000-0002-3078-9482} \and
           C.~L.~Brogan\inst{\ref{nraoVA}}\orcidlink{0000-0002-6558-7653} \and
           T.~R.~Hunter\inst{\ref{nraoVA}}\orcidlink{0000-0001-6492-0090} \and
           A.~Ahmadi\inst{\ref{leiden}}\orcidlink{0000-0003-4037-5248} \and
           A.~Avison\inst{\ref{skaUK}, \ref{manchester}, \ref{almaUK}}\orcidlink{0000-0002-2562-8609} \and
           M.~T.~Beltr\'an\inst{\ref{arcetri}}\orcidlink{0000-0003-3315-5626} \and
           H.~Beuther\inst{\ref{mpia}}\orcidlink{0000-0002-1700-090X} \and
           A.~Coletta\inst{\ref{rome}, \ref{sapienza}}\orcidlink{0000-0001-8239-8304} \and
           G.~A.~Fuller\inst{\ref{manchester}, \ref{unikoeln}}\orcidlink{0000-0001-8509-1818} \and
           K.~G.~Johnston\inst{\ref{lincoln}}\orcidlink{0000-0003-4509-1180} \and
           B.~Jones\inst{\ref{unikoeln}}\orcidlink{0000-0002-0675-0078} \and
           S.-Y.~Liu\inst{\ref{asiaa}} \and
           C.~Mininni\inst{\ref{rome}}\orcidlink{0000-0002-2974-4703} \and
           S.~Molinari\inst{\ref{rome}}\orcidlink{0000-0002-9826-7525} \and
           P.~Schilke\inst{\ref{unikoeln}}\orcidlink{0000-0002-1730-8832} \and
           E.~Schisano\inst{\ref{rome}}\orcidlink{0000-0003-1560-3958} \and
           Y.-N.~Su\inst{\ref{asiaa}} \and
           A.~Traficante\inst{\ref{rome}}\orcidlink{0000-0003-1665-6402} \and
           Q.~Zhang\inst{\ref{cfa}}\orcidlink{0000-0003-2384-6589} \and
           C.~Battersby\inst{\ref{connecticut}}\orcidlink{0000-0002-6073-9320} \and
           M.~Benedettini\inst{\ref{rome}}\orcidlink{0000-0002-0560-3172} \and
           D.~Elia\inst{\ref{rome}}\orcidlink{0000-0002-9120-5890} \and
           P.~T.~P.~Ho\inst{\ref{asiaa}, \ref{hawaii}}\orcidlink{0000-0002-3412-4306} \and
           P.~D.~Klaassen\inst{\ref{edinburgh}}\orcidlink{0000-0001-9443-0463} \and
           R.~S.~Klessen\inst{\ref{ita}, \ref{uniheidelberg}, \ref{cfa}, \ref{radcliffe}}\orcidlink{0000-0002-0560-3172} \and
           C.~Y.~Law\inst{\ref{arcetri}} \and
           D.~C.~Lis\inst{\ref{jplcaltech}}\orcidlink{0000-0002-0500-4700} \and
           T.~Liu\inst{\ref{shanghai}} \and
           L.~Maud\inst{\ref{eso}}\orcidlink{0000-0002-7675-3565} \and
           T.~M\"oller\inst{\ref{unikoeln}}\orcidlink{0000-0002-9277-8025} \and
           L.~Moscadelli\inst{\ref{arcetri}}\orcidlink{0000-0002-8517-8881} \and
           S.~Pezzuto\inst{\ref{rome}}\orcidlink{0000-0001-7852-1971} \and
           K.~L.~J.~Rygl\inst{\ref{bologna}}\orcidlink{0000-0003-4146-9043} \and
           P.~Sanhueza\inst{\ref{meguro}, \ref{naoj}}\orcidlink{0000-0002-7125-7685} \and
           J.~D.~Soler\inst{\ref{rome}}\orcidlink{0000-0002-0294-4465} \and
           G.~Stroud\inst{\ref{manchester}}\orcidlink{0000-0002-4935-2416} \and
           Y.~Tang\inst{\ref{asiaa}} \and
           F.~F.~S.~van der Tak\inst{\ref{sron}, \ref{kapteyn}}\orcidlink{0000-0002-8942-1594} \and
           D.~L.~Walker\inst{\ref{manchester}} \and
           J.~Wallace\inst{\ref{connecticut}} \and
           S.~Walch\inst{\ref{unikoeln}, \ref{datacologne}} \and
           M.~R.~A.~Wells\inst{\ref{mpia}}\orcidlink{0000-0002-3643-5554} \and
           F.~Wyrowski\inst{\ref{mpifr}} \and
           T.~Zhang\inst{\ref{zhejiang}, \ref{unikoeln}}\orcidlink{0000-0002-1466-3484} \and
           J.~Allande\inst{\ref{arcetri}, \ref{unifirenze}} \and
           L.~Bronfman\inst{\ref{unichile}}\orcidlink{0000-0002-9574-8454} \and
           E.~Dann\inst{\ref{unikoeln}, \ref{mpifr}} \and
           F.~De~Angelis\inst{\ref{rome}}\orcidlink{0009-0002-6765-7413} \and
           F.~Fontani\inst{\ref{arcetri}, \ref{mpe}, \ref{meudon}}\orcidlink{0000-0003-0348-3418} \and
           Th.~Henning\inst{\ref{mpia}} \and
           W.-J.~Kim\inst{\ref{unikoeln}}\orcidlink{0000-0003-0364-6715} \and
           R.~Kuiper\inst{\ref{duisburg}}\orcidlink{0000-0003-2309-8963} \and
           M.~Merello\inst{\ref{unichile}, \ref{pontificia}}\orcidlink{0000-0003-0709-708X} \and
           F.~Nakamura\inst{\ref{naoj}, \ref{sokendai}, \ref{todai}}\orcidlink{0000-0001-5431-2294} \and
           A.~Nucara\inst{\ref{rome}, \ref{torvergata}}\orcidlink{0009-0005-9192-5491} \and
           A.~J.~Rigby\inst{\ref{leeds}}\orcidlink{0000-0002-3351-2200}}

   \institute{
           \label{icecsic}Institut de Ci\`encies de l'Espai (ICE), CSIC, Campus UAB, Carrer de Can Magrans s/n, E-08193, Bellaterra (Barcelona), Spain\\ \email{asanchez@ice.csic.es}
           \and
           \label{ieec}Institut d'Estudis Espacials de Catalunya (IEEC), E-08860, Castelldefels (Barcelona), Spain
           \and
           \label{nraoVA}National Radio Astronomy Observatory, 520 Edgemont Road, Charlottesville VA 22903, USA
           \and
           \label{leiden}Leiden Observatory, Leiden University, PO Box 9513, 2300 RA Leiden, The Netherlands
           \and
           \label{skaUK}SKA Observatory, Jodrell Bank, Lower Withington, Macclesfield, SK11 9FT, UK
           \and
           \label{manchester}Jodrell Bank Centre for Astrophysics, Oxford Road, The University of Manchester, Manchester M13 9PL, UK
           \and
           \label{almaUK}UK ALMA Regional Centre Node, M13 9PL, UK
           \and
           \label{arcetri}INAF-Osservatorio Astrofisico di Arcetri, Largo E.\ Fermi 5, I-50125 Firenze, Italy
           \and
           \label{mpia}Max Planck Institute for Astronomy, K\"onigstuhl 17, D-69117 Heidelberg, Germany
           \and
           \label{rome}INAF-Istituto di Astrofisica e Planetologia Spaziale, Via Fosso del Cavaliere 100, I-00133 Roma, Italy
           \and
           \label{sapienza}Dipartimento di Fisica, Sapienza Universit\`a di Roma, Piazzale Aldo Moro 2, I-00185, Rome, Italy
           \and
           \label{unikoeln}I.\ Physikalisches Institut, Universit\"{a}t zu K\"{o}ln, Z\"{u}lpicher Str.\ 77, D-50937 K\"{o}ln, Germany
           \and
           \label{lincoln}School of Engineering and Physical Sciences, Isaac Newton Building, University of Lincoln, Brayford Pool, Lincoln, LN6 7TS, United Kingdom
           \and
           \label{asiaa}Institute of Astronomy and Astrophysics, Academia Sinica, 11F of ASMAB, AS/NTU No.\ 1, Sec.\ 4, Roosevelt Road, Taipei 10617, Taiwan
           \and
           \label{cfa}Center for Astrophysics $|$ Harvard \& Smithsonian, 60 Garden St, Cambridge, MA 02138, USA
           \and
           \label{connecticut}University of Connecticut, Department of Physics, 2152 Hillside Road, Unit 3046 Storrs, CT 06269, USA
           \and
           \label{hawaii}East Asian Observatory, 660 N.\ A'ohoku, Hilo, Hawaii, HI 96720, USA
           \and
           \label{edinburgh}UK Astronomy Technology Centre, Royal Observatory Edinburgh, Blackford Hill, Edinburgh EH9 3HJ, UK
           \and
           \label{ita}Universit\"at Heidelberg, Zentrum f\"ur Astronomie, Institut f\"ur Theoretische Astrophysik, Heidelberg, Germany
           \and
           \label{uniheidelberg}Universit\"at Heidelberg, Interdisziplin\"ares Zentrum f\"ur Wissenschaftliches Rechnen, Heidelberg, Germany
           \and
           \label{radcliffe}Elizabeth S.\ and Richard M.\ Cashin Fellow at the Radcliffe Institute for Advanced Studies at Harvard University, 10 Garden Street, Cambridge, MA 02138, USA
           \and
           \label{jplcaltech}Jet Propulsion Laboratory, California Institute of Technology, 4800 Oak Grove Drive, Pasadena, CA 91109, USA
           \and
           \label{shanghai}Shanghai Astronomical Observatory, Chinese Academy of Sciences, Shanghai 200030, People's Republic of China
           \and
           \label{eso}European Southern Observatory, Karl-Schwarzschild Str.\ 2, 85748 Garching bei M\"unchen, Germany
           \and
           \label{bologna}INAF-Istituto di Radioastronomia \& Italian ALMA Regional Centre, Via P. Gobetti 101, I-40129 Bologna, Italy
           \and
           \label{meguro}Department of Earth and Planetary Sciences, Institute of Science Tokyo, Meguro, Tokyo, 152-8551, Japan
           \and
           \label{naoj}National Astronomical Observatory of Japan, National Institutes of Natural Sciences, 2-21-1 Osawa, Mitaka, Tokyo 181-8588, Japan
           \and
           \label{sron}SRON Netherlands Institute for Space Research, Landleven 12, 9747 AD Groningen, The Netherlands
           \and
           \label{kapteyn}Kapteyn Astronomical Institute, University of Groningen, The Netherlands
           \and
           \label{datacologne}Center for Data and Simulation Science, University of Cologne, Germany
           \and
           \label{mpifr}Max-Planck-Institut f\"ur Radioastronomie, Auf dem H\"ugel 69, D-53121 Bonn, Germany
           \and
           \label{zhejiang}Zhejiang Laboratory, Hangzhou 311100, P.R.China
           \and
           \label{unifirenze}Dipartimento di Fisica e Astronomia, Universit\`a degli Studi di Firenze, Via G.\ Sansone 1,I-50019 Sesto Fiorentino, Firenze, Italy
           \and
           \label{unichile}Departamento de Astronomía, Universidad de Chile, Casilla 36-D, Santiago, Chile
           \and
           \label{mpe}Max-Planck-Institute for Extraterrestrial Physics (MPE), Garching bei M\"unchen, Germany
           \and
           \label{meudon}Laboratory for the study of the Universe and eXtreme phenomena (LUX), Observatoire de Paris, Meudon, France
           \and
           \label{duisburg}Faculty of Physics, University of Duisburg-Essen, Lotharstra{\ss}e 1, D-47057 Duisburg, Germany
           \and
           \label{pontificia}Centro de Astro-Ingeniería (AIUC), Pontificia Universidad Católica de Chile, Av. Vicuña Mackena 4860, Macul, Santiago, Chile
           \and
           \label{sokendai}Department of Astronomical Science, SOKENDAI (The Graduate University for Advanced Studies), 2-21-1 Osawa, Mitaka, Tokyo 181-8588, Japan
           \and
           \label{todai}Department of Astronomy, Graduate School of Science, The University of Tokyo, 7-3-1 Hongo, Bunkyo-ku, Tokyo 113-0033, Japan
           \and
           \label{torvergata}Dipartimento di Fisica, Universit\`a di Roma Tor Vergata, Via della Ricerca Scientifica 1, I-00133 Roma, Italy
           \and
           \label{leeds}School of Physics and Astronomy, University of Leeds, Leeds LS2 9JT, UK
           }

   \date{Received yesterday; accepted tomorrow}
      
   \titlerunning{ALMAGAL data processing and pipeline}
   \authorrunning{\'A.\ S\'anchez-Monge et al.}

% \abstract{}{}{}{}{}
% 5 {} token are mandatory
 
\abstract
{Stars form preferentially in clusters embedded inside massive molecular clouds, many of which contain high-mass stars. Thus, a comprehensive understanding of star formation requires a robust and statistically well-constrained characterization of the formation and early evolution of these high-mass star clusters. To achieve this, we designed the ALMAGAL Large Program that observed 1017 high-mass star-forming regions distributed throughout the Galaxy, sampling different evolutionary stages and environmental conditions.}
{In this work, we present the acquisition and processing of the ALMAGAL data. The main goal is to set up a robust pipeline that generates science-ready products, that is, continuum and spectral cubes for each ALMAGAL field, with a good and uniform quality across the whole sample.}
{ALMAGAL observations were performed with the Atacama Large Millimeter/submillimeter Array (ALMA). Each field was observed in three different telescope arrays, being sensitive to spatial scales ranging from $\approx1000$~au up to $\approx0.1$~pc. The spectral setup allows sensitive ($\approx0.1$~mJy~beam$^{-1}$) imaging of the continuum emission at 219~GHz (or 1.38~mm), and it covers multiple molecular spectral lines observed in four different spectral windows that span about $\approx4$~GHz in frequency coverage. We have designed a Python-based processing workflow to calibrate and image these observational data. This ALMAGAL pipeline includes an improved continuum determination, suited for line-rich sources; an automatic self-calibration process that reduces phase-noise fluctuations and improves the dynamical range by up to a factor $\approx5$ in about 15\% of the fields; and the combination of data from different telescope arrays to produce science-ready, fully combined images.}
{The final products are a set of uniformly generated continuum images and spectral cubes for each ALMAGAL field, including individual-array and combined-array products.
The fully combined products have spatial resolutions in the range 800--2000~au, and mass sensitivities in the range 0.02--0.07~$M_\odot$. We also present a first analysis of the spectral line information included in the ALMAGAL setup, and its potential for future scientific studies. As an example, specific spectral lines (e.g., SiO and CH$_3$CN) at $\approx1000$~au scales resolve the presence of multiple outflows in clusters and will help us to search for disk candidates around massive protostars. Moreover, the broad frequency bands provide information on the chemical richness of the different cluster members, which can be used to study the chemical evolution during the formation process of star clusters.}
%A first look at the line richness reveals its potential correlation with different evolutionary indicators.}
{}

   \keywords{instrumentation: interferometers --
             methods: observational --
             ISM: clouds --
             stars: formation --
             stars: protostars --
             stars: massive
             }

   \maketitle
%
%________________________________________________________________

%________________________________________________________________
%
\section{Introduction\label{sec:intro}}

Stars are the fundamental building blocks of galaxies and hosts of planetary systems. As such, understanding their formation is key not only to characterize stars themselves, but also to unveil the properties and evolution of galaxies as well as planetary systems. In this context, a relevant aspect is that most stars do not form in isolation, but in groups and clusters \citep[e.g.,][]{LadaLada2003, Adams2010, Krumholz2014, Krause2020, ReiterParker2022}. Moreover, the majority ($\approx90$\%) of stars born in groups and clusters form in very rich clusters of more than 100 stars containing at least one high-mass star (i.e., a star with a mass larger than $\approx10$~$M_\odot$; \citealt{LadaLada2003}). The presence of high-mass stars is expected to have a huge impact on the surrounding environment and early development of the stars forming within the cluster (e.g., due to large levels of feedback: radiation, winds, outflows, explosions). Thus, a comprehensive understanding of star formation and galaxy evolution requires a detailed study of the formation of high-mass star clusters.

Stars and their associated clusters form out of large molecular clouds ($>1$~pc and $>500$~$M_\odot$) that must undergo a process of collapse and fragmentation \citep[e.g.,][]{Palau2013, Palau2014, Dobbs2014, Zhang2015, KlessenGlover2016, Beuther2018, Chevance2023, Traficante2023, Morii2024, SchinnererLeroy2024}. At the low temperatures ($\approx20$~K) and high densities ($>10^{3}$~cm$^{-3}$) of molecular clouds, the most efficient observational strategy to study the fragmentation process and formation of clusters is to characterize the thermal dust emission, which is bright at millimeter wavelengths, together with the emission of molecular species, abundant in the dense gas. For this, the Atacama Large Millimeter/submillimeter Array \citep[ALMA,][]{ALMA2015} has enabled new opportunities thanks to its enhanced sensitivity and resolution \citep[see e.g.,][]{Cesaroni2017, SanchezMonge2017, Sanhueza2019, Liu2020, Motte2022}. We have taken advantage of these new capabilities to design a large observing project. ALMAGAL is an ALMA Large Program that studies the formation and early evolution of high-mass protoclusters throughout the Galaxy. 

The primary goal of ALMAGAL is to study the properties of high-mass star-forming clusters in a statistically relevant way, characterizing clusters at different evolutionary stages, with different physical properties, and in different environments. At a spatial resolution of $\approx1000$~au and mass sensitivity $\lesssim0.1$~$M_\odot$, the dust continuum emission probes the conditions for high-mass star formation, and the properties of clusters such as spatial distribution and mass segregation \citep[e.g.,][]{Plunkett2018, Pavlik2019, Sadaghiani2020, Morii2023}. Moreover, observations of different molecular species help us to study the dense gas dynamics and temperature, provide a comprehensive catalogue of molecular outflows to study feedback processes, as well as disk candidates around high-mass young stellar objects, and enable studies of chemical evolution throughout the whole star-formation process \citep[e.g.,][]{Johnston2015, Giannetti2017, Bonfand2019, Yang2022, Ahmadi2023}. The ALMAGAL sample includes 1017 high-mass star-forming regions distributed throughout the Galaxy, covering a broad range of evolutionary stages (e.g., from quiescent, infrared dark clouds to evolved regions with already developed \hii\ regions powered by high-mass stars), physical properties (e.g., masses, densities, temperatures), and environments (e.g., inner and outer Galaxy, inside and outside spiral arms). A description of the main scientific goals of the ALMAGAL project is presented in \citet{Molinari2024}.

In this paper, we present the data reduction strategy used to produce the first data release of the ALMAGAL project. The paper is organized as follows. In Sect.~\ref{sec:obs}, we briefly introduce the ALMAGAL sample and describe the observations and data acquisition. In Sect.~\ref{sec:workflow}, we introduce the main steps of the ALMAGAL processing workflow. In Sect.~\ref{sec:pipeline}, we describe in detail each of the analysis steps, including calibration, identification of line-free channels for continuum determination, automatic self-calibration, and image combination of the different telescope arrays. We also include a description of the final data products, including the achieved sensitivities and resolution, and their quality assessment. Section~\ref{sec:lines} demonstrates some preliminary science applications, focusing on the spectral line content and the potential of the data. Finally, we summarize the main aspects of this work in Sect.~\ref{sec:summary}.

%-------------------------------- Figure --------------------------------
\begin{figure}[t!]
\centering
\includegraphics[width=0.98\columnwidth]{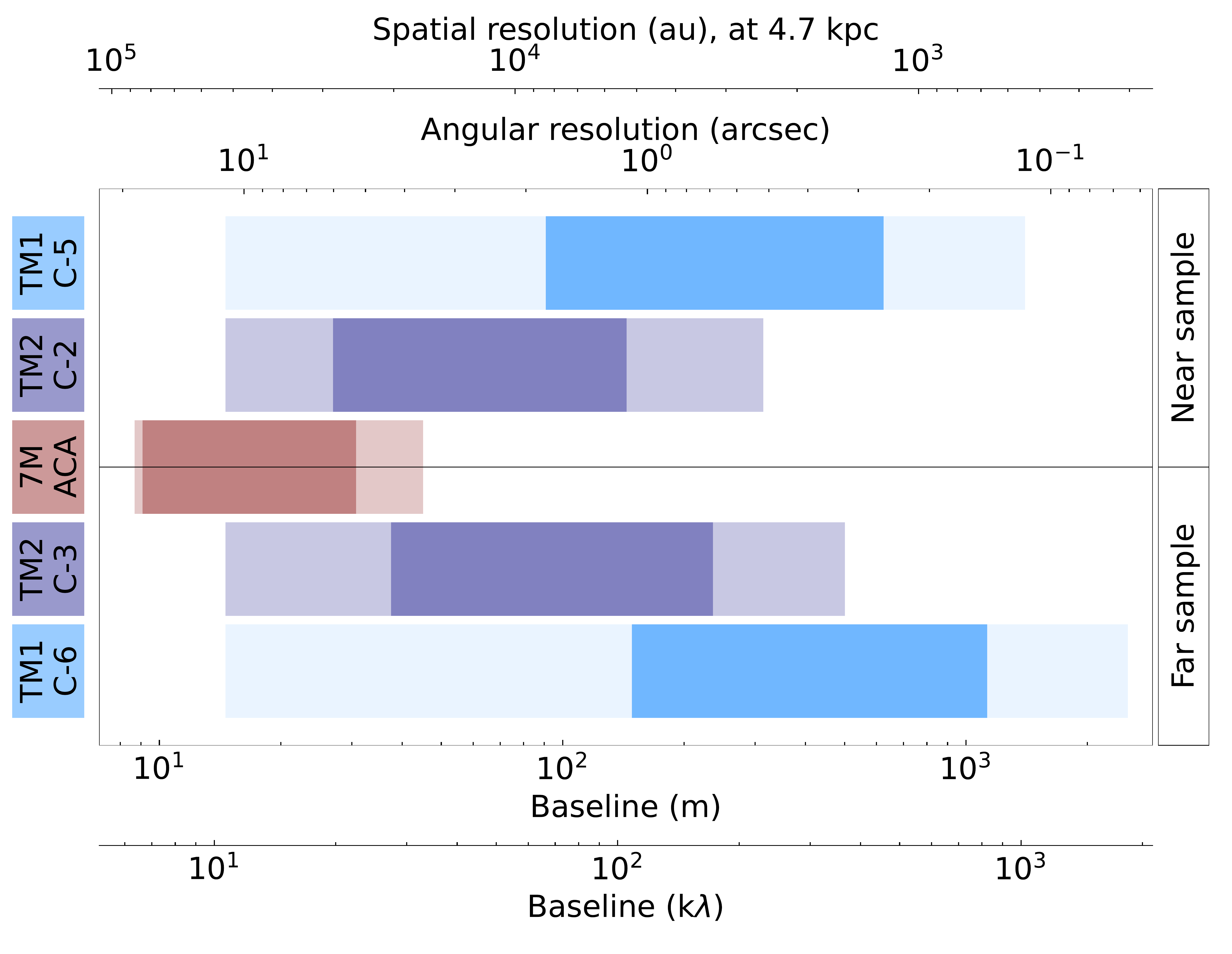}\\
\caption{ALMA configurations used in the ALMAGAL project. Four different nominal ALMA 12m (TM) arrays were used for the 1017 targets, plus the ACA 7m array. The targets are divided into two samples: ``Near'' for sources with distances $<4.7$~kpc, and ``Far'' for sources at $>4.7$~kpc. The bottom $x$-axis shows the baseline ranges, while the top $x$-axis shows the angular scales, and the corresponding spatial scales for a distance of 4.7~kpc. Light-colored rectangles show the range from minimum to maximum baseline for each array configuration, while the darker colors show the range from the 5$^\mathrm{th}$ to the 80$^\mathrm{th}$ percentile of array baselines.}
\label{fig:configuration-setup}
\end{figure} 
%------------------------------------------------------------------------

%________________________________________________________________
%
\section{Observations and data acquisition\label{sec:obs}}

In this section, we provide a brief description of the ALMAGAL sample (Sect.~\ref{sec:sample}) and a summary of the observations taken with the ALMA radio-interferometer (Sect.~\ref{sec:alma}).

%________________________________________________________________
%
\subsection{ALMAGAL sample\label{sec:sample}}

The ALMAGAL sample consists of 1017 targets with declination $<0^\circ$ (i.e., observable from the ALMA observatory at a high elevation) and distributed throughout the Galactic plane, from the near tip of the Galactic bar at $l\simeq30^\circ$ in the first quadrant to $l\simeq215^\circ$ in the third quadrant, including sources located both in the inner and outer Galaxy regions. The 1017 targets were selected from two complementary catalogues: 917 fields from the catalogue of dense clumps based on the Hi-GAL 70--500~$\mu$m images \citep{Elia2017}, and 100 fields from the RMS \citep[Red MSX Source survey,][]{Lumsden2013} catalogue of star-forming regions originally observed at 8--21~$\mu$m. The targets were selected to have distances $<7.5$~kpc from the Sun, clump masses $>500$~$M_\odot$ (or $>250$~$M_\odot$ in the outer Galaxy), and surface densities $>0.1$~g~cm$^{-2}$. Based on the luminosity and mass of the selected regions, we cover a broad range of evolutionary stages as defined by the luminosity over mass ($L/M$) ratio \citep[see][]{Molinari2008}. The selected fields include very early stage star-forming regions with $L/M\approx0.05$~$L_\odot/M_\odot$ up to clumps hosting active star formation and already developed \hii\ regions with $L/M\approx500$~$L_\odot/M_\odot$. See \citet{Molinari2024} for more details on the coordinates and properties of the ALMAGAL sample.

%________________________________________________________________
%
\subsection{ALMA observations\label{sec:alma}}

The observation strategy for the ALMAGAL program (project code: 2019.1.00195.L) is based on an homogeneous approach to image the 1017 selected regions achieving a common sensitivity of 0.1~mJy~beam$^{-1}$ in the continuum at a common spatial resolution of 1\,000~au, while being sensitive to large-spatial scales of about 0.1~pc. This range of spatial scales allows us to resolve the typical size of dense cores ($\approx0.01$~pc, equivalent to $\approx2\,000$~au; e.g., \citealt{SanchezMonge2013, Beuther2018, Sadaghiani2020}), while, at the same time, being sensitive to clump-scale structures ($\approx0.1$~pc, equivalent to $\approx20\,000$~au; e.g., \citealt{Koenig2017, Mottram2020, Motte2022}). To achieve this, we used different array configurations of the main ALMA array consisting of about 40 12m-size antennas (the so-called twelve-meter or TM array) together with the ten 7m-size antennas of the Morita array (Atacama Compact Array or ACA).

In more detail, the 1017 ALMAGAL targets were divided into two major groups: the ``Near sample'', which includes 538 sources with heliocentric distances $<4.7$~kpc, and the ``Far sample'', which includes 479 sources at distances $>4.7$~kpc \citep[see][]{Molinari2024}. Based on this major division of the full sample\footnote{We note that follow-up studies of the ALMAGAL targets have revisited their distance measurements resulting in some variations \citep[see][for details]{Molinari2024}. In the following, we use the most recent distance estimates, which result in 41 sources of the ``Near sample'' having distances above 4.7~kpc, and 30 sources of the ``Far sample'' with distances below 4.7~kpc.}, the observations were designed to use a combination of ALMA array configurations that results in a common final spatial resolution of $\approx1\,000$~au for both samples. This was achieved using the C-2 and C-5 configurations for the ``Near sample'', and the C-3 and C-6 configurations for the ``Far sample''. Figure~\ref{fig:configuration-setup} shows the typical baseline ranges of the array configurations and the angular scales to which they are sensitive. For consistency with ALMA nomenclature, the two most compact TM configurations (i.e., C-2 and C-3) are labeled as TM2 in the rest of the paper; while the two most extended configurations (i.e., C-5 and C-6) are labeled as TM1. Complementing the ALMA main array observations, all the 1017 ALMAGAL targets were also observed with the 7m-array (labeled as 7M throughout the paper), which ensures sensitivity to spatial scales of $\approx0.1$~pc.

%-------------------------------- Figure --------------------------------
\begin{figure}[t!]
\centering
\includegraphics[width=0.98\columnwidth]{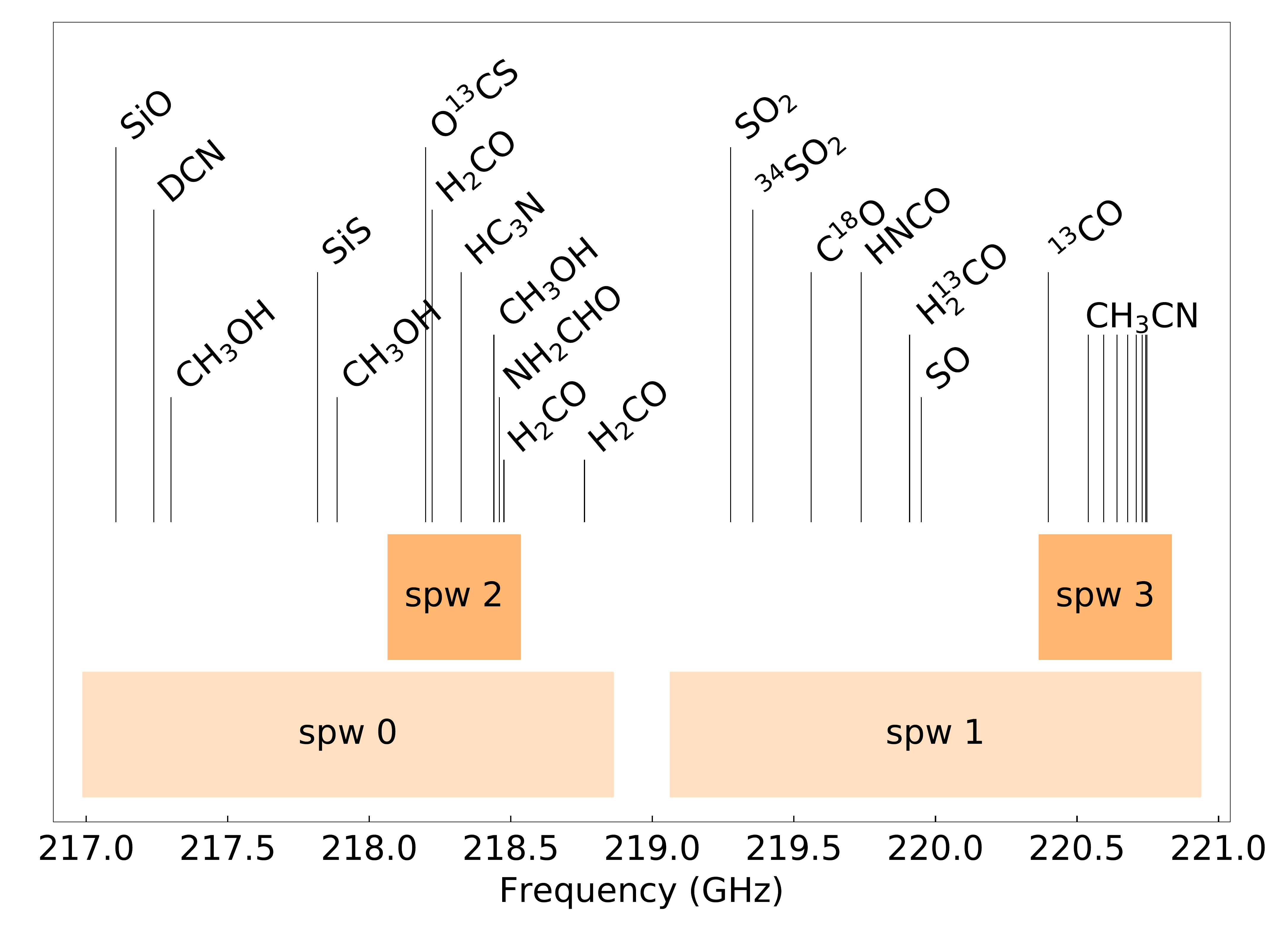}\\
\caption{ALMAGAL spectral setup. The four spectral windows are depicted with orange rectangles: the two broad spectral windows spw0 and spw1 are shown in light orange, while the two narrow, high-spectral resolution spectral windows spw2 and spw3 are shown in dark orange. The location of selected molecular transitions covered in the spectral setup are marked with vertical black lines.}
\label{fig:spectral-setup}
\end{figure} 
%------------------------------------------------------------------------

The ALMAGAL observations, approved during the Cycle~7 of ALMA, were conducted during two main periods: from October 2019 to March 2020, and from March 2021 to July 2022, the latter following the re-opening of the ALMA observatory after its closure due to the COVID-19 pandemic. All the 7M data and most of the TM2 observations were taken during the first observing run, while all the TM1 observations and a few remaining observing blocks of the TM2 were completed in the second run. Each source was observed for about 2--4~minutes in the 7M array, 1~minute in the TM2 array, and 2--3~minutes in the TM1 array. In addition to the two main subsamples: ``Near'' and ``Far'', the ALMAGAL targets were further divided and clustered into 15 and 16 groups, respectively for the two distance samples. The clustering was based on the sky distribution of the targets. These final groups (or Group OUS, observation unit sets, in ALMA terminology) contain different number of sources ranging from two in the smallest one up to 134 in the largest, with typically between 35 and 60 targets per group. During the observations, each Group OUS was divided into Execution Blocks (EBs) of about 1--2~hours in length. For the Group OUS with few sources, usually one single execution was enough to reach the required sensitivity. For those Group OUS with many sources, multiple executions were needed. The data of all EBs for each source are combined in the ALMAGAL pipeline to reach the aimed sensitivity and resolution (see Sect.~\ref{sec:pipeline}). The weblogs produced by the ALMA observatory during the observation and the QA2 (quality assessment) phase were inspected to search for major calibration problems. No obvious errors were identified and the standard calibration products generated by the ALMA pipeline \citep{Hunter2023} were taken as a starting point for the ALMAGAL pipeline for further processing and image generation.

%--------------------------------- Table --------------------------------
\begin{table}[t!]
\centering
\caption{\label{tab:almagal_summary}ALMAGAL data summary}
\begin{tabular}{l c c c}
\hline\hline \noalign{\smallskip} 
 Description &&& Value \\
\hline \noalign{\smallskip} 
 Targets                            &&& 1017 \\
 --- Near sample targets            &&& 538 \\
 --- Far sample targets             &&& 479 \\
 Field of view                      &&& 35$\farcs$4 \\
 \multicolumn{4}{l}{Median angular resolution} \\
 --- ACA (7M)                       &&& $7\farcs44\times4\farcs79$ \\
 --- C-2 (TM2)                      &&& $1\farcs31\times1\farcs07$ \\
 --- C-3 (TM2)                      &&& $0\farcs74\times0\farcs64$ \\
 --- C-5 (TM1)                      &&& $0\farcs35\times0\farcs29$ \\
 --- C-6 (TM1)                      &&& $0\farcs22\times0\farcs15$ \\
 \multicolumn{4}{l}{Median resolution for all arrays combined} \\
 --- Near sample                    &&& $0\farcs47\times0\farcs38$ \\
 --- Far sample                     &&& $0\farcs28\times0\farcs19$ \\
 \multicolumn{4}{l}{Continuum images} \\
 --- Typical effective bandwidth    &&& 2.8~GHz \\
 --- Median rms sensitivity        &&& 0.05~mJy~beam$^{-1}$ \\
 \multicolumn{4}{l}{Spectral line images} \\
 --- spw0/spw1 spectral resolution  &&& 1.4~km~s$^{-1}$ \\
 --- spw0/spw1 rms sensitivity      &&& 2.2~mJy~beam$^{-1}$ \\
 --- spw2/spw3 spectral resolution  &&& 0.34~km~s$^{-1}$ \\
 --- spw2/spw3 rms sensitivity      &&& 4.7~mJy~beam$^{-1}$ \\
\hline
\end{tabular}
\tablefoot{
These numbers correspond to median values for the first data release of ALMAGAL, produced as described in this paper.}
\end{table}
%------------------------------------------------------------------------

%-------------------------------- Figure --------------------------------
\begin{figure*}[t!]
\centering
\includegraphics[width=0.98\textwidth]{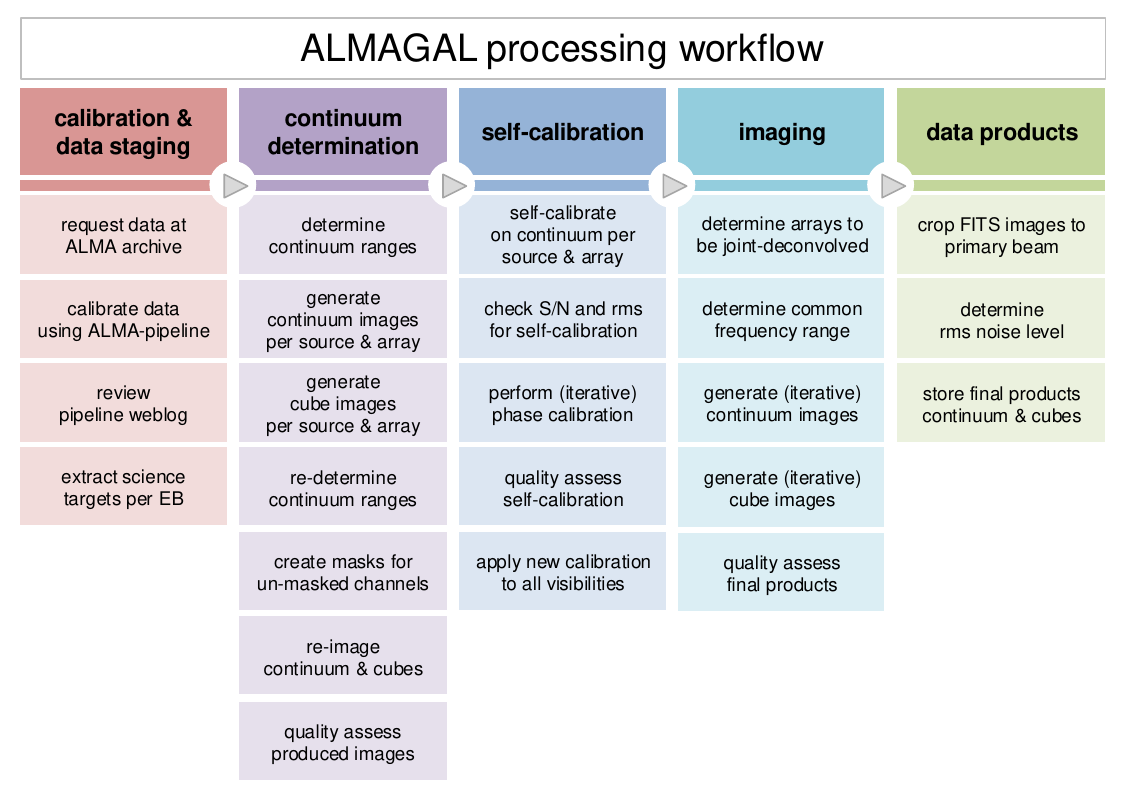}\\
\caption{ALMAGAL processing workflow. Schematic description of the five major steps. The process starts with ``calibration and data staging'', preparing the data for the following steps (Sect.~\ref{sec:calibration}). We then determine line-free frequency ranges for the continuum emission, and generate a first set of images for each source and individual array (Sect.~\ref{sec:continuum-determination}). The next step applies an automatized self-calibration process to each source and individual array (Sect.~\ref{sec:selfcalibration}). The self-calibrated data of each array are then joint-deconvolved together to produce final images of continuum emission and line cubes (Sect.~\ref{sec:joint-deconvolution}). Finally, the data products are quality assessed and stored for distribution to the consortium (Sect.~\ref{sec:data-products}).}
\label{fig:pipeline-workflow}
\end{figure*} 
%------------------------------------------------------------------------

The spectral setup chosen for the ALMAGAL program consists of four spectral windows at around 219~GHz (1.38~mm), within the Band~6 of ALMA \citep{Ediss2004, Kerr2004}. Out of the four spectral windows, two of them (spw0 and spw1) are set to cover a broad frequency range of 1.87~GHz each, with 3840~channels of 488~kHz each. The other two (spw2 and spw3) cover a narrower frequency range of 0.468~GHz with an enhanced resolution\footnote{The channel spacing of the two different sets of spectral windows is 488 and 122~kHz, corresponding to a spectral resolution of $\approx1.4$ and $\approx0.34$~km~s$^{-1}$, respectively, after Hanning smoothing is considered.} of 122~kHz per channel, for a total of 3840~channels in each window. The two high-spectral resolution windows were placed to cover key spectral line transitions of the molecular species H$_2$CO and CH$_3$CN, which provide information on the temperature of the dense gas in the ALMAGAL fields at a resolution of $\approx0.34$~km~s$^{-1}$. The two broad spectral windows provide the necessary frequency coverage to achieve good sensitivity in the continuum images. Complementary to this, they are also sensitive to spectral lines of species associated with shocks (e.g., SiO, SO), dense gas (e.g., HNCO, SO$_2$, CH$_3$OH), deuterated chemistry (e.g., DCN), as well as complex organic molecules (e.g., formamide: NH$_2$CHO, ethyl cyanide: C$_2$H$_5$CN, methylamine: CH$_3$NH$_2$), all of them observed at a resolution of $\approx1.4$~km~s$^{-1}$. Figure~\ref{fig:spectral-setup} shows a sketch of the ALMAGAL spectral setup, highlighting the frequency coverage of the different spectral windows, and relevant spectral lines covered in the setup.

In the following sections (Sects.~\ref{sec:workflow} and \ref{sec:pipeline}), we present and describe the processing workflow used by the ALMAGAL consortium to produce science-ready images and products. We note that both calibration and imaging were done using the Common Astronomy Software Applications (\texttt{CASA}) package \citep{CASA2022}. Two different versions were used: version 5.6.1 for calibration and most of the steps within the ALMAGAL processing pipeline (see Sects.~\ref{sec:calibration} to \ref{sec:selfcalibration}), and version 6.2.0 for the final joint-deconvolution imaging (see Sect.~\ref{sec:joint-deconvolution}). Table~\ref{tab:almagal_summary} shows an executive summary of the main parameters of the ALMAGAL final images. More details are provided in the following sections, figures and tables.

%________________________________________________________________
%
\section{ALMAGAL processing workflow\label{sec:workflow}}

In this section, we describe the main processing workflow of the ALMAGAL pipeline. The pipeline, including main functions and processing scripts, can be found on the GitHub repository\footnote{\url{https://github.com/betacygni/ALMAGAL}}. Our pipeline performs several essential steps such as continuum level determination, self-calibration, and array combination, all necessary to generate the final science-ready products. The main advantages of these products, compared to the publicly available data in the ALMA archive, are the improved continuum determination (see Sect.~\ref{sec:continuum-determination}), the self-calibration implementation which improves the signal-to-noise by up to a factor of five in several fields (see Sect.~\ref{sec:selfcalibration}), and the combination of the individual array configurations reaching the final required sensitivity and resolution (see Sect.~\ref{sec:joint-deconvolution}). Moreover, we note that for the Group OUS with many sources, the image products that are available in the public ALMA Science Archive may not be complete (i.e., not all sources may have been imaged) due to size-mitigation issues related to storage space in the archive.

The ALMAGAL processing workflow is divided into five major steps, which are summarized in Fig.~\ref{fig:pipeline-workflow}, and briefly described in the following:

\begin{itemize}

\item[1.] \textit{Calibration and data staging}:\newline The processing workflow starts with requesting data from the ALMA Science Archive. These data are calibrated following the standard ALMA interferometric calibration pipeline, and stored in ``measurement set'' files, which contain visibilities with calibrated phases and amplitudes. The ALMA-QA2 pipeline weblogs produced during calibration are investigated to rule out the presence of major problems in the data. Finally, the calibrated data are staged in a form appropriate for the following processing steps. The main output of this stage is the creation of individual files containing the calibrated visibilities for each ALMAGAL science target for each independent EB (Execution Block). See more details in Sect.~\ref{sec:calibration}.

\item[2.] \textit{Continuum determination and first imaging}:\newline The workflow continues with the generation of the first images using standard ALMA-pipeline CLEANing functions. This enables the identification of the frequency ranges to be used for continuum determination. Continuum images as well as spectral cubes for each of the four spectral windows are generated for each source and each array. The CLEANed cubes are inspected for possible errors in the continuum determination process. When necessary, the pipeline re-determines the continuum frequency ranges. CLEANing masks are stored for each source and array. The pipeline also identifies un-masked channels and creates CLEANing masks for them. A standard quality assessment is performed to evaluate the quality of the images. The main output of this stage are files containing the continuum frequency ranges, CLEANing masks for continuum and cubes, and a first set of CLEANing parameters and CLEANed images. See more details in Sect.~\ref{sec:continuum-determination}.

\item[3.] \textit{Self-calibration}:\newline The ALMAGAL pipeline automatically identifies which sources have strong-enough continuum emission that enables self-calibration. When possible, several iterations of phase-calibration are done on the continuum emission individually for each source and array. The self-calibration solutions are then applied to all the continuum and spectral cubes visibilities. The main output of this stage is a set of self-calibrated data files for each individual source and array. See more details in Sect.~\ref{sec:selfcalibration}.

\item[4.] \textit{Final joint-deconvolution imaging}:\newline We use the (self-)calibrated dataset to generate the final CLEANed images combining multiple arrays (e.g., 7M+TM2+TM1, 7M+TM2). First, the data of each array are regridded to a common frequency range that takes into account small Doppler shifts common in observations that span several days. We use the continuum frequency ranges determined in the first steps of the workflow, and the CLEANing masks of each individual array to generate the final images. These masks are used in an iterative CLEANing strategy for both the continuum and cube images. The main output of this stage is a set of science-ready images for each ALMAGAL target. See more details in Sect.~\ref{sec:joint-deconvolution}.

\item[5.] \textit{Data products}:\newline In the last stage of the workflow, standard FITS files are generated for each source and array. These final images are cropped to contain only the field of view, thus reducing the size of each individual file. The rms noise level of the images is calculated and stored in the header of the FITS image files. In the final step, we generate a catalog file that contains basic information on a number of image parameters (e.g., angular resolution, noise levels, channel widths) for each image, both continuum and cubes. See more details in Sect.~\ref{sec:data-products}.
\end{itemize}

We faced a number of challenges when running the ALMAGAL processing workflow and handling the corresponding large volumes of data in a timely manner. Similar to other ALMA large programs (e.g., ALMA-IMF: \citealt{Ginsburg2022}), while the raw data products are relatively modest ($\approx30$~TB), the data set increased up to $\approx500$~TB after intermediate data products and final images were created. Note that we removed all non-critical intermediate products to reduce the final volume of data to about 500~TB. Most of the data volume is due to products containing data from the most extended array configurations (i.e., TM1). For cubes with about 3840 channels and $900\times900$ pixels each, we end up having about 12\,000 million data points for each source (after considering the four spectral windows). For each source, spectral window and array (or array-combination), we produce a total of five different FITS files (see Sect.~\ref{sec:joint-deconvolution}): \texttt{image}, \texttt{residual}, \texttt{pb}, \texttt{psf} and \texttt{model}. This results in about 400~GB per source, with most of the data volume dominated by the 7M+TM2+TM1 products; and a total of $\approx500$~TB for the whole ALMAGAL sample.

Since the beginning of the project, the ALMAGAL consortium gained access to substantial computing resources at the JUWELS (J\"ulich Wizard for European Leadership Science) supercomputer of the J\"ulich Supercomputer Center (J\"ulich, Germany), including 10+ million CPU hours and enough storage for the data products (project codes: 17905, 20946, 23984, and 26560). This ensured that most of the data were processed using the same machine in a uniform manner. We also made use of the computing resources across the whole consortium for certain aspects of the workflow. In particular, the first set of images for the TM1-alone array, together with the files containing the corresponding continuum frequency ranges, were produced in a distributed manner across different members and institutes of the ALMAGAL consortium, which provided access to different computing facilities such as the supercomputer SURFsara (Amsterdam, Netherlands), and computers at the University of Cologne (Cologne, Germany), the Max-Planck Institute for Astronomy (Heidelberg, Germany), the UK ALMA Regional Centre Node at the University of Manchester (Manchester, UK), the INAF Institute of Space Astrophysics and Planetology (Rome, Italy), and the Academia Sinica Institute of Astronomy and Astrophysics (Taipei, Taiwan). For this, members at the different institutes used the ALMAGAL pipeline to generate the first set of images of the TM1-alone array for specific sets of sources. The products generated at this stage were sent back to the J\"ulich main storage center for the following ALMAGAL processing steps.

%-------------------------------- Figure --------------------------------
\begin{figure*}[t!]
\centering
\includegraphics[width=1.0\textwidth]{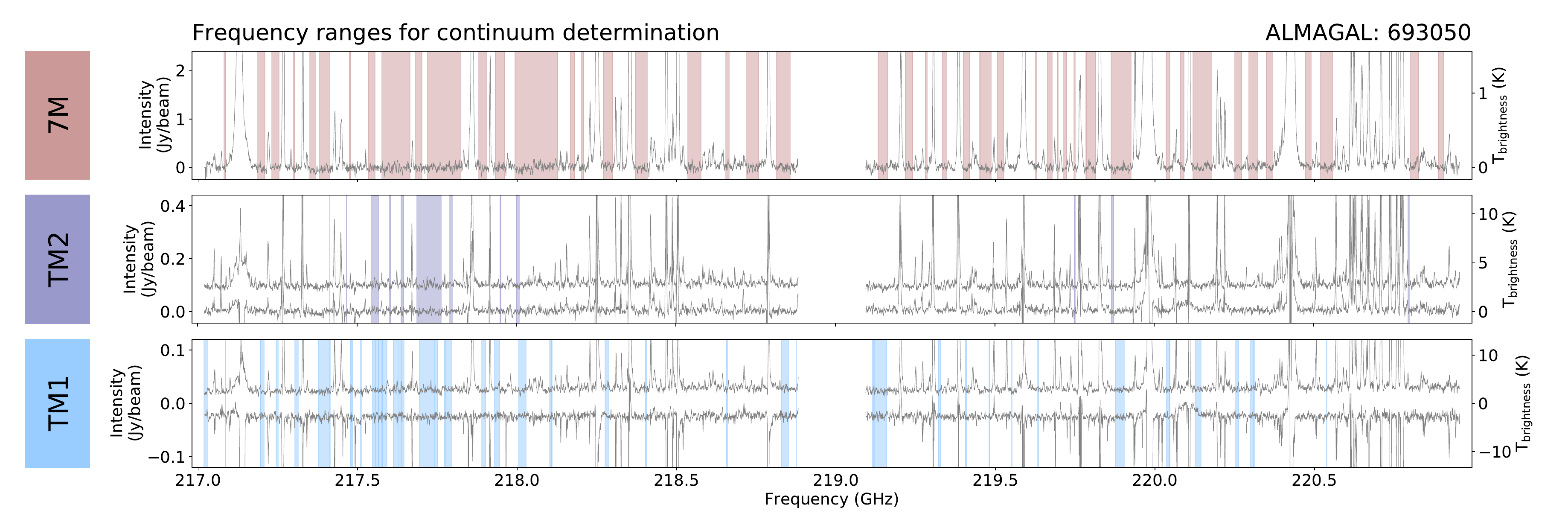}\\
\caption{Frequency ranges used for continuum determination for the ALMAGAL target 693050 (also labelled AG301.1365$-$0.2259 in \citealt{Molinari2024}). The continuum ranges identified by \texttt{findcont} are depicted as shaded color areas across the spectra. From top to bottom, the different rows show the spectra extracted in the different array configurations: 7M (top), TM2 (center), and TM1 (bottom). For TM1 and TM2, the spectra were extracted towards two different positions (not resolved in the 7M image) corresponding to the bright compact sources also seen in the TM2 panels of Fig.~\ref{fig:selfcalibration-example_v01}. These positions have been chosen as examples of complex line-rich cases with both emission and absorption features (as seen in the spectra of the TM1 panel). The frequency range correspond to spw0 and spw1 spectral windows (cf.\ Fig.~\ref{fig:spectral-setup}).}
\label{fig:continuum-level}
\end{figure*} 
%------------------------------------------------------------------------

%________________________________________________________________
%
\section{Detailed steps of the processing workflow\label{sec:pipeline}}
%\section{ALMAGAL pipeline processing\label{sec:pipeline}}

%________________________________________________________________
%
\subsection{Calibration and data staging\label{sec:calibration}}

The first steps of the ALMAGAL pipeline deal with data calibration and staging the calibrated files for further processing. First, the original data taken by the observatory can be retrieved and extracted from the ALMA archive. The standard ALMA-pipeline calibration procedures are accessible in the \texttt{scriptForPI.py} file, also available at the ALMA Science Archive. Running this script restores the measurement sets with the necessary calibrated visibilities. We note that for a more efficient process of this first step, the ALMAGAL consortium reached an agreement with the ALMA Regional Center (ARC) in Europe (ESO, Garching) to have direct access to the calibrated data files that were generated during the QA2 assessment process and later on ingested into the ALMA Science Archive. These calibrated datasets, delivered by the ARC and used by the ALMAGAL consortium, correspond to the same output that any user can generate when running the \texttt{scriptForPI.py} file on the data publicly available at the ALMA archive.

After calibration is executed, we end up having a calibrated measurement set for each individual observing execution block (EB). As stated earlier, these EBs may contain multiple sources (from a few to several hundred sources). The ALMAGAL script\footnote{All scripts used for processing the ALMAGAL data are available in the GitHub repository indicated in footnote 2, see Sect.~\ref{sec:workflow}} \texttt{scriptToSplitSources.py} splits the data contained in the calibrated measurement sets of each EB into files containing only data for each individual scientific target, for a total of 1017 ALMAGAL targets. The split calibrated measurement set files for each source and EB contain only the data of the four scientific spectral windows: spw0, spw1, spw2, and spw3 (see Fig.~\ref{fig:spectral-setup}). Note that the numbering of these spectral windows in the original data files available in the ALMA archive are spw16, spw18, spw20 and spw22 for the 7M observations, and spw25, spw27, spw29 and spw31 for the TM2 and TM1 observations, respectively.

%________________________________________________________________
%
\subsection{Continuum determination and first imaging\label{sec:continuum-determination}}

Following the calibration and staging of the data for each individual source, the next step is to process the data of each source for each array configuration (i.e., 7M, TM2, and TM1) through a series of \texttt{CASA} functions to generate: (1) first images of the continuum and data cubes for each source and array in the LSRK frequency reference frame, (2) a \texttt{cont.dat} file that contains the frequency ranges in LSRK that are free from line emission and are used to determine the continuum level, and (3) CLEANing masks and CLEANing parameters that will be used in the final imaging. This first imaging step in the ALMAGAL pipeline is carried out with the \texttt{scriptForImagingARRAY.py} file, with \texttt{ARRAY} corresponding to the three different configurations used in the project: \texttt{7M}, \texttt{TM2} and \texttt{TM1}. The functions of this script can be divided into five major steps that are described in the following.

\begin{figure*}[th!]
\centering
\includegraphics[width=1.0\textwidth]{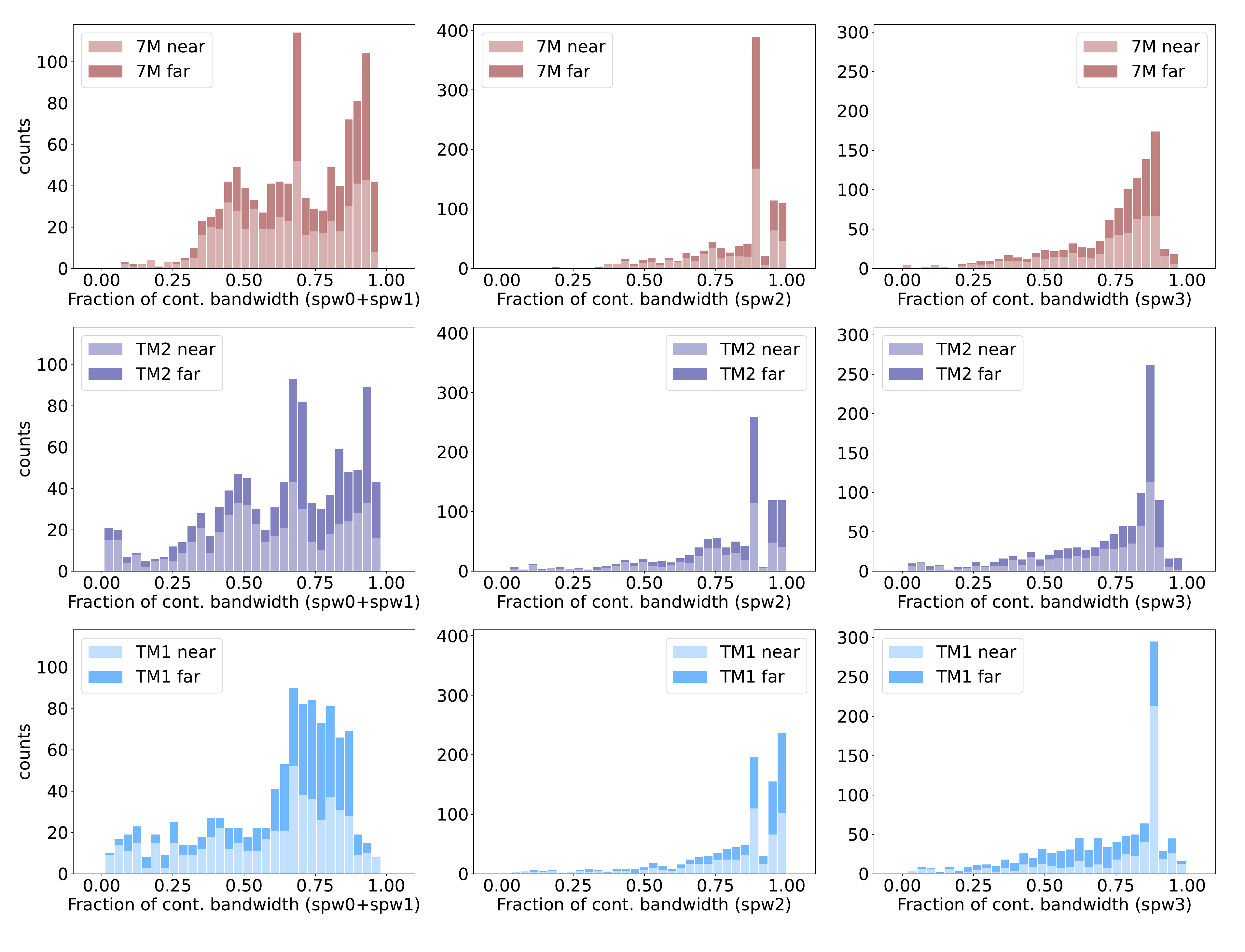}\\
\caption{Distribution of the fraction of final continuum bandwidth determined with the improved version of \texttt{findcont} (see Sect.~\ref{sec:continuum-determination}) for different spectral windows (columns) and array configurations (rows). The $y$-axis in all panels depicts the number of ALMAGAL fields. Lighter and darker histograms show data for sources in the ``Near'' and ``Far'' samples, respectively. The fraction of bandwidth indicated in the histograms is normalized to the total bandwidth, which is 3.75~GHz for spw0+spw1 (left column), 0.468~GHz for spw2 (middle column), and 0.468~GHz for spw3 (right column).}
\label{fig:continuum-bandwidth}
\end{figure*}

First, the frequency ranges for the continuum emission are determined using the standard ALMA-pipeline function \texttt{findcont}. The ALMAGAL consortium explored alternative approaches to determine the continuum level using other packages such as \texttt{STATCONT}\footnote{\url{https://github.com/radio-astro-tools/statcont}} \citep{SanchezMonge2018} and \texttt{LumberJack}\footnote{\url{https://github.com/adam-avison/LumberJack}}. We obtained consistent results on the continuum emission level, within 5\% differences, when using the three different software packages in a subset of a few sources with different levels of line-richness. The decision to use \texttt{findcont} in the ALMAGAL pipeline was essentially related to the scripting structure and functionality that was implemented in the processing workflow. Figure~\ref{fig:continuum-level} shows an example of the continuum frequency ranges that \texttt{findcont} defines for each one of the three array configurations. This example has been chosen as a complex case, with both line-rich emission spectra in some regions as well as strong absorption features in other regions (see e.g., TM1 spectra in the Fig.~\ref{fig:continuum-level}). As seen in the TM2 and TM1 examples of Fig.~\ref{fig:continuum-level}, we found that \texttt{findcont} is conservative when defining the frequency ranges in line-rich sources. While additional frequency ranges could be manually added, this would require an individual implementation for each source that was beyond the scope of the first data release. In Sect.~\ref{sec:QAgeneral}, we investigate the relation between the bandwidth used for continuum determination and the final rms noise level of the images. Figure~\ref{fig:continuum-bandwidth} shows the bandwidth fraction used for continuum determination in the different spectral windows and array configurations for all the targets. In most cases, about 50 to 100\% of the available bandwidth in spw0 and spw1 (corresponding to about 1.9--3.7~GHz) is used for determining the continuum emission.

%-------------------------------- Figure --------------------------------
%------------------------------------------------------------------------

In the second step, we use the frequency ranges for the continuum emission, stored in the \texttt{cont.dat} files, to generate the first images of both continuum and line-only spectral cubes for each spectral window. For the continuum images, we only use data from the two broad spectral windows (spw0 and spw1). Both continuum and cube images are generated using the \texttt{tclean} task in \texttt{CASA}, with a weighting robust Briggs parameter of $+0.5$, as a compromise between sensitivity and angular resolution. CLEANing masks are produced following the automasking multi-thresh functionality described by \citet{Kepley2020}. This first set of images is inspected to ensure that the expected angular resolutions and sensitivities of each individual array are reached.

%-------------------------------- Figure --------------------------------
\begin{figure}[t!]
\centering
\includegraphics[width=0.88\columnwidth]{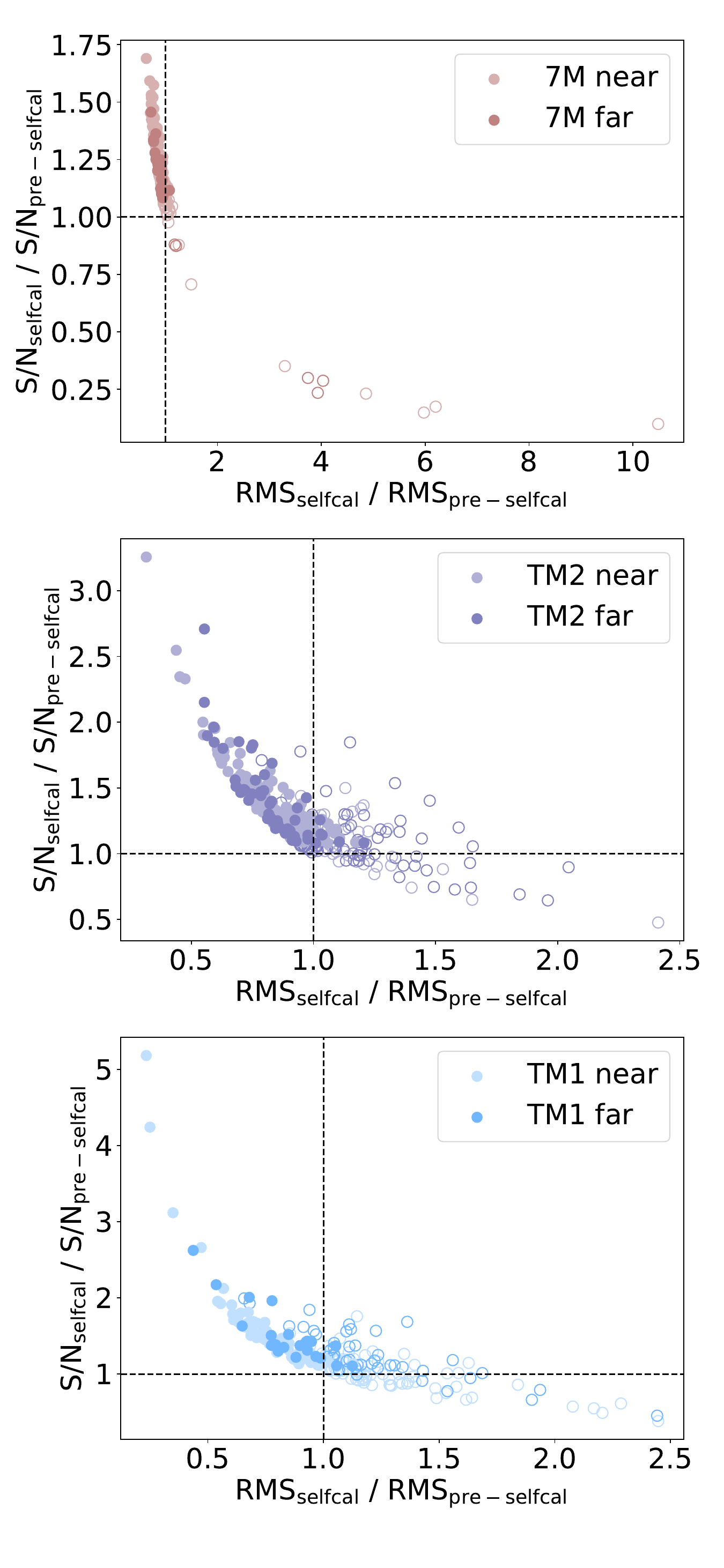}\\
\caption{Signal-to-noise ratio (S/N) and rms noise level (RMS) for those ALMAGAL fields where self-calibration is possible (i.e., $\mathrm{S/N}_\mathrm{selfcal}>3$). The $y$-axis shows the ratio of the S/N after self-calibration with respect to the original S/N, while the $x$-axis shows the corresponding ratio for the rms noise level. Filled circles correspond to those sources where self-calibration improved the final images and the solutions were applied. Empty circles show those sources where self-calibration, even if originally possible, did not improve the quality of the image (e.g., due to coarser angular resolution images, see Sect.~\ref{sec:selfcalibration}), and therefore the solutions were not applied. From top to bottom, the different panels show the results for the 7M, TM2, and TM1 arrays.}
\label{fig:selfcalibration-stats}
\end{figure}
%------------------------------------------------------------------------

Third, we inspect the quality of the defined continuum frequency ranges and re-define them if necessary. For some ALMAGAL targets, the default pipeline \texttt{findcont} task in \texttt{CASA} version~5.6.1 does not produce optimal results. In some cases, we found that there was still significant residual emission in the line-only cubes at the frequency channels that should have no emission after continuum subtraction (i.e., the continuum frequency channels). This emission was seen in the \texttt{mom8\_fc} images (following ALMA nomenclature) that show the maximum emission in continuum channels after continuum subtraction. The two reasons for this are: (1) for strong line-rich sources (typically hot cores), the \texttt{findcont} task should use a relatively compact spatial mask when generating the mean spectrum used to determine the line-free frequency ranges, to prevent the emission from compact hot cores being diluted in larger masks; and (2) sources can have spatially distinct weaker emission that is not included in the more restrictive mask that helps to overcome issue (1), and thus can never reach the threshold used in \texttt{findcont} to determine line-free channels. This latter aspect is especially true for lines with very extended emission such as the $^{13}$CO\,(2--1) transition.

To overcome these issues, the ALMAGAL pipeline includes the \texttt{REFIND.py} script, which runs a new instance of \texttt{findcont} on the CLEANed cube image. The script uses a higher intensity threshold when defining the mask, which results in a more compact mask from which the mean spectrum is extracted. The script then evaluates the residual emission in \texttt{mom8\_fc} images. If it is above a maximum allowed value (i.e., there is still line contamination in the continuum channels), the threshold used to exclude channels with line emission is lowered, resulting in the exclusion of more channel ranges that could still be contaminated by line emission. This proceeds iteratively until the criterion is met (i.e., no residual emission is identified in the \texttt{mom8\_fc} images), or until 5 iterations have been completed (see below). To address issue (2), after 3 iterations, any bright emission in the current \texttt{mom8\_fc} images that is outside the \texttt{findcont} mask is added to the mask, and a new mean spectrum is extracted. The iterations then continue. Based on our experience, all the line-contaminated channels can be identified and excluded from the continuum ranges if enough iterations (typically 12) are done. However, this process may result in a fair amount of ``good'' bandwidth being also removed and thus not available for the continuum imaging. Therefore, to mitigate this aspect, we stop at iteration 5, and for the 6$^\mathrm{th}$ iteration, we simply remove any remaining channels that have emission above the given threshold. We note that this improved approach for \texttt{findcont} has been implemented in the main ALMA pipeline functions and is now available to all users. Specific details on the task and how the continuum frequency ranges are refined using the moment difference analysis can be found in \citet{Hunter2023}. The outcome of this step in the ALMAGAL processing workflow is the generation of a new \texttt{cont.dat} file with the final continuum frequency ranges. The original file is renamed as \texttt{original.cont.dat}. Figure~\ref{fig:continuum-bandwidth} shows the final bandwidth fraction used for continuum determination after the above-mentioned corrections are applied.

In the fourth step, new continuum and cube images are generated with the \texttt{tclean} task, using the newly defined continuum frequency ranges. Finally, in step five, the ALMAGAL pipeline assesses whether there are channels with emission above the CLEANing threshold that lack a CLEANing mask. If there are, it then creates an amended mask that uses the maximum of $2\times$ the clean threshold or $0.4\times$ the peak line intensity to amend the mask. It then restarts the CLEANing for those specific channels. We note that for this re-cleaning we set the automasking multi-thresh parameter \texttt{growiterations} to 1 (the default is 75) to avoid channels with extended emission getting fully masked (including negative bowl regions, in particular in the 7M datasets).

After this process, the ALMAGAL pipeline has generated for each source and array configuration a \texttt{cont.dat} file that includes the continuum frequency ranges, CLEANing masks, and a set of continuum and line-only cube images. These products are later used to perform self-calibration (see Sect.~\ref{sec:selfcalibration}) and to perform the final joint-deconvolution imaging (see Sect.~\ref{sec:joint-deconvolution}).

%-------------------------------- Figure --------------------------------
\begin{figure*}[h!]
\centering
\includegraphics[width=0.95\textwidth]{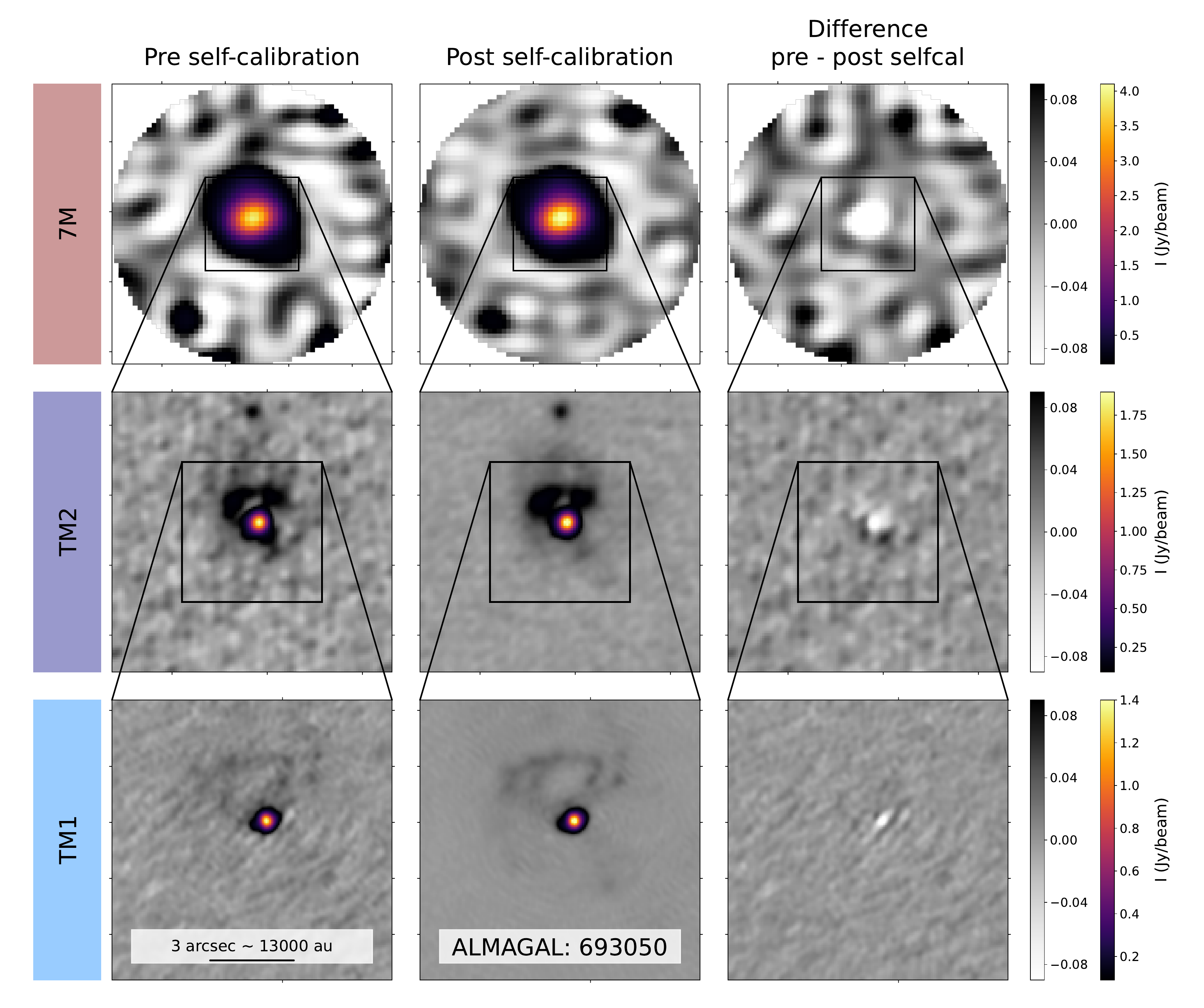}\\
\caption{ALMAGAL self-calibration results for target 693050 (also labelled AG301.1365$-$0.2259 in \citealt{Molinari2024}). From top to bottom, the different rows show the 7M, TM2, and TM1 images. For all three arrays, the left column shows the continuum emission image before self-calibration, the middle column shows the continuum image after self-calibration, and the right column shows the differences of the two images (i.e., pre self-calibrated minus post self-calibrated images). The intensity scales are fixed for each row (i.e., each array). This field is an example of a bright and compact source, where self-calibration improves the rms noise level removing phase-noise fluctuations throughout the image (see TM2 and TM1 panels). In all images throughout the paper we use a double color bar to better emphasize both the weak, extended emission (in grey colors) and bright, compact regions (in magma colors).}
\label{fig:selfcalibration-example_v01}
\end{figure*} 
%------------------------------------------------------------------------

%-------------------------------- Figure --------------------------------
\begin{figure*}[h!]
\centering
\includegraphics[width=0.95\textwidth]{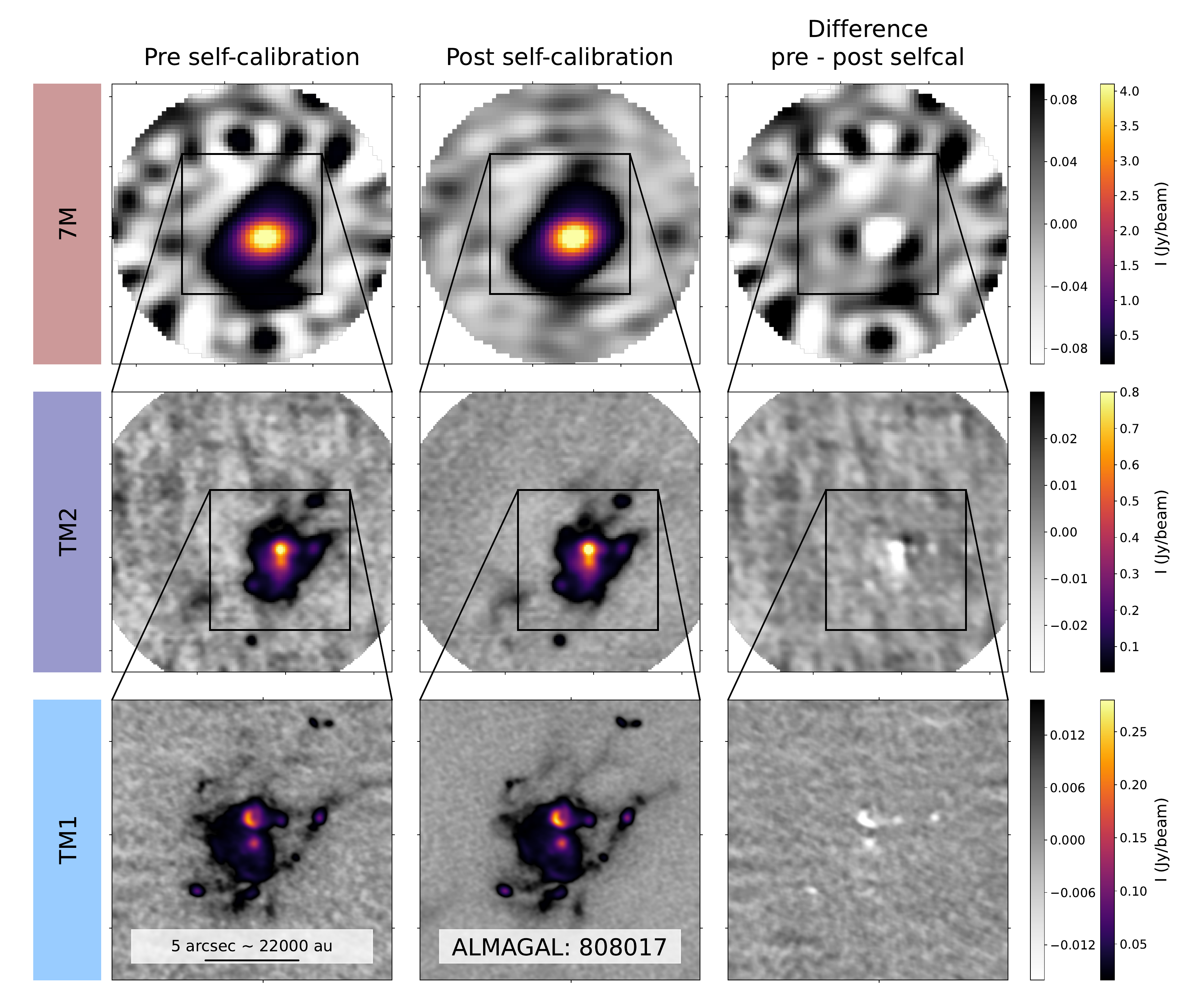}\\
\caption{ALMAGAL self-calibration results for target 808017 (also labelled AG332.8271$-$0.5492 in \citealt{Molinari2024}). Panels in the different rows and columns are the same as in Fig.~\ref{fig:selfcalibration-example_v01}. This field is an example of a moderately bright and extended source where self-calibration also improves the quality of the final images.}
\label{fig:selfcalibration-example_v02}
\end{figure*} 
%------------------------------------------------------------------------

%________________________________________________________________
%
\subsection{Self-calibration\label{sec:selfcalibration}}

The continuum images generated in Sect.~\ref{sec:continuum-determination} reveal bright continuum sources in several fields, some of them having intensities above 50~mJy~beam$^{-1}$ that may limit the dynamic range and create artifact features across the images. We used these continuum images to evaluate if the emission in each individual array configuration is strong enough for self-calibration. For this, we followed the standard interferometric guidelines that suggest that for phase-only self-calibration we need to detect the target with a signal-to-noise ratio (S/N) $>3$ in a solution time interval shorter than the time for significant phase variations for all baselines to a single antenna \citep[e.g.,][]{Taylor1999}. Based on this, the S/N necessary to do self-calibration is defined as
\begin{equation}
\mathrm{S/N}_\mathrm{selfcal} = \frac{I_\mathrm{peak,cont}}{\mathrm{rms}_\mathrm{cont} \times \sqrt{N_\mathrm{ant}-3} \times \sqrt{t_\mathrm{exp}/t_\mathrm{int}}},
\end{equation}
where $I_\mathrm{peak,cont}$ is the peak intensity of the continuum emission, $\mathrm{rms}_\mathrm{cont}$ is the rms noise level of the continuum image, $N_\mathrm{ant}$ is the number of antennas available during the observation, $t_\mathrm{exp}$ is the total on-source observing time, and $t_\mathrm{int}$ is the integration time, corresponding to 10~seconds for the 7M observations and 6~seconds for the TM2 and TM1 observations. If $\mathrm{S/N}_\mathrm{selfcal}>3$, the ALMAGAL pipeline initiates the self-calibration process running the \texttt{scriptForSelfCalibration.py} file. For most ALMAGAL targets, this condition approximately corresponds to $\mathrm{S/N}\geq50$ in the continuum images. Out of the 1017 ALMAGAL targets, the self-calibration condition ($\mathrm{S/N}_\mathrm{selfcal}>3$) is fulfilled in 177 fields for the 7M array, 342 fields for the TM2 array, and 225 fields for the TM1 array.

For all sources with $\mathrm{S/N}_\mathrm{selfcal}>3$, the ALMAGAL pipeline initiates an iterative process of self-calibration. In the first step, it generates a continuum image CLEANing down to a threshold of $4\times$ rms. This image is used as a model to self-calibrate the visibilities in phase-only mode (i.e., \texttt{calmode='p'}), and using a solution interval equal to the integration time $t_\mathrm{int}$ (i.e., \texttt{solint='int'}). The new calibration solutions are then applied to all visibilities (i.e., all four spectral windows). A new continuum image is generated using the self-calibrated data down to the same threshold (i.e., $4\times$ rms). Both images, before and after self-calibration, are compared and if the rms noise level has improved by more than a factor\footnote{In the ALMAGAL pipeline we use different factors (e.g., 7.5\% and 15\%) within the self-calibration processing steps. We note that considering slightly different values for these factors does not considerably change the results.} of 7.5\% and the beam size has not increased\footnote{We note that the beam may increase due to certain baselines, in particular the long ones, becoming flagged as bad data during the self-calibration process. This can happen because they do not meet the S/N criteria (\texttt{minsnr=2.5}) when determining new calibration solutions. If long baselines are flagged, the resulting image will have a coarser resolution, preventing us from reaching the requested angular resolution. Thus, if the beam increases by more than 15\% or the rms noise level does not improve by more than 7.5\%, the process of self-calibration stops, and the newly determined solutions are not applied to the final measurement set.} by more than 15\%, the solutions are stored and the next self-calibration iteration proceeds. In the next two iterations, the pipeline CLEANs down to a threshold of $3\times$ rms, determining new calibration solutions only in phase and with the same solution interval, and evaluating if the images before and after the new self-calibration iteration have improved (i.e., the rms has decreased by more than 7.5\% and the beam has not increased by more than 15\%). From the fourth iteration on, the process is repeated, CLEANing down to a threshold of $2\times$ rms. Out of the sources with $\mathrm{S/N}_\mathrm{selfcal}>3$, a total of 81\% for the 7M array, 66\% for the TM2 array and 45\% for the TM1 array, have self-calibration solutions that improve the final continuum images. Figure~\ref{fig:selfcalibration-stats} shows the sources for which self-calibration was possible, and how the rms noise level and the S/N improved in the new images after self-calibration. As stated earlier, the self-calibration solutions are applied to all the visibilities, making it possible to produce new images of both continuum and cubes with the new corrected data.

Figures~\ref{fig:selfcalibration-example_v01} and \ref{fig:selfcalibration-example_v02} show two examples of ALMAGAL targets where self-calibration improved the quality of the images for the three configuration arrays. Field 693050 (see Fig.~\ref{fig:selfcalibration-example_v01}) corresponds to a very bright and compact source where the new calibration helps to improve the rms noise level, removing phase-noise fluctuations that were clearly visible in the TM2 and TM1 arrays. Moreover, the intensity of the source in the new images increases, as shown in the right column panels. Another example is field 808017 (see Fig.~\ref{fig:selfcalibration-example_v02}), which shows extended emission features. Even in fields with extended emission, the self-calibration method used in the ALMAGAL pipeline improves the final images, removing phase-noise fluctuations and recovering more flux in the brightest sources. Overall (cf.\ Fig.~\ref{fig:selfcalibration-stats}), self-calibration has improved the S/N of the images by a factor $>1.5$ in about 16\% of the ALMAGAL fields. Although self-calibration procedures specifically designed for each individual source may further improve the rms and S/N of the images, the procedures currently implemented in the ALMAGAL processing workflow enable an automatic and robust approach for self-calibration consistently across all images.

After this processing step, each ALMAGAL target has new measurement set files for each EB with self-calibrated visibilities when possible, or with standard calibration visibilities otherwise. These are the measurement set files that are used to generate the final images (see Sect.~\ref{sec:joint-deconvolution}).

%--------------------------------- Table --------------------------------
\begin{table*}[t!]
\centering
\caption{\label{tab:tclean_parameters}Parameters used in the CLEANing process of the different ALMAGAL arrays combinations (see Sect.~\ref{sec:joint-deconvolution})}
\begin{tabular}{l c c c c c}
\hline\hline \noalign{\smallskip} 
 \texttt{tclean} parameters & 7M & TM2 & TM1 & 7MTM2 & 7MTM2TM1 \\
\hline \noalign{\smallskip} 
 General parameters &&&&& \\
 --- robust (Briggs)   & +0.5 & +0.5 & +0.5 & +0.5 & +0.5 \\
 --- pbmask            & 0.3  & 0.3  & 0.3  & 0.3  & 0.3  \\
 --- pblim             & 0.3  & 0.3  & 0.3  & 0.3  & 0.3  \\
 Auto-multithresh parameters &&&&& \\
 --- sidelobeThreshold & 1.25 & 2.0  & 3.0  & 3.0  & 1.25 \\
 --- noiseThreshold    & 5.0  & 4.25 & 5.0  & 4.5  & 4.5  \\
 --- minBeamFrac       & 0.1  & 0.3  & 0.3  & 0.3  & 0.3  \\
 --- lowNoiseThreshold & 2.0  & 1.5  & 1.5  & 1.5  & 1.5  \\
 --- negativeThreshold & 0.0  & 0.0  & 0.0  & 0.0  & 0.0  \\
 Multiscale deconvolver &&&&& \\
 --- scales            & [0, 30, 60] & [0, 6, 18, 30] & [0, 6, 18, 30] & [0, 6, 18, 30, 60] & [0, 6, 18, 30, 60] \\
\hline
\end{tabular}
\end{table*}
%------------------------------------------------------------------------

%-------------------------------- Figure --------------------------------
\begin{figure*}[t!]
\centering
\includegraphics[width=0.95\textwidth]{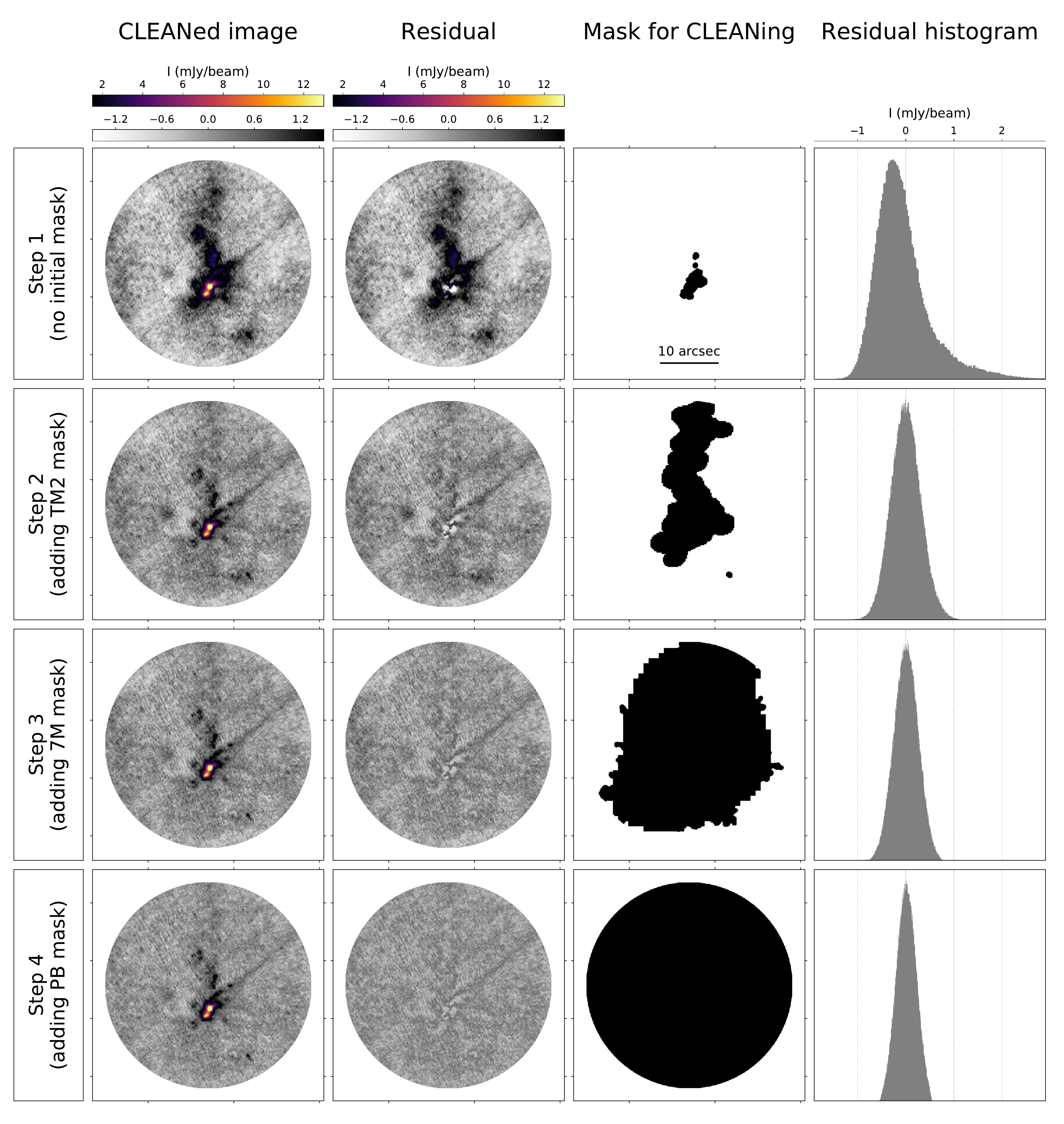}\\
\caption{Joint-deconvolution imaging strategy used in the ALMAGAL processing workflow. The target used as an example is the ALMAGAL field 101899 (also labelled AG023.0108$-$0.4102 in \citealt{Molinari2024}). The different columns show from left to right: CLEANed image, residual image, CLEANing mask, and histogram of residual emission. The different rows show the four main steps of the iterative CLEANing process (see details in Sect.~\ref{sec:joint-deconvolution}). Note that in Step~1, there is no starting CLEANing mask and the output (shown in the panel) is generated from the multi-thresh automasking functionalities of \texttt{tclean}.}
\label{fig:imaging-strategy}
\end{figure*} 
%------------------------------------------------------------------------

%________________________________________________________________
%
\subsection{Final joint-deconvolution imaging\label{sec:joint-deconvolution}}

At this stage, the ALMAGAL processing workflow has all the required files and information to generate the final science-ready images. For this, we jointly image the different array configurations using the \texttt{tclean} task in \texttt{CASA} following the strategy described below. We note that we not only implement this strategy for the combined arrays (e.g., 7M+TM2+TM1, 7M+TM2) but also to generate new sets of images for the individual arrays (i.e., 7M, TM2, and TM1), thus enabling a direct comparison between all image products. All images were produced using the \textit{multiscale} deconvolver and the multithresh automasking routines available in \texttt{tclean}. Table~\ref{tab:tclean_parameters} lists the main \texttt{tclean} parameters used to produce both continuum and spectral cube images.

%________________________________________________________________
%
\subsubsection{Continuum images\label{sec:joint-deconvolution_continuum}}

Figure~\ref{fig:imaging-strategy} summarizes the iterative process used to generate the final set of continuum ALMAGAL images. The ALMAGAL pipeline uses \texttt{tclean} (CASA version~6.2.0; see Table~\ref{tab:tclean_parameters} for the specific parameters) throughout the whole process, which is divided into four major steps (see rows in Fig.~\ref{fig:imaging-strategy}).

In the first step, it does not set any initial mask and lets the multi-thresh auto-masking functionality of \texttt{tclean} generate its own (similarly to what is done for the first set of images generated in Sect.~\ref{sec:continuum-determination}). This first step runs for only 100 CLEANing iterations or until the noise reaches a threshold of $4\times$ the rms. Throughout all the processes described in this section, the rms noise level is determined as the median absolute deviation (MAD) of the residual image (see also Sect.~\ref{sec:data-products}). For targets with emission, the output is usually a very shallow CLEANed image (see first column of Fig.~\ref{fig:imaging-strategy}) with large amounts of unCLEANed emission in the residual (see second column) and a relatively compact mask around the brightest intensity peaks (see third column).

In the second step, the pipeline merges the mask created in the first step with the TM2 CLEANed mask produced during the generation of the first continuum images for each individual array (see Sect.~\ref{sec:continuum-determination}). This ensures that extended and relatively faint emission present in the TM2-array data is included in the CLEANing process. With this new mask, 
a first round of CLEAN with 500 iterations (or until a threshold of $4\times$ the rms) is performed using the multi-thresh auto-masking functionalities of \texttt{tclean} to allow the mask to continue growing. Following this, additional runs of \texttt{tclean} are executed iteratively, in rounds of 500 iterations, until one of these two conditions are met: (1) the flux in the model (or CLEAN components) does not increase by more than 1\%, or (2) the emission in the residual is less than $2.5\times$ the expected theoretical noise (e.g., 0.1~mJy~beam$^{-1}$ in the 7M+TM2+TM1 images). This iterative approach, in rounds of 500 iterations each, was found to be useful to avoid divergences that could result in large CLEANing artifacts in the final images (particularly in the line cubes, see Sect.~\ref{sec:joint-deconvolution_cubes}). As shown in the second row of Fig.~\ref{fig:imaging-strategy}, the CLEANing process results in less contaminated residuals (see also the histogram in the fourth column) and a larger mask, including elongated features present in the image. Note that there seems to be a decrease in flux between the CLEANed images of the first and second steps. This is due to the different intensity units in the CLEANed (units of Jy/cleaned-beam) and dirty (units of Jy/dirty-beam) images. During the process of imaging, we explored the possibility of applying a correction factor that takes into account the different units. We followed the strategy described in the works by \citet{JorsaterVanMoorsel1995} and \citet{Czekala2021}. However, the ``snapshot'' nature of our observations (i.e., short integration times) together with the combination of multiple arrays result in relevant substructures in the point spread functions. In Appendix~\ref{app:JvMcorrection}, we discuss the potential limitations of applying this correction factor to the ALMAGAL dataset. Based on this, we discarded the application of such a correction factor to the first ALMAGAL data release.

In the third step, the ALMAGAL pipeline merges the output mask of the second step with the 7M CLEANed mask produced in the generation of the first continuum images (similar to the addition of the TM2 mask). Similar to the previous step, this ensures that extended emission detectable in the 7M-array data is also properly CLEANed. We CLEAN for 500 iterations (or until a threshold of $4\times$ the rms). This process is repeated twice, and then iteratively until the same conditions as in the second step are met.

Finally, in the fourth step, the pipeline uses a mask that is equal to the whole primary beam (PB) and repeats the CLEANing process of the second and third steps: 500 iterations or until a threshold of $4\times$ the rms, and iteratively until the flux in the CLEANed model does not increase more than 1\% or the residual peaks are be $2.5\times$ the theoretical noise level. The resulting image is shown in the last row of Fig.~\ref{fig:imaging-strategy}.

The output of this process is a joint-deconvolved CLEANed \texttt{image} for the continuum emission, together with additional files generated with \texttt{tclean}: \texttt{residual}, that can be used to determine the noise level in the image; \texttt{pb}, that contains information about the primary beam response; \texttt{psf}, that can be used to explore the structure and presence of sidelobes in the point spread function; and \texttt{model}, that contains the CLEANed components used to generate the final image.

%-------------------------------- Figure --------------------------------
\begin{figure*}[h!]
\centering
\includegraphics[width=0.95\textwidth]{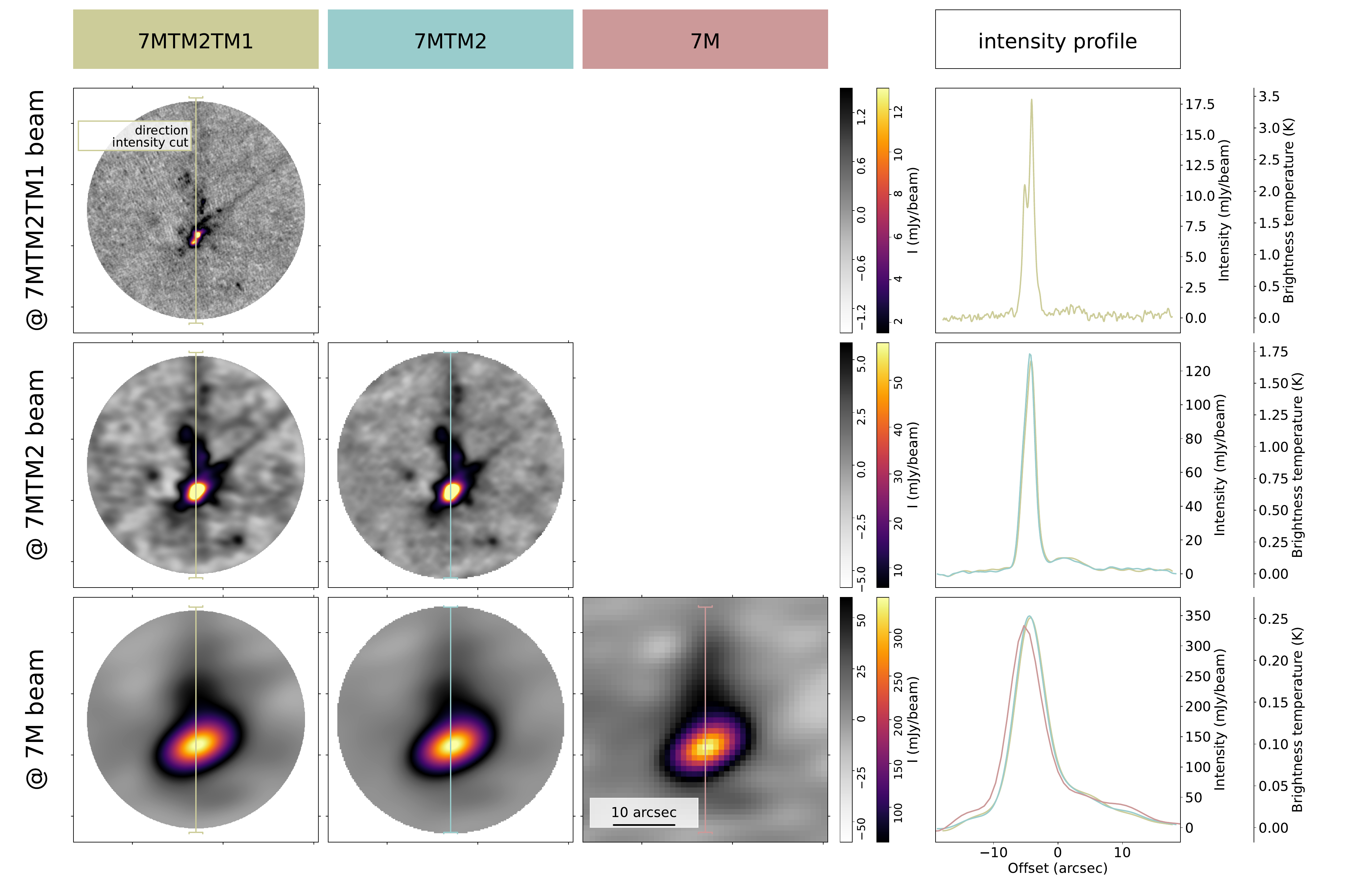}\\
\caption{Comparison of the ALMAGAL images for field 101899 (also labelled AG023.0108$-$0.4102 in \citealt{Molinari2024}) to explore the quality of the data combination. The images are convolved to different angular resolutions in the different rows: The top row shows images convolved to the 7M+TM2+TM1 beam ($0\farcs38\times0\farcs34$), the middle row corresponds to images convolved to the 7M+TM2 beam ($1\farcs64\times1\farcs17$), and the bottom row shows the images convolved to the 7M-only beam ($8\farcs0\times4\farcs4$). The first three columns correspond to the images of the arrays 7M+TM2+TM1, 7M+TM2 and 7M, from left to right, respectively. The rightmost column shows the intensity profiles (both in intensity, mJy~beam$^{-1}$, and in brightness temperature, K) of the vertical intensity cuts indicated in the different images. The good agreement between the images and the intensity profiles ensures a correct process in combining all arrays.}
\label{fig:smoothing-beams}
\end{figure*} 
%------------------------------------------------------------------------

%________________________________________________________________
%
\subsubsection{Line-only cube images\label{sec:joint-deconvolution_cubes}}

For the line data cubes, we followed the same strategy as for the continuum images (see Sect.~\ref{sec:joint-deconvolution_continuum} and Fig.~\ref{fig:imaging-strategy}), with the main difference being that we performed the joint-deconvolution CLEANing in the $\approx3840$ channels of each spectral window. For this, the ALMAGAL pipeline first starts regridding the frequency channels to a common frame for all the measurement set (ms) files of all three array configurations. While differences in the frequency frame are easily handled by the \texttt{tclean} task, the main reason for realigning the ms files to a common number of channels and starting frequency is to facilitate the division of the ms files into chunks of a few hundred channels each that can later be processed in parallel throughout multiple computers\footnote{Note that this produces a similar result as the \texttt{parallel=True} option in \texttt{tclean} within the MPI version of CASA. Unfortunately, due to the architecture of the JUWELS supercomputer, we were not able to use the MPI version of CASA across multiple computing nodes, and thus, we developed a strategy that enables the parallel processing of large cubes.}.

Following the same CLEANing strategy described in Fig.~\ref{fig:imaging-strategy}, the processing pipeline produces 38 line-only cubes of about 100~channels each for each spectral window (i.e., four in total per source). The CLEANing strategy consists, as for the continuum images (see Sect.~\ref{sec:joint-deconvolution_continuum}), of iterative rounds of \texttt{tclean}, with 500 iterations each and with the CLEANing mask expanded step-wise by adding the TM2 and 7M masks. The CLEANing process stops when the flux in the model does not increase more than 1\% or the emission in the residual is less than $2.5\times$ the expected theoretical noise (e.g., 2.5 and 5~mJy~beam$^{-1}$ for the broad and narrow-channel spectral windows, respectively). This ensures that all sub-cubes are CLEANed down to a similar rms noise level. These 100-channel cubes, each with a synthesized beam that varies with frequency across the different channels, are merged together to produce the final CLEANed cube. The final stitched cube is smoothed to a common beam, corresponding to the largest beam in the originally stitched cube. Similar to the processing of the continuum images, five cube files are generated: CLEANed \texttt{image}, \texttt{residual}, \texttt{pb}, \texttt{psf} and \texttt{model}, for each spectral window.

%________________________________________________________________
%
\subsection{Data products and quality assessment\label{sec:data-products}}

The iterative CLEANing strategy used in the ALMAGAL processing workflow (see Sect.~\ref{sec:joint-deconvolution}) generates final science-ready products for both continuum and line cubes. This strategy was used to joint-deconvolve multiple arrays (e.g., 7M+TM2+TM1, 7M+TM2) as well as individual arrays (e.g., 7M, TM2, and TM1) making it possible to carry out one-to-one comparisons of images sensitive to different spatial scales. We note that the line cubes for the TM1-only data using the new strategy are not generated in this first data release due to computing time restrictions.

The science-ready images have been converted to the standard FITS format common in astronomy, and cropped to contain only the size of the field of view (which extends down to the 30\% primary beam response). A total of five files were generated for each ALMAGAL target and array configuration: \texttt{image}, \texttt{residual}, \texttt{pb}, \texttt{psf} and \texttt{model}. The \texttt{image} files are not primary-beam corrected, which results in a homogeneous noise level across the whole image, enabling an easier process of source and structure identification \citep[see][]{Coletta2024}. The primary-beam correction can be applied using the \texttt{pb} files, thus obtaining reliable flux measurements for sources across the whole field of view. The \texttt{residual} images are used to determine the noise levels (see Sect.~\ref{sec:QAintensity}) and to confirm that no major residual emission is left unCLEANed. Finally, the \texttt{psf} and \texttt{model} files provide the dirty beam and CLEAN components that can be used to reconstruct new images if necessary. In the following sections, we assess the quality of the ALMAGAL scientific products.

%________________________________________________________________
%
\subsubsection{Quality assessment of data combination\label{sec:QAcombination}}

In this section we check if the strategy followed to combine the data from multiple array configurations is able to recover all the emission that each array is sensitive to. For this, we compare the properties of different images in Fig.~\ref{fig:smoothing-beams} (see Appendix~\ref{app:QAcombinationCubes}, and Figs.~\ref{fig:smoothing-beams_spw0} and \ref{fig:smoothing-beams_spw1}, for similar examples of spectral lines and cubes). The panels in the main diagonal of the figure show the final products for the 7M+TM2+TM1, 7M+TM2 and 7M-alone datasets. Although at first glance, the 7M+TM2 and 7M-alone images seem to be sensitive to fainter and more extended emission compared to the 7M+TM2+TM1 image, this is only due to the different beam sizes (see also Sect.~\ref{sec:QAbeam}). To confirm this, we convolved the images to the same angular resolutions. In the first column, the 7M+TM2+TM1 image is convolved to the final resolution of the 7M+TM2 (middle row) and 7M-alone (bottom-row) images. Similarly, in the second column, we convolve the 7M+TM2 image to the resolution of the 7M-alone image (bottom-row). The convolved images confirm that we recover the same structure and intensities. The rightmost column shows intensity profiles for vertical cuts along the different images. The similarity in the images and intensity profiles confirms a successful combination strategy, recovering all the emission detectable in all the relevant array configurations.

%-------------------------------- Figure --------------------------------
\begin{figure}[ht!]
\centering
\includegraphics[width=0.85\columnwidth]{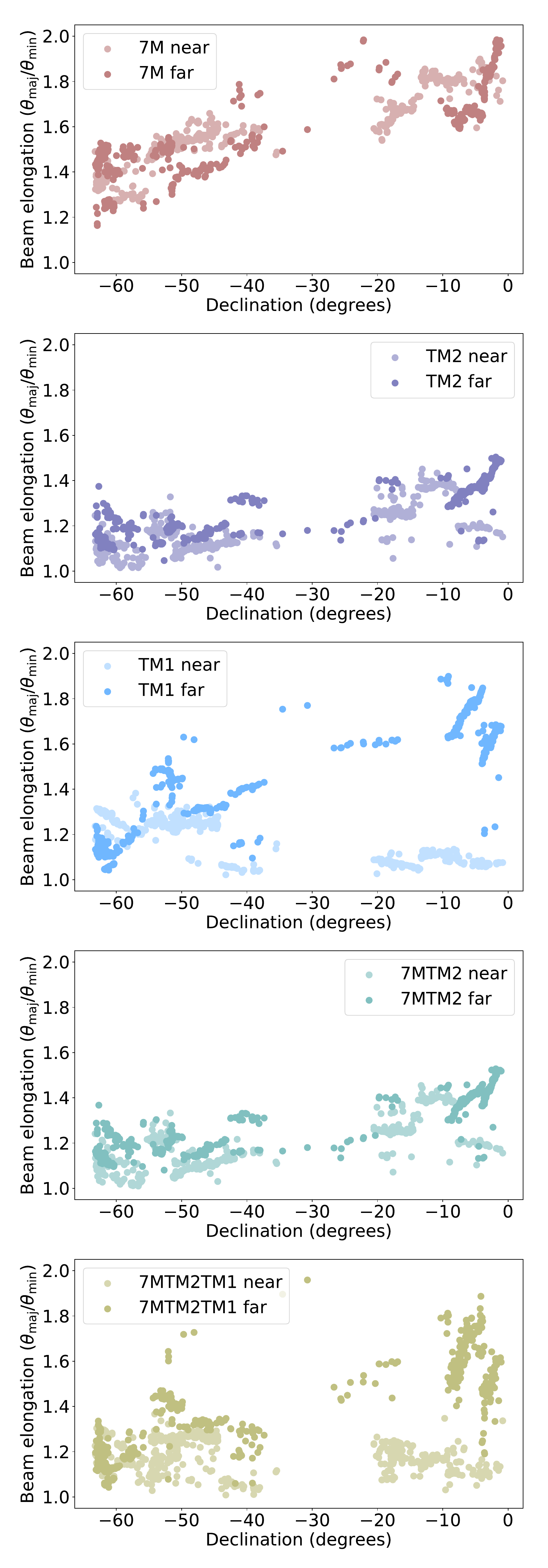}\\
\caption{Beam elongation of the ALMAGAL images as a function of the target declination. From top to bottom, the panels show the data for the 7M, TM2, TM1, 7M+TM2, and 7M+TM2+TM1 images.}
\label{fig:beam-declination}
\end{figure}
%------------------------------------------------------------------------

%________________________________________________________________
%
\subsubsection{Angular resolution in final products\label{sec:QAbeam}}

The target spatial resolution for ALMAGAL is $\approx1000$~au, which corresponds to angular resolutions of $0\farcs15$ and $0\farcs30$ for the ``far'' and ``near'' samples respectively (corresponding to the more extended configurations C-6 and C-5 of the ALMA main array). Figure~\ref{fig:beam-properties} shows the distribution of beam properties (major axis, minor axis and beam elongation) for the continuum images. From top to bottom it shows the beam properties of the 7M, TM2, TM1, 7M+TM2, and 7M+TM2+TM1 images. The requested angular resolutions are achieved in the TM1 data images, with median beam sizes of $0\farcs22\times0\farcs15$ and $0\farcs35\times0\farcs29$ for the ``far'' and ``near'' samples, respectively. Table~\ref{tab:almagal_summary} has a summary of the beam sizes for the different images and arrays, whereas Table~\ref{tab:almagal_observations} lists the minimum and maximum beams together with the median and the 16$^\mathrm{th}$ and 84$^\mathrm{th}$ percentiles. The combined 7M+TM2+TM1 images have slightly larger beam sizes with a median of $0\farcs28\times0\farcs19$ and $0\farcs47\times0\farcs38$ for the ``far'' and ``near'' samples, respectively. This coarser resolution (about 30\% compared to the TM1-alone images) is expected due to the weights of the TM2 and 7M data when generating the \texttt{psf} image that describes the beam size.

The right column of Fig.~\ref{fig:beam-properties} shows the distributions of the beam elongation for the different images. The beam elongation is defined as the ratio between the major and minor beam axis (i.e., $\theta_\mathrm{maj}/\theta_\mathrm{min}$). The images generated with only the 7M data are those having the most elongated beams, with a median of about 1.5. For all the other images, including the fully combined 7M+TM2+TM1 images, the beams have elongations in the range 1.0--1.4. We note that a certain fraction of ``far'' sample sources have beam elongations between 1.4 and 1.8 (see bottom-right panel of Fig.~\ref{fig:beam-properties}). The origin of these elongated beams is in the TM1 observations. In Fig.~\ref{fig:beam-declination} we explore the relation of the beam elongation with the declination of the targets. We can see that ALMAGAL targets with declinations between $-20^\circ$ and $0^\circ$ (i.e., close to the equatorial celestial plane) have on average larger beam elongations. Despite these effects, and based on the well-behaved beam sizes for the majority of the ALMAGAL images, we do not circularize the beams in the final images, leaving them with the native maximum possible angular resolution.

%-------------------------------- Figure --------------------------------
\begin{figure}[t!]
\centering
\includegraphics[width=0.82\columnwidth]{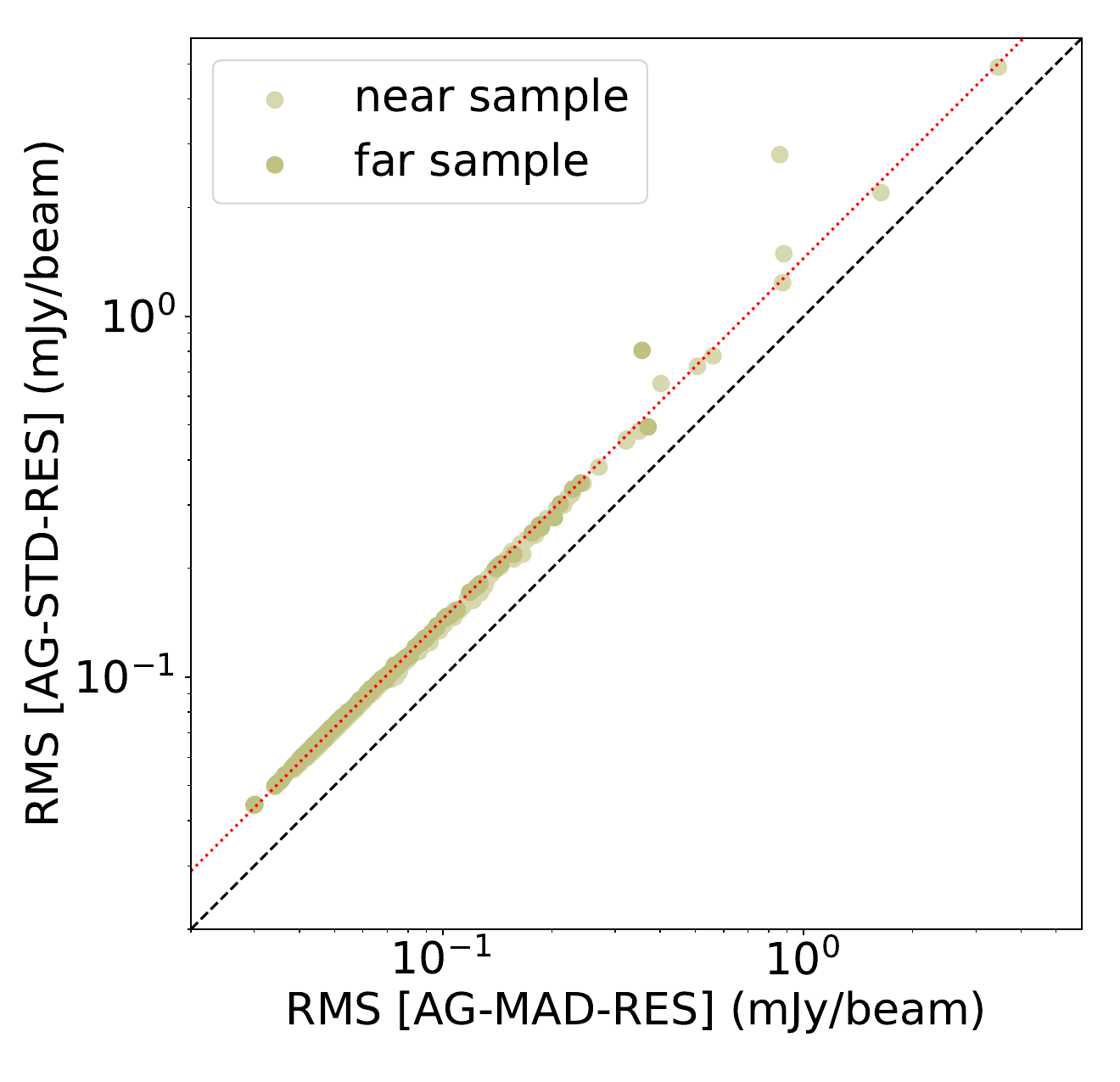}\\
\caption{Comparison of different rms noise level estimates for the final ALMAGAL 7M+TM2+TM1 continuum images. The noise is calculated as the standard deviation of the residual image (\texttt{AG-STD-RES}) as a function of the noise calculated as the median absolute deviation of the residual image (\texttt{AG-MAD-RES}). The black dashed lines corresponds to $y=x$, while the red dotted line corresponds to $y=1.45\,x$ (i.e., the rms noise level estimated with the standard deviation is about 45\% larger compared to the median absolute deviation estimate).}
\label{fig:rms-estimates}
\end{figure}
%------------------------------------------------------------------------

%________________________________________________________________
%
\subsubsection{Intensity and noise levels in final products\label{sec:QAintensity}}

Figure~\ref{fig:continuum-properties} shows the distributions of the rms noise levels, the peak intensity, and the peak dynamic range (or signal-to-noise ratio) for the continuum images of the different array configurations. Table~\ref{tab:almagal_observations} shows the minima, maxima, median, and 16$^\mathrm{th}$ and 84$^\mathrm{th}$ percentiles for both the continuum and line-only cube images.

The noise level in the ALMAGAL images is calculated using six different methods. The derived values are stored in the header of the FITS image files using the following keywords: \texttt{AGMADRES}, the noise is determined as the MAD (median absolute deviation) of the emission in the residual image; \texttt{AGSTDRES}, the noise is determined as the STD (standard deviation) of the emission in the residual image; \texttt{AGMADREM} and \texttt{AGSTDREM}, the noise is determined as the MAD and STD, respectively, of the residual image after masking\footnote{The masked regions are defined as those regions with emission above $5\,\sigma$ level, where $\sigma$ is determined from the median absolute deviation of the residual image. These regions are then expanded down to a $3\,\sigma$ level.} out bright  regions in the final image; \texttt{AGMADIMM} and \texttt{AGSTDIMM}, the noise is determined as the MAD and STD, respectively, of the final image after masking out bright regions. Figure~\ref{fig:rms-estimates} shows a comparison between the different noise estimates. Overall, the rms derived with STD is about 45\% larger compared to MAD-derived rms noise level. There are no major differences between the analysis performed directly in the residual images, or in the final images, since the bright emission in the latter are masked out and the remaining regions are expected to be those with contribution only from the residual image. The rms noise level shown in Fig.~\ref{fig:continuum-properties} and throughout the paper correspond to the \texttt{AGMADRES} values. The final MAD noise level in the fully combined (7M+TM2+TM1) images ranges from 0.04 to 0.1~mJy~beam$^{-1}$ (or from 0.06 to 0.15~mJy~beam$^{-1}$ if the STD values are considered), in agreement with the expected noise level originally requested in the project. For the spectral cubes (see Table~\ref{tab:almagal_observations}), the rms noise level is in the range 2.1--2.4~mJy~beam$^{-1}$ for the broad spectral windows (spw0 and spw1) with a spectral resolution of 1.4~km~s$^{-1}$, and 4.3--5.1~mJy~beam$^{-1}$ for the narrow spectral windows (spw2 and spw3) with a spectral resolution of 0.34~km~s$^{-1}$.

The peak intensities in the fully combined images are typically in the range from 0.44 to 17~mJy~beam$^{-1}$ for the continuum images, although some sources are as bright as 10~Jy~beam$^{-1}$ (see central column of Fig.~\ref{fig:continuum-properties} and Table~\ref{tab:almagal_observations}). This results in typical signal-to-noise ratios of about 9 to 200 for most of the cases. There are a few cases with large dynamics ranges of several thousand (e.g., ALMAGAL target 693050, see also Fig.~\ref{fig:selfcalibration-example_v01}). For the line cubes, the dynamic range is lower with values typically between 10 and 80.

%-------------------------------- Figure --------------------------------
\begin{figure}[t!]
\centering
\includegraphics[width=0.85\columnwidth]{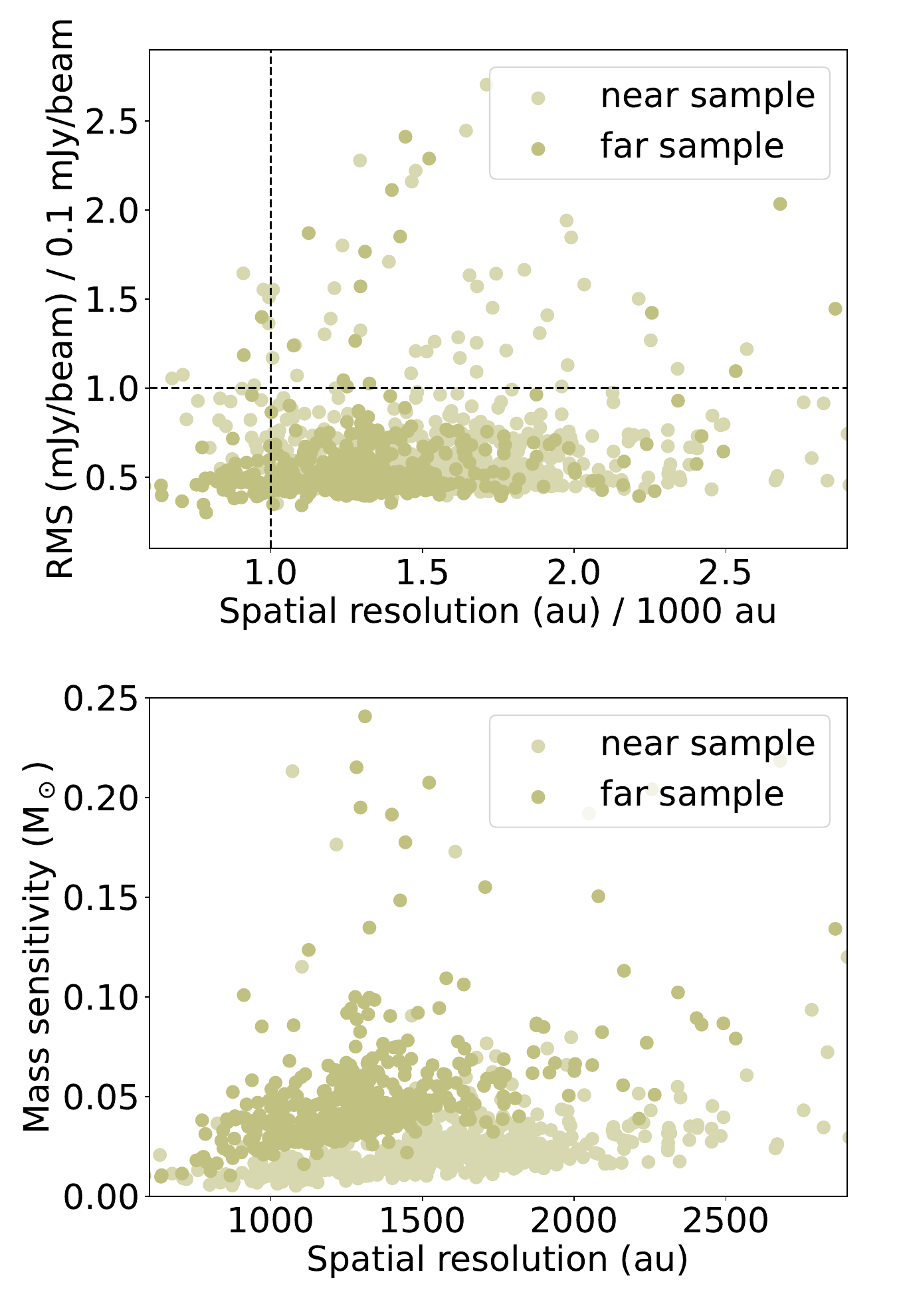}\\
\caption{Top panel: rms noise level of the 7M+TM2+TM1 continuum images (in units of 0.1~mJy~beam$^{-1}$ against the spatial resolution (in units of 1000~au). The dashed horizontal and vertical lines mark the target sensitivity and spatial resolutions of the ALMAGAL project. Bottom panel: Point source mass sensitivity (in units of $M_\odot$) as function of the spatial resolution as in the top panel. The mass sensitivity is derived from the rms noise level (top panel) assuming a dust temperature of 20~K, a dust opacity of 0.899~cm$^{2}$~g$^{-1}$ at 1.3~mm, and a gas-to-dust ratio of 100. For most of the ALMAGAL fields, the mass sensitivity is better than the nominal 0.1~$M_\odot$ aimed in the design of the ALMAGAL project.}
\label{fig:massSensitivity-spatialResolution}
\end{figure}
%------------------------------------------------------------------------

%________________________________________________________________
%
\subsubsection{ALMAGAL science products\label{sec:QAgeneral}}

The ALMAGAL project aimed for a sensitivity of $\approx0.1$~mJy~beam$^{-1}$ in the continuum images, and a spatial resolution of $\approx1000$~au. The top panel of Fig.~\ref{fig:massSensitivity-spatialResolution} shows the final sensitivities and resolutions for the fully combined (7M+TM2+TM1) images. Most of our targets have rms noise levels up to a factor of two better than the requested sensitivity. However, only a small fraction of fields have reached a spatial resolution better than 1000~au, with many fields having resolutions of up to 2500~au. The bottom panel of Fig.~\ref{fig:massSensitivity-spatialResolution} shows the sensitivity of the images converted to a gas and dust mass sensitivity, assuming a conservative temperature of 20~K, a dust opacity of 0.899~cm$^{2}$~g$^{-1}$ \citep{OssenkopfHenning1994}, and a gas-to-dust ratio of 100. For most of the fields our $1\,\sigma$ point source mass sensitivity is better than 0.1~$M_\odot$, making the ALMAGAL images well-suited to search for low-mass dense cores across a large number of high-mass star-forming regions.

%-------------------------------- Figure --------------------------------
\begin{figure}[t!]
\centering
\includegraphics[width=0.79\columnwidth]{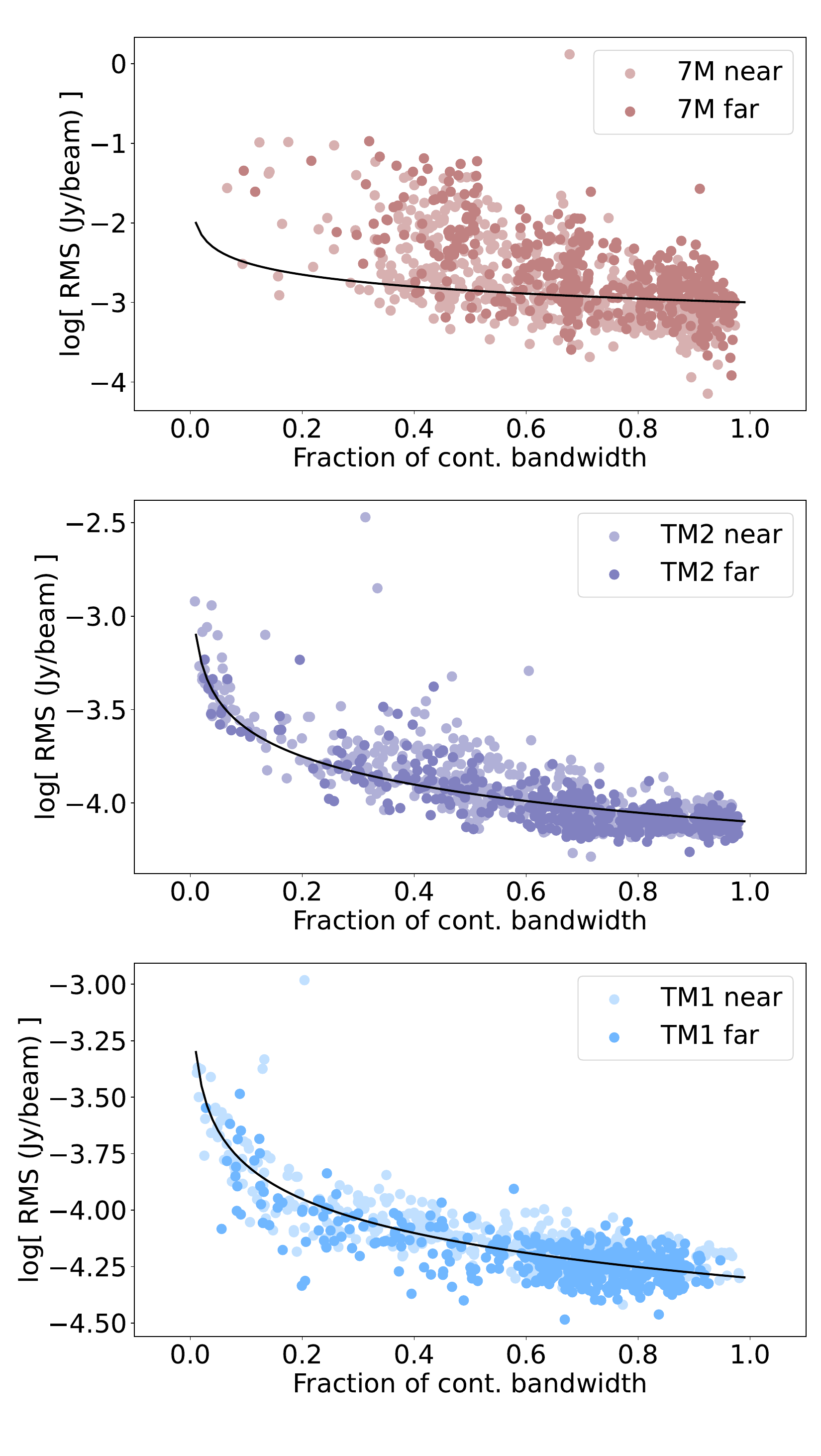}\\
\caption{Relation between rms noise level and available fraction of frequency bandwidth used to derive the continuum emission. From top to bottom, the panels show the results for the images of the 7M, TM2, and TM1 individual arrays. The rms noise level increases by a factor of three for narrow continuum bandwidths. The black solid lines depict the theoretical relation of the rms noise level as function of the available bandwidth (i.e., $\mathrm{rms}\propto 1/\sqrt{\mathrm{bandwidth}}$). Note that the continuum bandwidth is normalized to the total available bandwidth, corresponding to 3.75~GHz (i.e., spw0 and spw1 combined).}
\label{fig:rms-bandwidth}
\end{figure}
%------------------------------------------------------------------------

%-------------------------------- Figure --------------------------------
\begin{figure}[ht!]
\centering
\includegraphics[width=0.79\columnwidth]{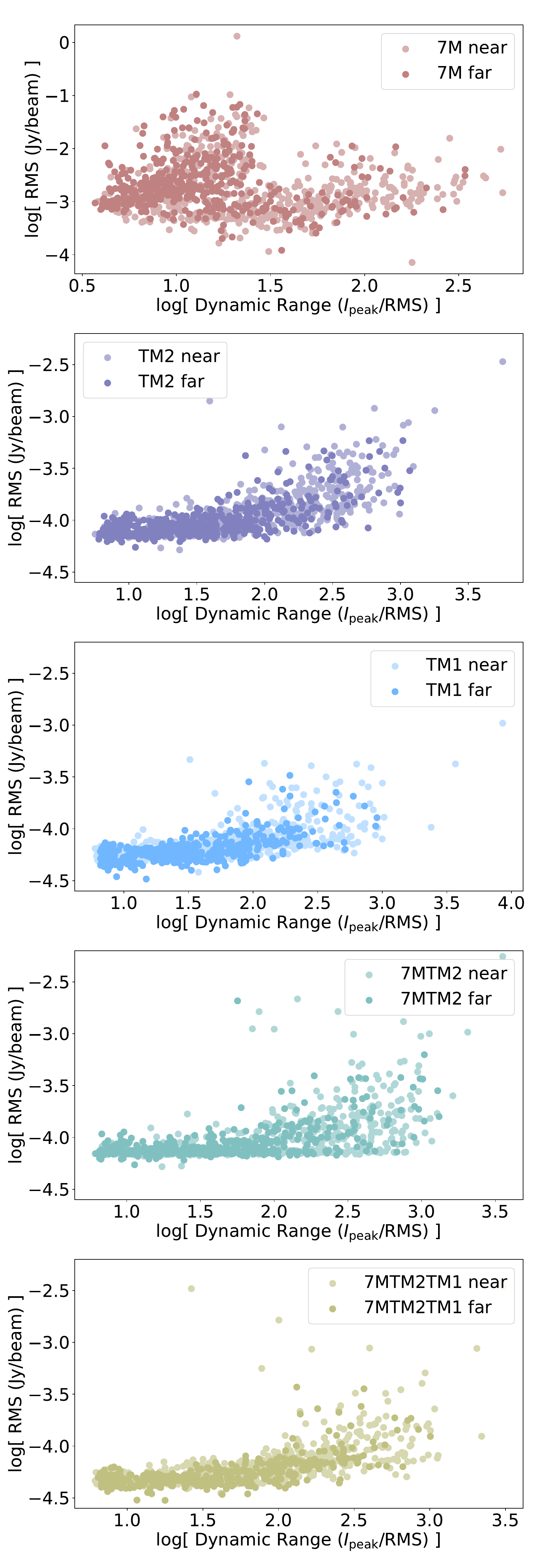}\\
\caption{Relation between rms noise level and the dynamic range of the continuum images. From top to bottom, the panels show the results for the images of the 7M, TM2, TM1, 7M+TM2, and 7M+TM2+TM1 arrays. The dynamic range is determined as the ratio between the peak intensity to the rms noise level. We see that for dynamic ranges $\gtrsim300$ the rms noise level starts to increase.}
\label{fig:rms-snr}
\end{figure}
%------------------------------------------------------------------------

%-------------------------------- Figure --------------------------------
\begin{figure*}[t!]
\centering
\includegraphics[width=0.98\textwidth]{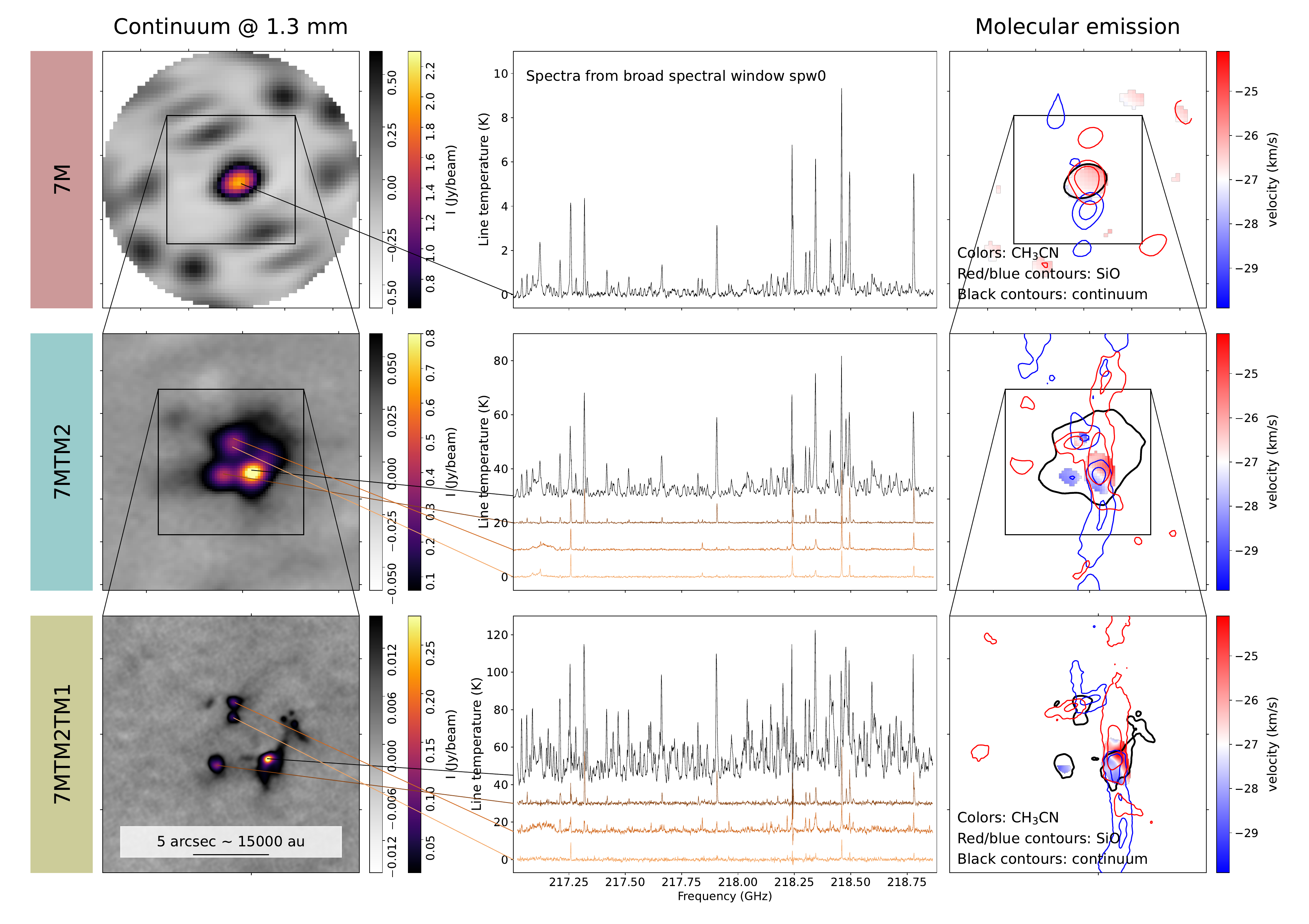}\\
\caption{Continuum and spectral line results towards the ALMAGAL field 865468 (also labelled AG345.0035$-$0.2240 in \citealt{Molinari2024}). From top to bottom, results for the 7M (beam size of $7\farcs3\times4\farcs7$), 7M+TM2 (beam size of $2\farcs1\times1\farcs7$), and 7M+TM2+TM1 (beam size of $0\farcs36\times0\farcs34$) images. The left-column panels show the dust continuum emission at 219~GHz (or 1.38~mm). The middle-column panels show the spectra towards different compact sources identified in the continuum images. The right-column panels show the emission for two molecular spectral lines: SiO\,(5--4) and CH$_3$CN\,(12$_\mathrm{K=3}$--11$_\mathrm{K=3}$). The color image shows the first order moment map, or velocity field of the CH$_3$CN line over a velocity range from $-37$ to $-17$~km~s$^{-1}$ (the systemic velocity of the region is $-27$~km~s$^{-1}$). The blue and red contours show zero-th order moment maps (or integrated emission) of the blue-shifted (from $-100$ to $-37$~km~s$^{-1}$) and red-shifted (from $-17$ to 46~km~s$^{-1}$) SiO emission, respectively. The black contour depicts the spatial distribution of the continuum emission as shown in the left-column panels.}
\label{fig:exampleLinesScience}
\end{figure*}
%------------------------------------------------------------------------

Figures~\ref{fig:rms-bandwidth} and \ref{fig:rms-snr} show the potential relation of the limited bandwidth used for the continuum determination (see Sect.~\ref{sec:continuum-determination} and Fig.~\ref{fig:continuum-bandwidth}) and the presence of bright sources in the field that limit the dynamic range, with the final rms noise level of the image. From the results on the TM1-alone images (see Fig.~\ref{fig:rms-bandwidth}), we see that the rms noise level is below 0.1~mJy~beam$^{-1}$ when the bandwidth for continuum determination was at least 50\% of the total broad bandwidth (i.e., at least 2.8~GHz bandwidth). The rms noise level increases for smaller bandwidths, approaching values of 0.3~mJy~beam$^{-1}$ or higher in some cases. The black solid lines shown in the three panels of Fig.~\ref{fig:rms-bandwidth} depict the theoretical relation of the rms noise level with the available bandwidth\footnote{The rms noise level is inversely proportional to the squared-root of the bandwidth.}. For the TM2 and TM1 datasets, the increase of rms noise level can be explained by the reduced bandwidth used to produce the continuum emission images. This relation is less clear for the 7M data where other effects (e.g., reduced image fidelity of the 7M array) may contribute to the increase of rms noise level. The sensitivity can also be limited by the presence of bright sources in the observed fields. Although self-calibration helps to improve the rms noise level in these cases, there is still a correlation between the final sensitivity and the achieved dynamic range that is related to the maximum intensity in the field of view. As shown in Fig.~\ref{fig:rms-snr} (see panel for the 7M+TM2+TM1 images), the rms noise level is lower than 0.1~mJy~beam$^{-1}$ for dynamic ranges  $\lesssim100$, and increases up to 0.3~mJy~beam$^{-1}$ for images with dynamic ranges $\gtrsim300$.

In addition to the FITS images of the continuum and line cubes\footnote{The FITS files will be made available at the ALMAGAL website \url{https://www.almagal.org/}}, the ALMAGAL consortium has compiled a table with information about relevant observational parameters of all the images. These include information on the central coordinates of the images, pixel size and number of pixels, number of channels and channel resolution for the different line cubes, as well as rms noise levels for both continuum and line cubes. This table is available in electronic format at the GitHub repository\footnote{See footnote 2, in Sect.~\ref{sec:workflow}}.

%-------------------------------- Figure --------------------------------
\begin{figure*}[t!]
\centering
\includegraphics[width=0.98\textwidth]{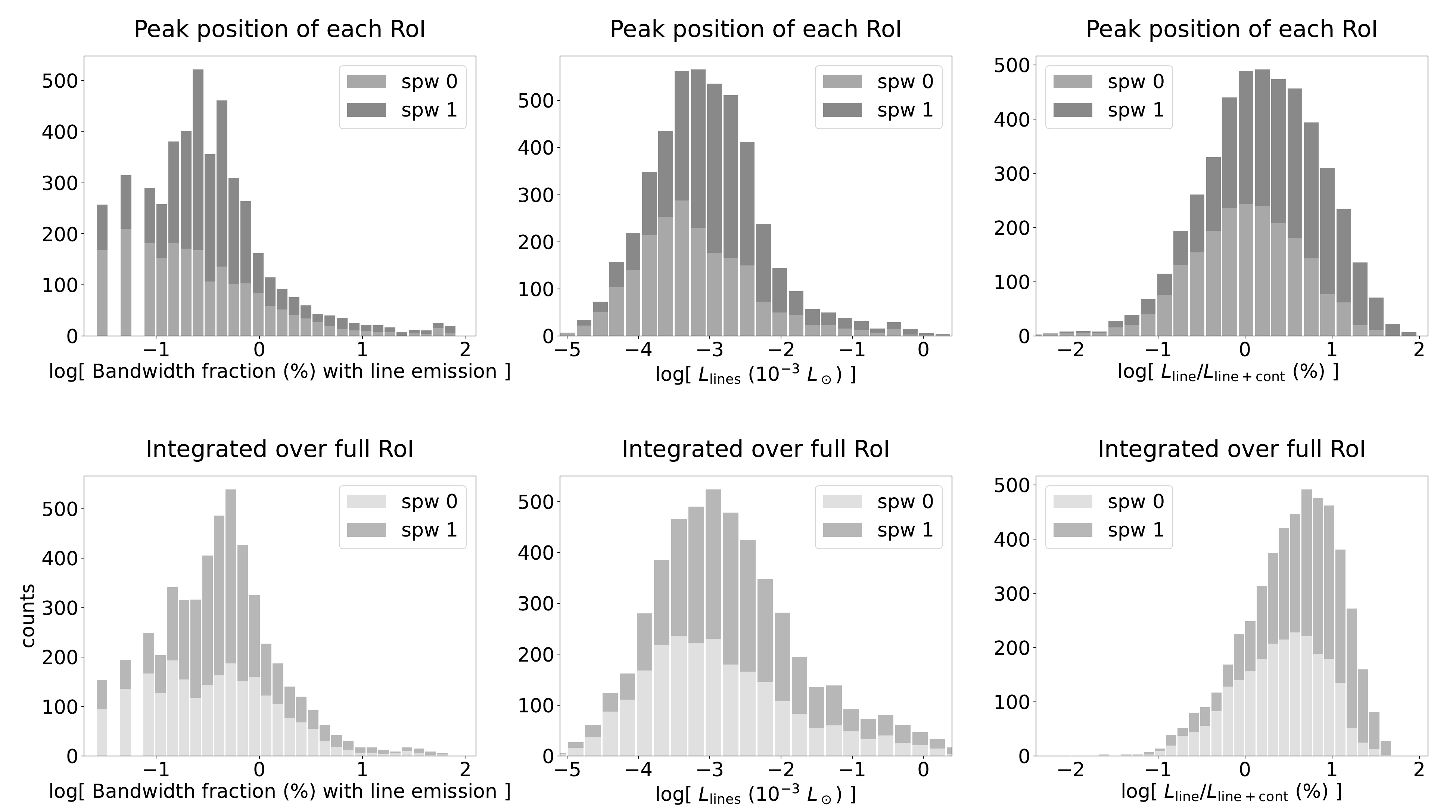}\\
\caption{Distribution of spectral line properties towards the regions of interest (RoIs) of the ALMAGAL fields. The $y$-axis in all panels depicts the number of RoIs. From left to right columns, the panels show the fraction of bandwidth with line emission, the luminosity contained in spectral lines, and the fraction of line luminosity to total luminosity (continuum and lines). The top panels show the distributions for the spectra extracted towards the continuum intensity peak of each RoI, while the bottom panels show the results for the spectra extracted integrated over the whole RoI. See Sect.~\ref{sec:chemical-richness} for more details.}
\label{fig:spectra-RoIs_v01}
\end{figure*}
%------------------------------------------------------------------------

%-------------------------------- Figure --------------------------------
\begin{figure*}[t!]
\centering
\includegraphics[width=0.98\textwidth]{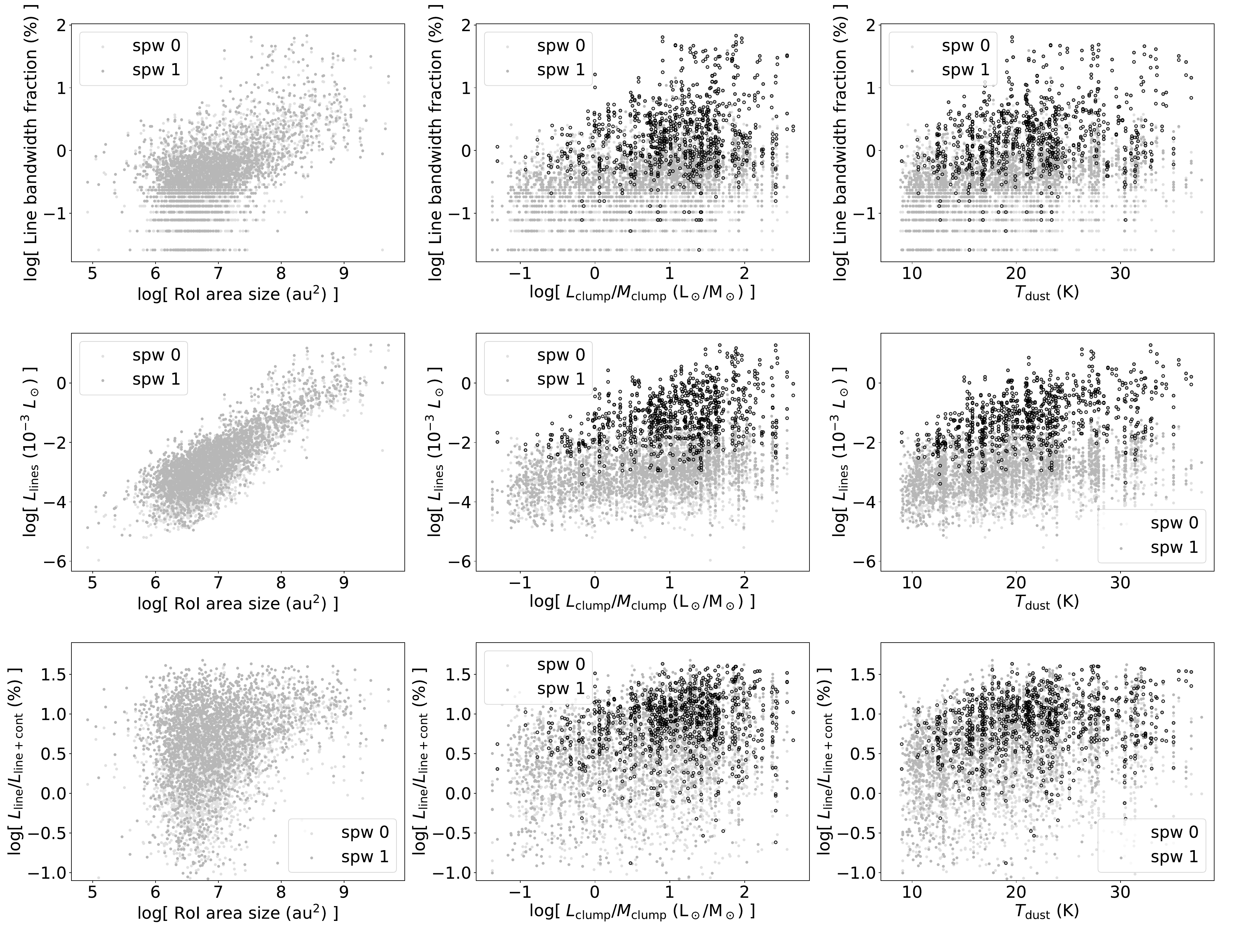}\\
\caption{Relation between the parameters describing the line richness of the RoIs and physical properties of the hosting ALMAGAL clumps. From top to bottom, the different rows show the fraction of bandwidth with line emission, the line luminosity, and the fraction of line luminosity to total luminosity. From left to right, the different columns show the size of the RoI area, the clump luminosity-to-mass ratio ($L_\mathrm{clump}/M_\mathrm{clump}$) as an indicator of evolutionary stage, and the clump dust temperature ($T_\mathrm{dust}$). No significant difference is seen for the spw~0 (lighter grey) and spw~1 (darker grey) data. Black symbols in the middle and right-column panels correspond to those RoIs with area sizes larger than $3.2\times10^{7}$~au$^{2}$, corresponding to the 84$^\mathrm{th}$ percentile of the RoI area sizes shown in the left-column panels.}
\label{fig:spectra-RoIs_v02}
\end{figure*}
%------------------------------------------------------------------------

%________________________________________________________________
%
\section{Science cases: Spectral line properties\label{sec:lines}}

A comprehensive description of the different scientific topics that can be explored with the ALMAGAL dataset is presented in \citet{Molinari2024}, which also introduces the scientific potential of the continuum images. In this current work, we highlight the potential that the spectral line images have to offer. We note that detailed studies on specific aspects related to topics such as fragmentation level, cluster properties, temperature and mass determination, outflows, rotating disks, accreting structures, association with masers or chemical content, among others, are deferred to ongoing studies and publications by the ALMAGAL consortium \citep[e.g.,][Jones et al.\ in prep., Law et al.\ in prep., Stroud et al.\ in prep.]{Coletta2024, Mininni2024, Wells2024}.

%________________________________________________________________
%
\subsection{Studies of specific spectral lines\label{sec:specific-lines}}

The spectral setup of the ALMAGAL observations (cf.\ Fig.~\ref{fig:spectral-setup}) covers a number of molecular spectral lines that enable the determination of the gas temperature in the dense gas (e.g., H$_2$CO, CH$_3$CN), the study of molecular outflows and shocked gas (e.g., SiO), the kinematics of the dense gas (e.g., HNCO, CH$_3$CN), and the presence of a rich complex chemistry. Figure~\ref{fig:exampleLinesScience} shows the ALMAGAL field 865468 as an example of the scientific potential of some of these lines thanks to the achieved spatial resolution and sensitivity of the ALMAGAL images.

The left-column panels in Fig.~\ref{fig:exampleLinesScience} show the distribution of the dust continuum emission at 219~GHz at different spatial scales as observed with the different ALMA array configurations. Complementary to the continuum emission, the middle and right-column panels show the spectral line content of the region together with the spatial distribution of two molecular species: the SiO\,(5--4) line as a tracer of shocks and molecular outflows, and the CH$_3$CN\,(12$_\mathrm{K=3}$--11$_\mathrm{K=3}$) line as a tracer of kinematics of the dense gas. From top to bottom, the panels reveal how the high angular resolution of the final ALMAGAL images allow us to resolve multiple structures in this selected field. The low-resolution ($7\farcs3\times4\farcs7$) 7M-only image panels show an unresolved compact dust core with a very rich chemistry. The CH$_3$CN velocity image does not reveal special kinematic signatures in the dense gas. The SiO emission line is found to be extremely broad in the spectra, with spatially separated, although unresolved, blue and red-shifted outflow lobes (cf.\ top-right panel). In the middle row, the 7M+TM2 images (angular resolution of $2\farcs1\times1\farcs7$) allow us to resolve the object into a small cluster of objects, each one with a very different chemistry. Interestingly, the velocity field determined from the CH$_3$CN line shows a velocity gradient in the north-west to south-east direction, almost perpendicular to a very collimated, north-south molecular outflow seen in SiO. The bottom row shows the same region as seen in the 7M+TM2+TM1 images, at an angular resolution of $0\farcs36\times0\farcs34$. The cluster is better resolved in at least eight well-identified members. The high-resolution images also better resolve the velocity gradient seen in the CH$_3$CN line emission, to distinguish precession in the SiO molecular outflow, and to clearly detect an additional molecular outflow in a secondary, and less massive dense core to the north of the brightest one.

Taking again field 865468 as an example, the ALMAGAL science-ready products towards 1017 high-mass star-forming regions distributed throughout the Galaxy allow us, among multiple research studies, to compile a large catalogue of molecular outflows, to identify compact dense cores with associated velocity gradients that constitute excellent targets to search for and resolve protostellar disks around massive stars, and to study the chemistry in a large, statistically significant number of dense cores to derive lifetimes (e.g., based on fractions of source types) of the evolutionary phase associated with a rich chemistry. These are some of the science topics that are studied in forthcoming publications of the ALMAGAL consortium.

%________________________________________________________________
%
\subsection{Statistical characterization of the chemical richness\label{sec:chemical-richness}}

A study on the spectral line richness of the different compact sources identified in the ALMAGAL fields is the main topic of a forthcoming paper (S\'anchez-Monge et al., in prep), which is based on the first ALMAGAL catalogue of compact cores \citep{Coletta2024}. Here, we explore the potential of studying the chemical richness towards the 1017 ALMAGAL targets by studying the line content of bright regions detected in each field. We follow the approach introduced in \citet{Molinari2024} and define these bright regions, or regions of interest (hereafter RoI), based on the continuum emission maps at 219~GHz as regions with a closed contour level at an intensity of $8\,\sigma_\mathrm{rms}$, with $\sigma_\mathrm{rms}$ the rms noise level of the continuum images.  We have identified a total of 2757 RoI in 744 of the 1017 ALMAGAL fields. For each one of these regions we have extracted the spectrum integrated over the whole RoI and the spectrum towards the continuum emission peak within the RoI. The spectra are extracted for the two broad spectral windows spw0 and spw1, which cover the frequency ranges 216.98--218.86~GHz and 219.06--220.94~GHz, respectively. We note that these RoIs can cover large spatial extents and contain multiple compact sources, thus resembling regions of star-forming clusters.

We characterize the line richness by defining three complementary parameters: the fraction of bandwidth with line emission, the luminosity contained in the lines, and the fraction of the line luminosity to the total luminosity (i.e., line plus continuum). Figure~\ref{fig:spectra-RoIs_v01} shows the distribution of these parameters, distinguishing between spectral windows spw0 and spw1, for the spectra extracted towards the peak of the RoI (top panels) and integrated over the whole RoI (bottom panels). The fraction of bandwidth with line emission is determined as the frequency coverage of channels with emission above $5\,\sigma_\mathrm{rms}$ after continuum subtraction (i.e., we consider a channel to have contribution from lines if its emission is $5\,\sigma_\mathrm{rms}$ above the continuum level), divided by the total spectral window bandwidth (1.87~GHz). The left-column panels in Fig.~\ref{fig:spectra-RoIs_v01} show that for most RoIs the fraction bandwidth with line emission is about 0.5\% (or 10~MHz), with some cases having a line frequency coverage of about 1~GHz (out of the 1.875~GHz bandwidth of each spectral window). We compare this line richness with that measured towards the hot core population of the Sagittarius\,B2 complex \citep[hereafter Sgr\,B2; e.g.,][]{SanchezMonge2017, Moeller2023}. Assuming that each spectral line feature has a typical width of $\approx5$~km~s$^{-1}$ \citep[e.g.,][]{Mininni2020, Mininni2021, Liu2020}, we convert the line bandwidth fraction of the RoIs to a density of lines per GHz. The common line emission range of 10~MHz corresponds to $\approx1.4$~lines per GHz, while the line richest RoIs with line-emission ranges of 1~GHz would result in a density of 140~lines per GHz. These line densities are comparable to those determined in the Sgr\,B2 region, where the richest hot cores have about 100~lines per GHz (cf.\ Table~3 in \citealt{SanchezMonge2017}).

The luminosity contained in spectral lines, $L_\mathrm{lines}$, is derived as
\begin{equation}
L_\mathrm{lines} = 4 \pi D^2 \sum_{S_\nu>5\sigma}{S_\nu \Delta v},
\end{equation}
where $D$ is the distance to the different ALMAGAL fields \citep[][Benedettini et al.\ in prep.]{Molinari2024}, and the last term is the sum of the product of the intensity times the channel width for all those channels with an intensity above $5\,\sigma$ (above the continuum level), with $\sigma$ being the rms of the extracted spectrum. The middle-column panels of Fig.~\ref{fig:spectra-RoIs_v01} show the distribution of the line luminosities. Most of the RoIs have very low line luminosities, well below a thousandth of the solar luminosity. The right-column panels show the fraction of line luminosity to total (continuum plus lines) luminosity. Most of the RoIs have about 1 to 10\% of their total luminosity coming from spectral lines, with some particular cases having more than 30\% of the luminosity originated in lines. These large line luminosity fractions are similar to the richest hot cores in the Sgr\,B2 region, which have values in the range 30--40\%.

A comparison between the top and bottom panels of Fig.~\ref{fig:spectra-RoIs_v01} reveals that there are more cases of extreme line rich spectra towards the RoI peak positions compared to the RoI integrated emission (cf.\ higher end of the frequency range and line luminosity fraction distributions). This suggests that the line richness is likely to be higher towards specific locations within the RoIs (e.g., dense and compact chemically rich hot cores embedded within the RoIs). As such, the results on the line richness towards the whole extent of the RoIs should be taken as lower limits.

In Fig.~\ref{fig:spectra-RoIs_v02} we compare the three indicators of line richness with different physical properties such as the RoI area size, the clump luminosity to mass ratio ($L_\mathrm{clump}/M_\mathrm{clump}$) that is used as an evolutionary indicator \citep[e.g.,][]{Molinari2008}, and the clump dust temperature $T_\mathrm{dust}$. The clump parameters (i.e., luminosity, mass and temperature) are determined from the \textit{Herschel} Hi-GAL measurements at wavelengths 70~$\mu$m to 500~$\mu$m \citep[see][for more details]{Elia2017, Elia2021, Molinari2024}. The three line-richness indicators increase with the RoI size. This is expected if larger RoIs include extended line emission (e.g., from species such as $^{13}$CO, C$^{18}$O and H$_2$CO) together with multiple compact sources with spectral line emission at slightly different velocities. The two other relations also reveal a correlation of the line richness with the evolutionary indicator $L_\mathrm{clump}/M_\mathrm{clump}$ of the clump and its dust temperature. These results are consistent with previous observations of high-mass star forming regions at clump scales ($\approx0.1$~pc; \citealt{Gerner2014, Giannetti2017, Mininni2021}). More evolved regions are expected to host already developed (proto)stars that heat up the surrounding gas either thermally (e.g., radiation feedback) or through shocks (e.g., outflows), enabling new chemical reactions and releasing molecular compounds, originally locked in the ice dust mantles, to the gas phase \citep[e.g.,][]{Palau2017, Gieser2019, Sabatini2021, Garrod2022}. A detailed characterization of the chemical content and line richness of the compact cores identified in the 1017 ALMAGAL fields (e.g., S\'anchez-Monge et al., in prep) enables the study of these correlations beyond the clump scales going down to the scales of individual cores at $\approx1000$~au, thus setting observational constraints on the chemical evolution of dense cores within stellar clusters.

%________________________________________________________________
%
\section{Summary\label{sec:summary}}

We have presented the ALMAGAL observational strategy, processing workflow and data reduction pipeline that was developed and implemented to generate the first set of science-ready products of the ALMAGAL Large Program. 

ALMAGAL has observed 1017 high-mass star-forming regions distributed throughout the Galaxy. The ALMAGAL sample was divided in two groups to account for different distances from the Sun: the ``Near sample'' contains 538 fields with distances mostly $<4.7$~kpc, and the ``Far sample'' contains 479 fields with distances mostly $>4.7$~kpc. As presented in \citet{Molinari2024}, new distances to the ALMAGAL fields have been re-derived affecting some of the ``near'' and ``far'' sources. Each ALMAGAL field was observed using three ALMA array configurations, two of them from the main 12m array and the third one corresponding to the ACA 7m array. This results in observations sensitive to a common range of spatial scales (from $\approx1\,000$~au up to $\approx0.1$~pc) for all the targets. The spectral setup has been designed to be sensitive to the continuum emission at 219~GHz (or 1.38~mm), and to a series of molecular species and spectral lines that are used as tracers of gas temperature (e.g., H$_2$CO, CH$_3$CN), dense gas (e.g., C$^{18}$O, HNCO), and shocked gas (e.g., SiO, SO), as well as deuterated (e.g., DCN) and several complex molecular species.

The ALMAGAL processing workflow and pipeline were designed and used to perform several essential steps in the production of the final products. Relevant steps, customized for the ALMAGAL project, are an improved determination of the continuum level in line-rich sources (now implemented in the ALMA pipeline, see \citealt{Hunter2023}), the implementation of automatic procedures for self-calibration that improves the signal-to-noise by up to a factor five in about 15\% of the fields (which are now also implemented in the ALMA pipeline\footnote{More information can be found in the ALMA Science Pipeline User's Guide for Release 2024.1.0.8, see \citet{ALMAPipelineTeam2024} available at \url{https://zenodo.org/records/14502284}.}), and an iterative strategy of imaging when combining the data for the three different telescope configurations. The final products are a set of images and spectral cubes for each 1017 ALMAGAL field, including individual-array and combined-array products. The fully combined 7M+TM2+TM1 images have a typical angular resolution of $0\farcs47\times0\farcs38$ for ``Near sample'' fields, and $0\farcs28\times0\farcs19$ for ``Far sample'' fields. The corresponding spatial resolution is in the range 800 to 2\,000~au. The continuum sensitivity is $\approx0.05$~mJy~beam$^{-1}$ for most of the fields, corresponding to a gas and dust mass sensitivity in the range 0.02--0.07~$M_\odot$ (assuming a temperature of 20 K, a dust opacity of 0.899~cm$^{2}$~g$^{-1}$ at 1.3~mm, and a gas-to-dust ratio of 100).

We have performed a preliminary analysis on the spectral line content of the ALMAGAL fields to demonstrate its scientific potential and the possible use of the chemical and line richness when studying high-mass star-forming regions in different evolutionary stages. We found that the line richness, determined with three different parameters, is correlated with the luminosity to mass ratio of the hosting clump and with the dust temperature, which are indicators of the evolutionary stage.

Overall, the ALMAGAL dataset serves as the basis of several ongoing and planned studies on the evolutionary study of high-mass star cluster formation throughout the Galaxy.

\begin{acknowledgements}
We thank the referee, Adele Plunkett, for her comments that have helped to improve the clarity of the manuscript.
A.S.-M.\ is grateful to Adam Ginsburg for multiple discussions on how to deal with the adventure of calibrating and imaging large amounts of radio interferometry data.
A.S.-M.\ acknowledges support from the RyC2021-032892-I grant funded by MCIN/AEI/10.13039/501100011033 and by the European Union `Next GenerationEU’/PRTR, as well as the program Unidad de Excelencia María de Maeztu CEX2020-001058-M, and support from the PID2023-146675NB-I00 (MCI-AEI-FEDER, UE).
M.T.B.\ acknowledges financial support through the INAF Large Grant The role of MAGnetic fields in MAssive star formation (MAGMA). 
G.A.F.\ gratefully acknowledges the Deutsche Forschungsgemeinschaft (DFG) for funding through SFB~1601 ``Habitats of massive stars across cosmic time'' (sub-project B2), the University of Cologne and its Global Faculty Programme. The UK ALMA Regional Centre (ARC) Node is supported by the Science and Technology Facilities Council [grant numbers ST/T001488/1 and ST/Y004108/1].
R.S.K.\ acknowledges financial support from the German Excellence Strategy via the Heidelberg Cluster of Excellence (EXC 2181-390900948) `STRUCTURES', and from the German Ministry for Economic Affairs and Climate Action in project `MAINN' (funding ID 50OO2206). R.S.K.\ is grateful for computing resources provided by the Ministry of Science, Research and the Arts (MWK) of the State of Baden-W\"{u}rttemberg through bwHPC and the German Science Foundation (DFG) through grants INST 35/1134-1 FUGG and 35/1597-1 FUGG, and also for data storage at SDS@hd funded through grants INST 35/1314-1 FUGG and INST 35/1503-1 FUGG. R.S.K.\ also thanks the Harvard-Smithsonian Center for Astrophysics and the Radcliffe Institute for Advanced Studies for their hospitality during his sabbatical, and the 2024/25 Class of Radcliffe Fellows for highly interesting and stimulating discussions.
D.C.L.\ acknowledges that part of this research was carried out at the Jet Propulsion Laboratory, California Institute of Technology, under a contract with the National Aeronautics and Space Administration (80NM0018D0004).
T.L.\ acknowledges the supports by the National Key R\&D
Program of China (No.\ 2022YFA1603101), National Natural Science Foundation of China (NSFC) through grants No.\ 12073061 and No.\ 12122307, the international partnership program of Chinese Academy of Sciences through grant No.\ 114231KYSB20200009, and the Tianchi Talent Program of Xinjiang Uygur Autonomous Region.
T.M.\ acknowledges financial support from BMBF/Verbundforschung through the projects ALMA-ARC 05A14PK1 and ALMA-ARC 05A20PK1.
P.S.\ was partially supported by a Grant-in-Aid for Scientific Research (KAKENHI Number JP22H01271 and JP23H01221) of JSPS. P.S.\ was also supported by Yoshinori Ohsumi Fund (Yoshinori Ohsumi Award for Fundamental Research).
T.Z.\ acknowledges financial support of the Bonn–Cologne Graduate School, which is funded through the German Excellence Initiative as well as funding by the Deutsche Forschungsgemeinschaft (DFG) via the Collaborative Research Center SFB 956 `High-mass Star Formation' (subproject A6); and China Postdoctoral Science Foundation No.\ 2023TQ0330.
L.B.\ gratefully acknowledges support by the ANID BASAL project FB210003.
R.K.\ acknowledges financial support via the Heisenberg Research Grant funded by the Deutsche Forschungsgemeinschaft (DFG, German Research Foundation) under grant no.~KU 2849/9, project no.~445783058
The teams at INAF-IAPS and Heidelberg University acknowledge financial support from the European Research Council via the ERC Synergy Grant ECOGAL (project ID 855130).
The authors gratefully acknowledge the Gauss Centre for Supercomputing e.V.\ (www.gauss-centre.eu) for funding this project by providing computing time through the John von Neumann Institute for Computing (NIC) on the GCS Supercomputer JUWELS at J\"ulich Supercomputing Centre (JSC). Project codes: 17905, 20946, 23984 and 26560. \\

\indent
This research made significant use of \texttt{Astropy}, a community-developed core Python package and an ecosystem of tools and resources for astronomy \citep{Astropy2013, Astropy2018, Astropy2022}, NumPy \citep{Harris2020} and matplotlib \citep{Hunter2007}. All figures in this publication are done in Python. \\

\indent
This paper makes use of the following ALMA data: ADS/JAO.ALMA\#2019.1.00195.L ALMA is a partnership of ESO (representing its member states), NSF (USA) and NINS (Japan), together with NRC (Canada), MOST and ASIAA (Taiwan), and KASI (Republic of Korea), in cooperation with the Republic of Chile. The Joint ALMA Observatory is operated by ESO, AUI/NRAO and NAOJ. The National Radio Astronomy Observatory is a facility of the National Science Foundation operated under cooperative agreement by Associated Universities, Inc.

\end{acknowledgements}


%________________________________________________________________
%
\begin{appendix}

%________________________________________________________________
%
\section{JvM correction factor on ALMAGAL\label{app:JvMcorrection}}

The process of CLEANing used to process interferometric data results in images where it is not straightforward to measure correct fluxes. Any CLEANed image consists of the sum of two images: one that contains the restored CLEAN components and another that contains the residuals. The difficulty to measure correct fluxes arises because these two images have different units: the former has units of Jansky per clean beam, and the latter has units of Jansky per dirty beam.

As first noted by \citet{JorsaterVanMoorsel1995}, it is possible to calculate correct fluxes after converting the units of the residual image to Jansky per clean beam. In general, the flux in the residual image is overestimated by a factor $\epsilon=\Omega_\mathrm{dirty}/\Omega_\mathrm{clean}$, where $\Omega_\mathrm{dirty}$ and $\Omega_\mathrm{clean}$ are the dirty and clean beam sizes\footnote{In short, the ``dirty beam'' (or PSF, point-spread function) is the response of the interferometer to an impulse sky brightness distribution (i.e., a Dirac $\delta$ function). Thus, the dirty beam is directly determined by the visibility-plane (hereafter $uv$-plane) coverage, which in turn is related to the number of antennas included in the array and the time span of the interferometric observations. The dirty beam usually consists of a well-defined Gaussian-like central feature surrounded by a sidelobe pattern. The amplitude of this sidelobe pattern decreases and becomes less relevant by increasing the $uv$-plane coverage. On the other hand, the ``clean beam'' (or synthesized-beam) is obtained by fitting a Gaussian to the central Gaussian-like feature of the dirty beam. As a result, the clean beam is a perfect Gaussian with no sidelobe pattern around.}, respectively. The $\epsilon$ factor (also known in literature as the JvM-factor) can be determined from the ratio of volumes of the beams as recently presented by \citet[][see also e.g., \citealt{Walter2008, Leroy2021, Cunningham2023}]{Czekala2021}. Based on this, final images with units of only Jansky per clean beam can be obtained by adding the restored clean components (i.e., model image convolved with the clean beam) and the residuals scaled by the $\epsilon$ factor.

%-------------------------------- Figure --------------------------------
\begin{figure}[h!]
\centering
\includegraphics[width=0.95\columnwidth]{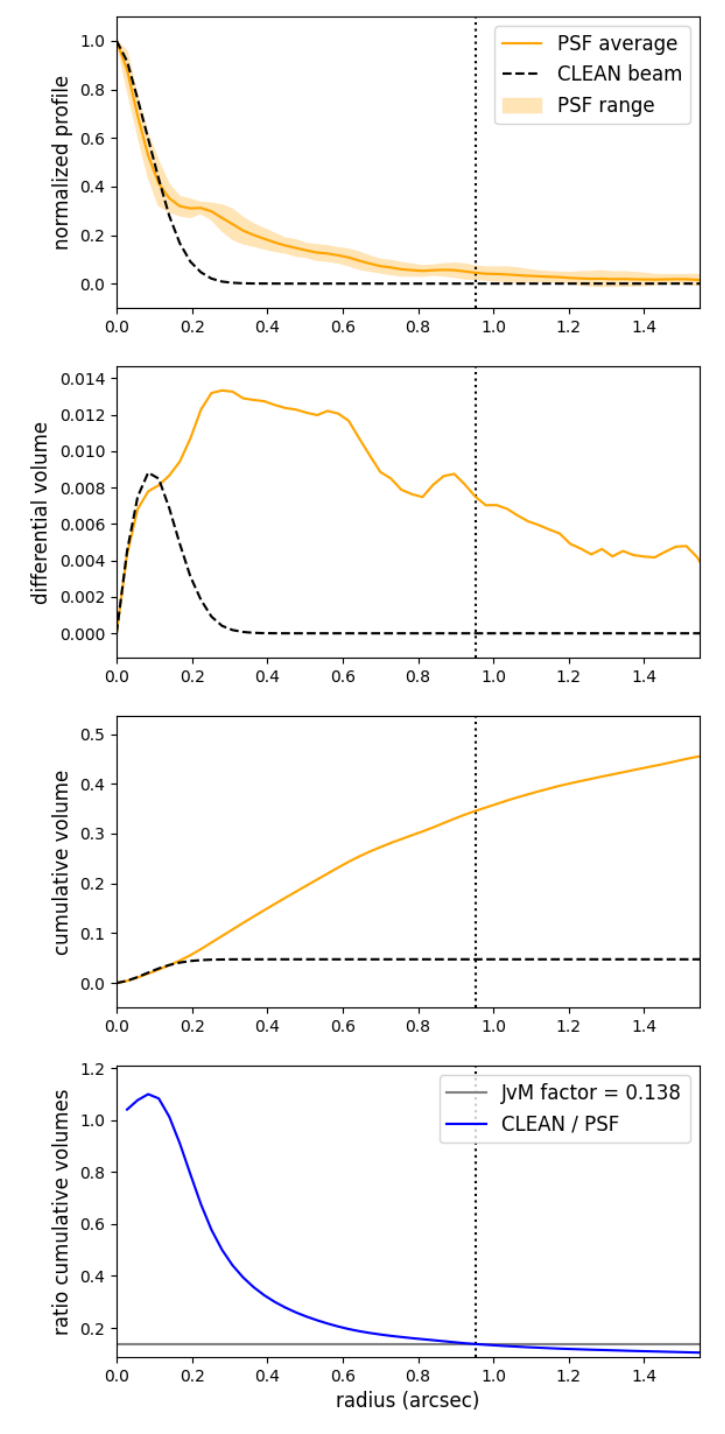}
\caption{Set of images describing the procedure used to determine the $\epsilon$ correction factor (or JvM-factor) based on \citet{Czekala2021}. The plots refer to the 7M+TM2+TM1 continuum data of the ALMAGAL field 644284 (also labelled AG285.2633$-$0.0501 in \citealt{Molinari2024}). \textit{Top panel}: normalized, elliptically averaged profile of the dirty beam (or PSF, in orange) and clean beam (black dashed line). The vertical dotted line marks the ``zero-crossing'' point of the dirty beam. Note that the PSF range of the elliptically averaged dirty beam is considered to determine the crossing point, thus resulting in a more conservative correction factor. \textit{Second and third panels}: differential and cumulative volumes of the dirty and clean beams (see \citealt{Czekala2021} for details in the calculation). \textit{Bottom panel}: ratio of the clean and dirty beam volumes. The value of this ratio at the distance of the dirty-beam zero-crossing point provides the $\epsilon$ (or JvM) correction factor (0.138 in this example).}
\label{fig:644284_JvM-correction}
\end{figure}
%------------------------------------------------------------------------

Figure~\ref{fig:644284_JvM-correction} shows the process of determination of the $\epsilon$ factor (or JvM factor) for one of the ALMAGAL fields, based on the strategy described in \citet{Czekala2021}. The ALMAGAL consortium has determined the $\epsilon$ correction factors and produced JvM-corrected images for the continuum emission maps of all fields. The derived $\epsilon$ correction factors for the combined 7M+TM2+TM1 continuum images cover a range between 0.1 and 0.3. We note that these values are smaller than the $\epsilon$ correction factors determined in other ALMA programs \citep[see e.g.,][]{Czekala2021, Cunningham2023} that typically range between 0.6 and 0.7.

Despite having generated JvM-corrected continuum images, the ALMAGAL consortium does not consider them yet suitable for scientific use due to different aspects. One potential issue of applying the JvM correction to the ALMAGAL data is the snapshot nature of the observations, together with the combination of three different array configurations. The short integration times (up to 2--4~minutes per field) result in a limited $uv$-plane coverage, and a dirty beam image with a still-significant sidelobe pattern. The sidelobe pattern is present in the dirty beams of each individual array configuration, which when combined may result in a larger plateau or shelf in the dirty beam compared to the clean beam (cf.\ top panel in Fig.~\ref{fig:644284_JvM-correction}). The presence of this larger plateau results in smaller $\epsilon$ correction factors, as also noted by \citet{Cunningham2023}. A detailed study on the effects of multi-array short-integration observations on the $\epsilon$ factor is beyond the goal of the first ALMAGAL data release.

We also explored the rms noise level that is measured in the JvM-corrected images. After applying the $\epsilon$ factor, the residuals are scaled down by a factor of 0.1--0.3, which result in measured rms noise levels of $\lesssim0.01$~mJy~beam$^{-1}$ (i.e., ten times better than the theoretical expected rms noise level). As discussed by \citet{Cunningham2023}, a more conservative or realistic estimate of the rms noise level can be obtained after multiplying the noise value derived in the JvM-corrected images by the factor $1/\epsilon$. This is equivalent to using the Jansky per dirty beam units (i.e., not-scaled residual image) when determining the rms noise level. \citet{Walter2008} provide a different strategy to avoid unrealistic low rms noise levels. This consists in blanking out those regions in the image that do not contain emission, resulting in images that do not contain regions with no emission (i.e., noise). A complementary strategy would consist in applying the $\epsilon$ correction factor only to those regions in the residual that still contain real emission that has not been CLEANed. Finally, we note that, based on the CLEANing strategy described in Sect.~\ref{sec:joint-deconvolution} (see also Fig.~\ref{fig:imaging-strategy}), the residuals of the ALMAGAL images do not contain significant emission that would need to be scaled down with the $\epsilon$ correction factor.

In summary, we consider that the JvM-correction currently tested for the ALMAGAL dataset is not adequate due to the nature of the observations (i.e., multi-array, short-integration observations). Based on this, the first data release of ALMAGAL science products do not include the JvM correction.

%________________________________________________________________
%
\section{Data quality assessment\label{app:QA}}

%________________________________________________________________
%
\subsection{Quality assessment of the data combination for cubes\label{app:QAcombinationCubes}}

\begin{landscape}
\begin{figure*}[h!]
\centering
\includegraphics[height=0.61\textwidth]{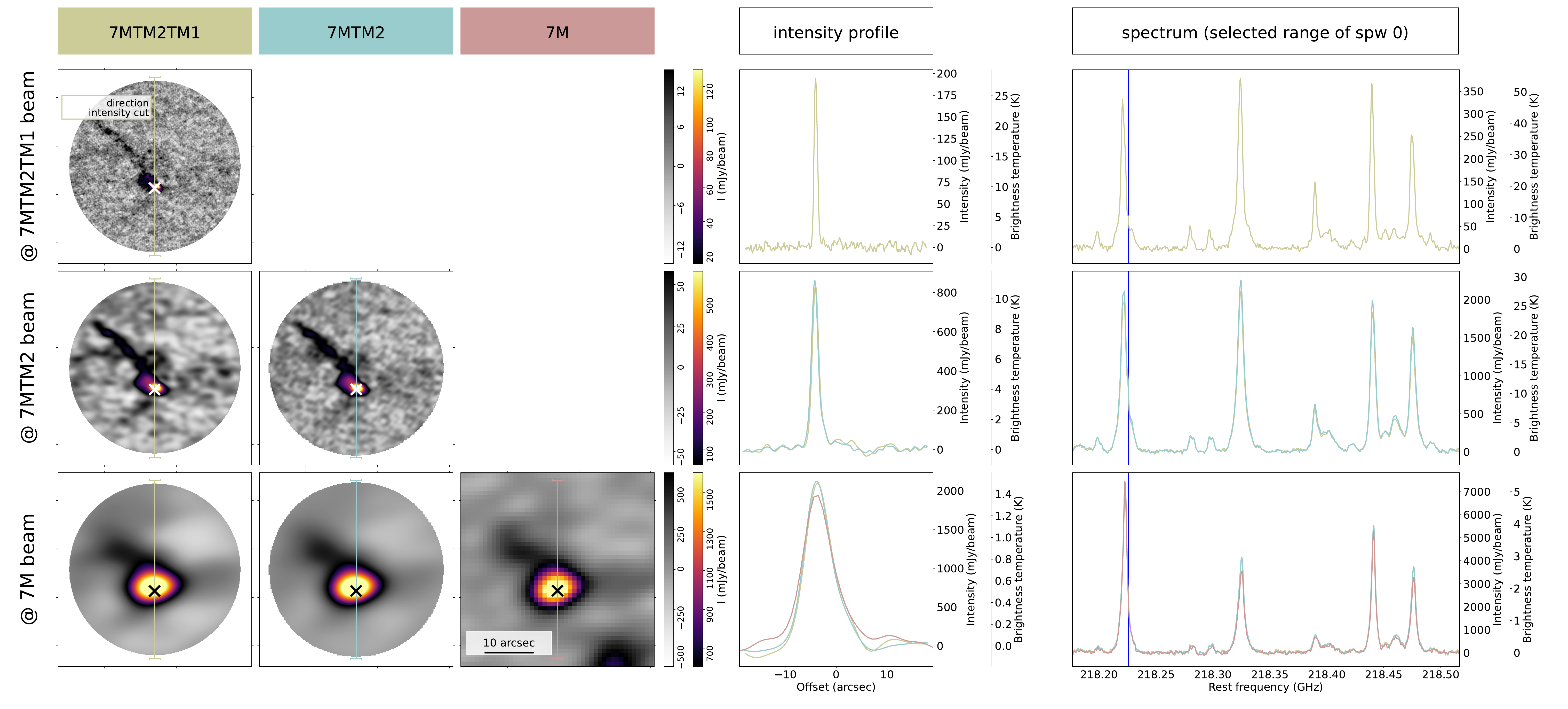}
\caption{Comparison of the ALMAGAL images for field 101899 (also labelled AG023.0108$-$0.4102 in \citealt{Molinari2024}) to explore the quality of the data combination in data cubes. The figure is divided in three main sections: left columns (with titles 7MTM2TM1, 7MTM2 and 7M), middle column  (with title ``intensity profile''), and right column (with title ``spectrum'').
\textit{Left columns}: The images shown in the panels of the left columns correspond to the emission of a specific channel in the spw0 data cube. The channel is highlighted with a blue vertical solid line in the spectra shown in the right column panels. The images have been convolved to different angular resolutions in the different rows: The top row shows images convolved to the 7M+TM2+TM1 beam ($0\farcs46\times0\farcs39$), the middle row corresponds to images convolved to the 7M+TM2 beam ($1\farcs65\times1\farcs20$), and the bottom row shows the images convolved to the 7M-only beam ($7\farcs9\times4\farcs6$). The first three columns correspond to the images of the arrays 7M+TM2+TM1, 7M+TM2 and 7M, from left to right, respectively.
\textit{Middle column}: The panels in this column show the intensity profiles (both in intensity, mJy~beam$^{-1}$, and in brightness temperature, K) of the vertical intensity cuts indicated in the different images.
\textit{Right column}: The spectra shown in these panels are extracted towards the position marked with a white/black cross in the images shown in the left columns panels. They depict a portion of the frequency coverage of the spectral window spw0.
}
\label{fig:smoothing-beams_spw0}
\end{figure*}
\end{landscape}

\begin{landscape}
\begin{figure*}[h!]
\centering
\includegraphics[height=0.61\textwidth]{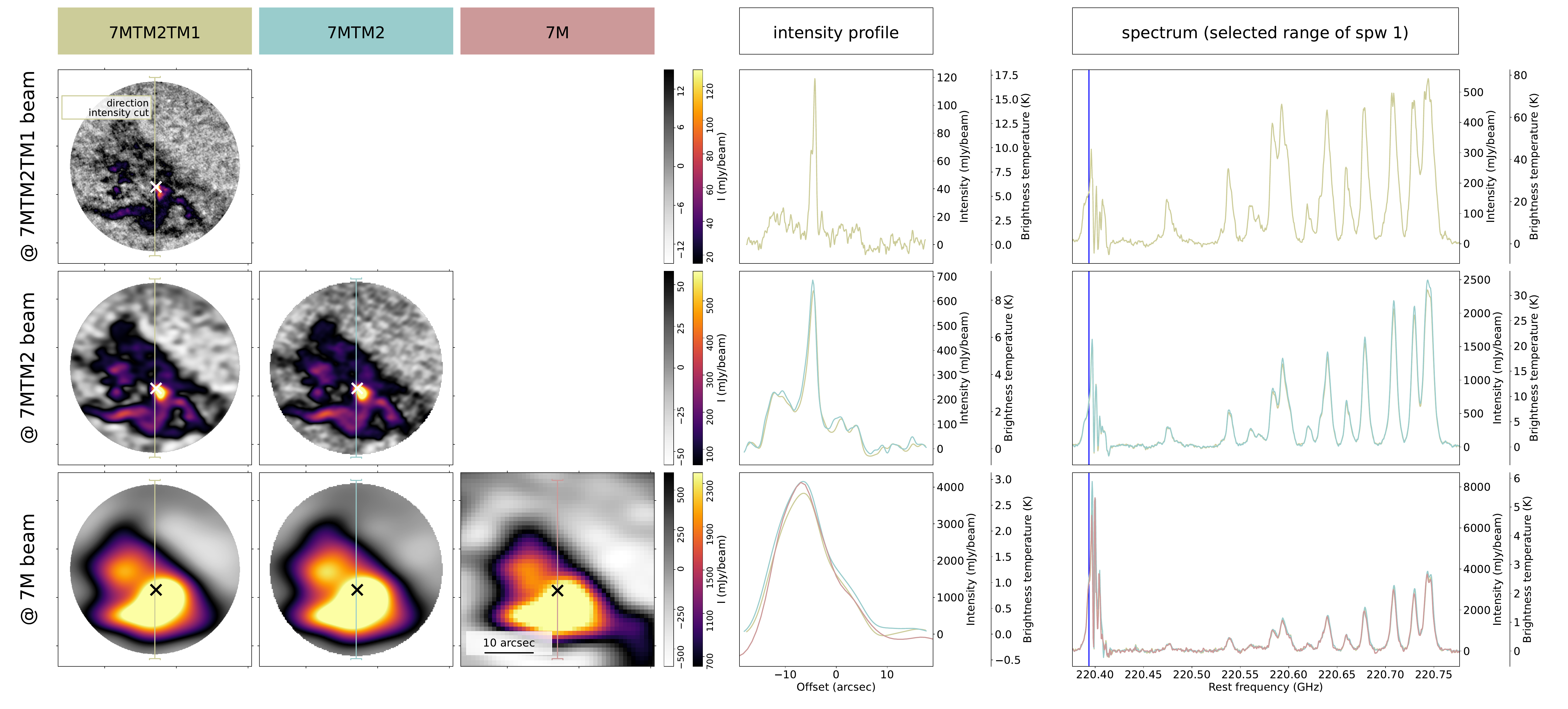}
\caption{Same as Fig.~\ref{fig:smoothing-beams_spw0} for a different channel of the spectral window spw1. The selected channel is highlighted in the spectra on the rightmost column panels. The images have been convolved to different angular resolutions in the different rows: The top row shows images convolved to the 7M+TM2+TM1 beam ($0\farcs46\times0\farcs39$), the middle row corresponds to images convolved to the 7M+TM2 beam ($1\farcs64\times1\farcs18$), and the bottom row shows the images convolved to the 7M-only beam ($7\farcs9\times4\farcs5$)}
\label{fig:smoothing-beams_spw1}
\end{figure*} 
\end{landscape}

In Figs.~\ref{fig:smoothing-beams_spw0} and \ref{fig:smoothing-beams_spw1}, we evaluate the results of the data combination process for spectral line cubes by comparing the different array configuration images towards the ALMAGAL field 101899. This complements the results of the continuum emission shown in Fig.~\ref{fig:smoothing-beams}. The left panels of these figures show the 7M+TM2+TM1, 7M+TM2 and 7M-alone images for a specific channel (highlighted in the spectrum shown in the right-most panels). We have selected channels with potentially complicated emission structure. In Fig.~\ref{fig:smoothing-beams_spw0}, we have selected a channel with relatively extended and faint emission associated with the O$^{13}$CS\,(18--17) spectral line, while in Fig.~\ref{fig:smoothing-beams_spw1} we have selected a channel with very extended and partially filtered-out emission associated with the $^{13}$CO\,(2--1) spectral line feature.

Similar to Fig.~\ref{fig:smoothing-beams}, the combined images are convolved to the synthesize beams of lower-resolution images. The spatial distribution of the emission, as well as the intensity cuts, confirm that the combined images recover the same structure and intensity. In addition to the spatial distribution, the right-most panels in Figs.~\ref{fig:smoothing-beams_spw0} and \ref{fig:smoothing-beams_spw1} show the spectra extracted towards the position highlighted with a cross in the left panels. The agreement of the intensities across different frequency channels for the different combined images confirms that the combination process is also successful for the ALMAGAL data cubes.

\subsubsection{Resolution, intensity and noise levels in final products\label{sec:QAadditional}}

Table~\ref{tab:almagal_observations} lists several parameters of the final ALMAGAL images, including single-array and combined products. For each product we list the major and minor beam axis, together with the beam elongation. We also list the intensity, noise level and peak dynamic range for the continuum and spectral window images. The median and the 16$^\mathrm{th}$ and 84$^\mathrm{th}$ percentiles are highlighted for an easier read.

\begin{figure*}[h!]
\centering
\includegraphics[width=0.92\textwidth]{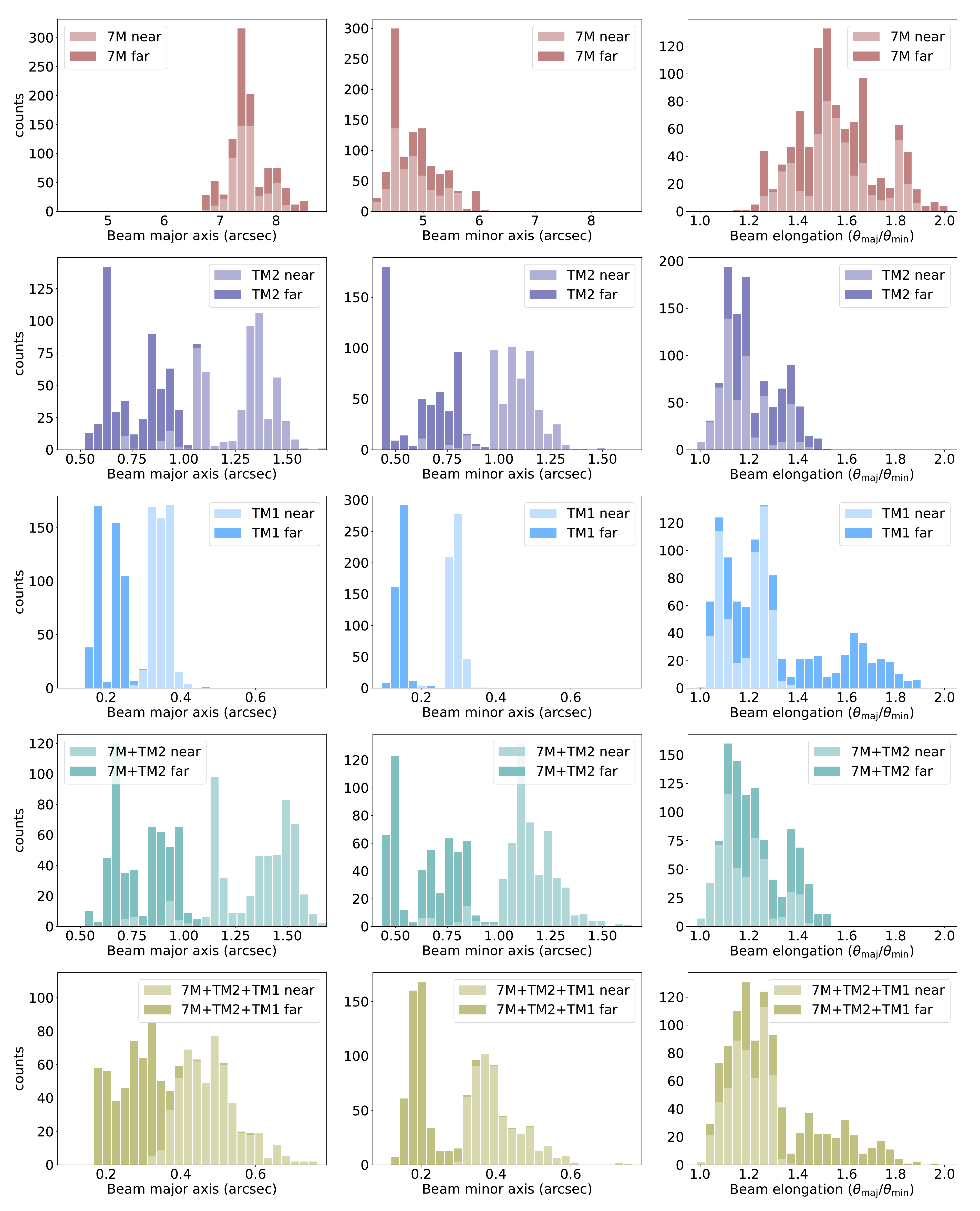}\\
\caption{Distribution of the beam properties for the ALMAGAL images. The $y$-axis in all panels depicts the number of images. From top to bottom, each row corresponds to the images for the 7M, TM2, TM1, 7M+TM2, and 7M+TM2+TM1 array configurations. The columns show from left to right, the beam major axis ($\theta_\mathrm{maj}$), the beam minor axis ($\theta_\mathrm{min}$), and the beam elongation ($\theta_\mathrm{maj}/\theta_\mathrm{min}$). Light and dark histograms correspond to the ``near'' and ``far'' samples, respectively.}
\label{fig:beam-properties}
\end{figure*}

\begin{figure*}[h!]
\centering
\includegraphics[width=0.92\textwidth]{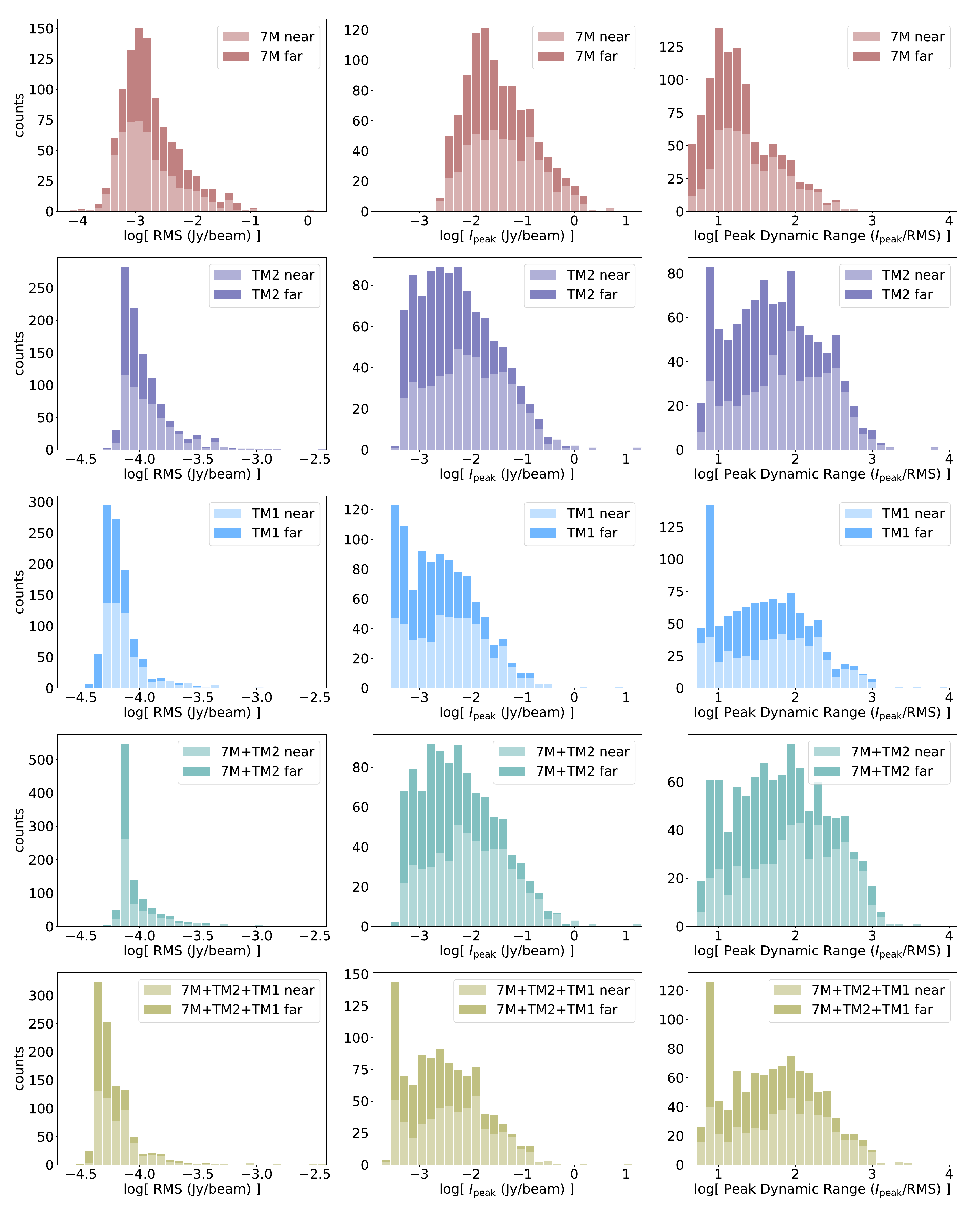}\\
\caption{Distribution of the noise level (left column), peak intensity (central column), and dynamic range (right column) for the continuum ALMAGAL images. The $y$-axis in all panels depicts the number of images. From top to bottom, each row corresponds to the images for the 7M, TM2, TM1, 7M+TM2, and 7M+TM2+TM1 array configurations. The dynamic range (or signal-to-noise ratio) is determined as the ratio between the peak intensity and the noise level, with the noise level derived as the median absolute deviation (MAD) of the residual image (see Sect.~\ref{sec:QAintensity}).}
\label{fig:continuum-properties}
\end{figure*}

Figures~\ref{fig:beam-properties} and \ref{fig:continuum-properties} complement the results listed in Table~\ref{tab:almagal_observations}. The distribution of the synthesized beam parameters (i.e., major and minor axis and beam elongation) are shown in Fig.~\ref{fig:beam-properties}, while the distributions of the rms noise level, peak intensity and dynamic range for the continuum images are shown in Fig.~\ref{fig:continuum-properties}. See Section~\ref{sec:QAbeam} and \ref{sec:QAintensity} for more details.

\onecolumn
%-------------------------------- Figure --------------------------------
%------------------------------------------------------------------------

%-------------------------------- Figure --------------------------------
%------------------------------------------------------------------------
\twocolumn

%--------------------------------- Table --------------------------------
\begin{sidewaystable*}[h!]
\centering
\caption{\label{tab:almagal_observations}Image parameters of the ALMAGAL science-ready products}
\begin{tabular}{P{5cm}P{5.85cm}P{5.85cm}P{5.85cm}}
\hline\hline \noalign{\smallskip} 

&\multicolumn{3}{c}{Parameter values for different ALMAGAL arrays}
\\
Parameter
&\multicolumn{3}{c}{(minimum --- 16$^\mathrm{th}$~percentile --- median --- 84$^\mathrm{th}$~percentile --- maximum)}
\\
\hline\hline \noalign{\smallskip} 

& 7M
& TM2
& TM1
\\
\hline \noalign{\smallskip}
% Major axis
\multicolumn{1}{l}{Major axis [$\arcsec$] --- near sample}
 & \multicolumn{1}{c}{6.79 --- {\bf 7.26} --- {\bf 7.45} --- {\bf 7.80} --- 8.15}
 & \multicolumn{1}{c}{0.71 --- {\bf 1.07} --- {\bf 1.31} --- {\bf 1.44} --- 1.70}
 & \multicolumn{1}{c}{0.27 --- {\bf 0.32} --- {\bf 0.35} --- {\bf 0.36} --- 0.42} \\
\multicolumn{1}{l}{\phantom{Major axis [$\arcsec$]} --- far sample}
 & \multicolumn{1}{c}{6.73 --- {\bf 7.12} --- {\bf 7.43} --- {\bf 8.02} --- 8.77}
 & \multicolumn{1}{c}{0.55 --- {\bf 0.62} --- {\bf 0.74} --- {\bf 0.91} --- 1.06}
 & \multicolumn{1}{c}{0.15 --- {\bf 0.17} --- {\bf 0.22} --- {\bf 0.24} --- 0.48} \\
% Minor axis
\multicolumn{1}{l}{Minor axis [$\arcsec$] --- near sample}
 & \multicolumn{1}{c}{4.18 --- {\bf 4.45} --- {\bf 4.76} --- {\bf 5.28} --- 5.88}
 & \multicolumn{1}{c}{0.63 --- {\bf 0.97} --- {\bf 1.07} --- {\bf 1.18} --- 1.48}
 & \multicolumn{1}{c}{0.20 --- {\bf 0.28} --- {\bf 0.29} --- {\bf 0.31} --- 0.36} \\
\multicolumn{1}{l}{\phantom{Minor axis [$\arcsec$]} --- far sample}
 & \multicolumn{1}{c}{4.16 --- {\bf 4.45} --- {\bf 4.82} --- {\bf 5.34} --- 6.05}
 & \multicolumn{1}{c}{0.44 --- {\bf 0.46} --- {\bf 0.64} --- {\bf 0.80} --- 0.93}
 & \multicolumn{1}{c}{0.10 --- {\bf 0.14} --- {\bf 0.15} --- {\bf 0.15} --- 0.39} \\
% Beam elongation
\multicolumn{1}{l}{Beam elongation --- near sample}
 & \multicolumn{1}{c}{1.27 --- {\bf 1.39} --- {\bf 1.54} --- {\bf 1.76} --- 1.90}
 & \multicolumn{1}{c}{1.01 --- {\bf 1.09} --- {\bf 1.15} --- {\bf 1.26} --- 1.45}
 & \multicolumn{1}{c}{1.02 --- {\bf 1.07} --- {\bf 1.22} --- {\bf 1.26} --- 1.38} \\
\multicolumn{1}{l}{\phantom{Beam elongation} --- far sample}
 & \multicolumn{1}{c}{1.16 --- {\bf 1.40} --- {\bf 1.51} --- {\bf 1.74} --- 2.07}
 & \multicolumn{1}{c}{1.05 --- {\bf 1.14} --- {\bf 1.21} --- {\bf 1.37} --- 1.50}
 & \multicolumn{1}{c}{1.04 --- {\bf 1.13} --- {\bf 1.42} --- {\bf 1.68} --- 1.90} \\
\hline \noalign{\smallskip}
% Continuum noise level
\multicolumn{1}{l}{Cont.\ noise level [mJy~beam$^{-1}$]}
 & \multicolumn{1}{c}{0.07 --- {\bf 0.64} --- {\bf 1.5} --- {\bf 6.1} --- 1300}
 & \multicolumn{1}{c}{0.05 --- {\bf 0.07} --- {\bf 0.10} --- {\bf 0.16} --- 3.4}
 & \multicolumn{1}{c}{0.03 --- {\bf 0.05} --- {\bf 0.06} --- {\bf 0.09} --- 1.1} \\
% Continuum intensity peak
\multicolumn{1}{l}{Cont.\ intensity peak [mJy~beam$^{-1}$]}
 & \multicolumn{1}{c}{2.4 --- {\bf 8.3} --- {\bf 27} --- {\bf 167} --- 27000}
 & \multicolumn{1}{c}{0.39 --- {\bf 0.96} --- {\bf 5.0} --- {\bf 35} --- 19000}
 & \multicolumn{1}{c}{0.29 --- {\bf 0.45} --- {\bf 2.3} --- {\bf 14} --- 9000} \\
% Continuum S/N
\multicolumn{1}{l}{Cont.\ peak dynamic range (S/N)}
 & \multicolumn{1}{c}{3.7 --- {\bf 7.4} --- {\bf 16} --- {\bf 57} --- 540}
 & \multicolumn{1}{c}{5.6 --- {\bf 12} --- {\bf 51} --- {\bf 230} --- 5000}
 & \multicolumn{1}{c}{5.9 --- {\bf 7.9} --- {\bf 38} --- {\bf 160} --- 8000} \\
\hline \noalign{\smallskip}
% SPW0 noise level
\multicolumn{1}{l}{spw~0 noise level [mJy~beam$^{-1}$]}
 & \multicolumn{1}{c}{18 --- {\bf 28} --- {\bf 35} --- {\bf 43} --- 85}
 & \multicolumn{1}{c}{2.7 --- {\bf 4.3} --- {\bf 4.7} --- {\bf 5.9} --- 8.9}
 & \multicolumn{1}{c}{1.3 --- {\bf 2.1} --- {\bf 2.2} --- {\bf 2.4} --- 3.1} \\
% SPW0 intensity peak
\multicolumn{1}{l}{spw~0 intensity peak [mJy~beam$^{-1}$]}
 & \multicolumn{1}{c}{180 --- {\bf 340} --- {\bf 1150} --- {\bf 4650} --- 28900}
 & \multicolumn{1}{c}{28 --- {\bf 47} --- {\bf 125} --- {\bf 520} --- 3700}
 & \multicolumn{1}{c}{14 --- {\bf 23} --- {\bf 46} --- {\bf 180} --- 1820} \\
% SPW0 S/N
\multicolumn{1}{l}{spw~0 peak dynamic range (S/N)}
 & \multicolumn{1}{c}{6.9 --- {\bf 8.9} --- {\bf 34} --- {\bf 135} --- 830}
 & \multicolumn{1}{c}{7.7 --- {\bf 9.0} --- {\bf 26} --- {\bf 100} --- 930}
 & \multicolumn{1}{c}{8.7 --- {\bf 11} --- {\bf 20} --- {\bf 78.9} --- 780} \\
\hline \noalign{\smallskip}
% SPW3 noise level
\multicolumn{1}{l}{spw~3 noise level [mJy~beam$^{-1}$]}
 & \multicolumn{1}{c}{32 --- {\bf 59} --- {\bf 74} --- {\bf 90} --- 125}
 & \multicolumn{1}{c}{5.4 --- {\bf 9.0} --- {\bf 9.7} --- {\bf 12} --- 18}
 & \multicolumn{1}{c}{4.8 --- {\bf 7.7} --- {\bf 8.4} --- {\bf 9.5} --- 11} \\
% SPW3 intensity peak
\multicolumn{1}{l}{spw~3 intensity peak [mJy~beam$^{-1}$]}
 & \multicolumn{1}{c}{530 --- {\bf 3900} --- {\bf 9200} --- {\bf 20400} --- 69700}
 & \multicolumn{1}{c}{79 --- {\bf 200} --- {\bf 470} --- {\bf 1300} --- 4200}
 & \multicolumn{1}{c}{94 --- {\bf 150} --- {\bf 190} --- {\bf 310} --- 820} \\
% SPW3 S/N
\multicolumn{1}{l}{spw~3 peak dynamic range (S/N)}
 & \multicolumn{1}{c}{7.8 --- {\bf 51} --- {\bf 120} --- {\bf 280} --- 950}
 & \multicolumn{1}{c}{8.4 --- {\bf 19} --- {\bf 45} --- {\bf 130} --- 380}
 & \multicolumn{1}{c}{17 --- {\bf 19} --- {\bf 22} --- {\bf 36} --- 87} \\
\end{tabular}
%-----------------------------------------------------------------------
\begin{tabular}{P{5cm}P{9cm}P{9cm}}
\hline\hline \noalign{\smallskip} 

& 7MTM2
& 7MTM2TM1
\\
\hline \noalign{\smallskip}
% Major axis
\multicolumn{1}{l}{Major axis [$\arcsec$] --- near sample}
 & \multicolumn{1}{c}{0.71 --- {\bf 1.15} --- {\bf 1.40} --- {\bf 1.53} --- 2.13}
 & \multicolumn{1}{c}{0.33 --- {\bf 0.40} --- {\bf 0.47} --- {\bf 0.55} --- 0.84} \\
\multicolumn{1}{l}{\phantom{Major axis [$\arcsec$]} --- far sample}
 & \multicolumn{1}{c}{0.55 --- {\bf 0.66} --- {\bf 0.78} --- {\bf 0.95} --- 1.07}
 & \multicolumn{1}{c}{0.17 --- {\bf 0.19} --- {\bf 0.28} --- {\bf 0.33} --- 0.59} \\
% Minor axis
\multicolumn{1}{l}{Minor axis [$\arcsec$] --- near sample}
 & \multicolumn{1}{c}{0.64 --- {\bf 1.05} --- {\bf 1.13} --- {\bf 1.27} --- 1.75}
 & \multicolumn{1}{c}{0.30 --- {\bf 0.34} --- {\bf 0.38} --- {\bf 0.48} --- 0.75} \\
\multicolumn{1}{l}{\phantom{Minor axis [$\arcsec$]} --- far sample}
 & \multicolumn{1}{c}{0.45 --- {\bf 0.48} --- {\bf 0.65} --- {\bf 0.81} --- 0.97}
 & \multicolumn{1}{c}{0.12 --- {\bf 0.17} --- {\bf 0.19} --- {\bf 0.22} --- 0.49} \\
% Beam elongation
\multicolumn{1}{l}{Beam elongation --- near sample}
 & \multicolumn{1}{c}{1.01 --- {\bf 1.08} --- {\bf 1.16} --- {\bf 1.26} --- 1.45}
 & \multicolumn{1}{c}{1.01 --- {\bf 1.11} --- {\bf 1.20} --- {\bf 1.27} --- 1.37} \\
\multicolumn{1}{l}{\phantom{Beam elongation} --- far sample}
 & \multicolumn{1}{c}{1.08 --- {\bf 1.15} --- {\bf 1.22} --- {\bf 1.41} --- 1.53}
 & \multicolumn{1}{c}{1.04 --- {\bf 1.15} --- {\bf 1.35} --- {\bf 1.61} --- 1.96} \\
\hline \noalign{\smallskip}
% Continuum noise level
\multicolumn{1}{l}{Cont.\ noise level [mJy~beam$^{-1}$]}
 & \multicolumn{1}{c}{0.05 --- {\bf 0.07} --- {\bf 0.07} --- {\bf 0.13} --- 5.55}
 & \multicolumn{1}{c}{0.03 --- {\bf 0.04} --- {\bf 0.05} --- {\bf 0.08} --- 3.46} \\
% Continuum intensity peak
\multicolumn{1}{l}{Cont.\ intensity peak [mJy~beam$^{-1}$]}
 & \multicolumn{1}{c}{0.41 --- {\bf 0.98} --- {\bf 5.3} --- {\bf 37} --- 19100}
 & \multicolumn{1}{c}{0.27 --- {\bf 0.44} --- {\bf 2.6} --- {\bf 17} --- 10500} \\
% Continuum S/N
\multicolumn{1}{l}{Cont.\ peak dynamic range (S/N)}
 & \multicolumn{1}{c}{6.1 --- {\bf 13} --- {\bf 67} --- {\bf 320} --- 3500}
 & \multicolumn{1}{c}{6.1 --- {\bf 9.2} --- {\bf 51} --- {\bf 200} --- 3000} \\
\hline \noalign{\smallskip}
% SPW0 noise level
\multicolumn{1}{l}{spw~0 noise level [mJy~beam$^{-1}$]}
 & \multicolumn{1}{c}{2.5 --- {\bf 3.9} --- {\bf 4.3} --- {\bf 5.5} --- 8.4}
 & \multicolumn{1}{c}{1.3 --- {\bf 2.1} --- {\bf 2.2} --- {\bf 2.4} --- 3.1} \\
% SPW0 intensity peak
\multicolumn{1}{l}{spw~0 intensity peak [mJy~beam$^{-1}$]}
 & \multicolumn{1}{c}{27 --- {\bf 48} --- {\bf 136} --- {\bf 590} --- 4900}
 & \multicolumn{1}{c}{14 --- {\bf 23} --- {\bf 46} --- {\bf 180} --- 1820} \\
% SPW0 S/N
\multicolumn{1}{l}{spw~0 peak dynamic range (S/N)}
 & \multicolumn{1}{c}{7.9 --- {\bf 10} --- {\bf 31} --- {\bf 120} --- 1100}
 & \multicolumn{1}{c}{8.7 --- {\bf 11} --- {\bf 20} --- {\bf 80} --- 780} \\
\hline \noalign{\smallskip}
% SPW3 noise level
\multicolumn{1}{l}{spw~3 noise level [mJy~beam$^{-1}$]}
 & \multicolumn{1}{c}{5.3 --- {\bf 8.4} --- {\bf 9.0} --- {\bf 11} --- 17}
 & \multicolumn{1}{c}{2.9 --- {\bf 4.3} --- {\bf 4.7} --- {\bf 5.1} --- 6.5} \\
% SPW3 intensity peak
\multicolumn{1}{l}{spw~3 intensity peak [mJy~beam$^{-1}$]}
 & \multicolumn{1}{c}{82 --- {\bf 230} --- {\bf 570} --- {\bf 1600} --- 12000}
 & \multicolumn{1}{c}{34 --- {\bf 67} --- {\bf 130} --- {\bf 360} --- 3100} \\
% SPW3 S/N
\multicolumn{1}{l}{spw~3 peak dynamic range (S/N)}
 & \multicolumn{1}{c}{9.2 --- {\bf 26} --- {\bf 57} --- {\bf 170} --- 1300}
 & \multicolumn{1}{c}{9.2 --- {\bf 15} --- {\bf 29} --- {\bf 74} --- 650} \\
\hline \noalign{\smallskip} 
\end{tabular}
\tablefoot{
These numbers correspond to the first ALMAGAL data release, produced as described in this paper. We provide numbers for the full set of 1017 sources, distinguishing between the near and far samples (see Sect.~\ref{sec:data-products}) for the beam parameters. The distributions of the different parameters are shown in Figs.~\ref{fig:beam-properties} and \ref{fig:continuum-properties}. We note that the values for the spectral windows ``spw~0'' and ``spw~3'' are good representations of the broad and narrow bandwidth images, respectively, with spectral resolutions of 1.4~km~s$^{-1}$ and 0.34~km~s$^{-1}$. Boldface numbers highlight the 16$^\mathrm{th}$ percentile, the median and the 84$^\mathrm{th}$ percentile values as the best to describe the properties of the full sample.}
\end{sidewaystable*}
%------------------------------------------------------------------------

%-------------------------------- Figure --------------------------------
%------------------------------------------------------------------------

%-------------------------------- Figure --------------------------------
%------------------------------------------------------------------------

\end{appendix}

\end{document}